\documentclass[a4paper]{amsart}
\usepackage{latexsym}
\usepackage{amsmath}
\usepackage{amssymb}
\usepackage{amsfonts}
\usepackage{esint}
\usepackage{graphicx}
\usepackage{accents}

\usepackage{color}
\definecolor{Blue}{rgb}{0.,0.,1.}
\definecolor{Red}{rgb}{1.,0.,0.}
\newcounter{smallarabics}
\newenvironment{arabicenumerate}
{\begin{list}{{\normalfont\textrm{(\arabic{smallarabics})}}}
  {\usecounter{smallarabics}\setlength{\itemindent}{0cm}
   \setlength{\leftmargin}{5ex}\setlength{\labelwidth}{4ex}
   \setlength{\topsep}{0.75\parsep}\setlength{\partopsep}{0ex}
   \setlength{\itemsep}{0ex}}}
{\end{list}}

\newcounter{smallroman}

\newcommand{\ben}{\begin{arabicenumerate}}  
\newcommand{\een}{\end{arabicenumerate}}

\def\init{\setcounter{equation}{0}}

\newtheorem{theoreme}{Theorem }[section]
\newtheorem{proposition}[theoreme]{Proposition}
\newtheorem{lemma}[theoreme]{Lemma}
\newtheorem{definition}[theoreme]{Definition}
\newtheorem{corollary}[theoreme]{Corollary}
\newtheorem{remark}[theoreme]{Remark}
\newtheorem{example}[theoreme]{Example}
\newcommand{\beq}{\begin{equation}}
\newcommand{\eeq}{\end{equation}}

\newcommand{\bex}{\begin{example}}
\newcommand{\eex}{\end{example}}
\def\bel{\begin{lemma}}
\def\eel{\end{lemma}}
\def\bet{\begin{theoreme}}
\def\eet{\end{theoreme}}
\def\bed{\begin{definition}}
\def\eed{\end{definition}}
\def\ber{\begin{remark}}
\def\eer{\end{remark}}

\def\rr{{\mathbb R}}

\def\cc{{\mathbb C}}
\def\nn{{\mathbb N}}

\def\Re{{\rm Re}}
\def\slim{{\rm s-}\lim}
\def\wlim{{\rm w-}\lim}

\def\bar{\overline}
\def\pp{{\rm pp}}
\def\cinf{C^\infty}
\def\coinf{C_0^\infty}

\def\cY{{\mathcal Y}}

\def\cD{{\mathcal D}}

\def\cM{{\mathcal M}}

\def\qed{$\Box$\medskip}
\def \p{ \partial}

\def\12{\frac{1}{2}}
\def\x{\langle x \rangle}
\def\s{\langle s \rangle}

\def\supp{{\rm supp}}

\def\ad{{\rm ad}}
\def\e{{\rm e}}

\def\bbbone{{\mathchoice {\rm 1\mskip-4mu l} {\rm 1\mskip-4mu l}
{\rm 1\mskip-4.5mu l} {\rm 1\mskip-5mu l}}}
\def\one{\bbbone}
\def\cH{{\mathcal H}}

\def\Ker{{\rm Ker}}

\def\cF{{\mathcal F}}
\def\cG{{\mathcal G}}

\def\cK{{\mathcal K}}

\def\bep{\begin{proposition}}
\def\eep{\end{proposition}}

\newcommand{\mat}[4]{\left(\begin{array}{cc}#1 &#2  \\ #3 &#4 \end{array}\right)}
\newcommand{\lin}[2]{\left(\begin{array}{c}#1 \\#2\end{array}\right)}

\begin{document}
\def\4{\frac{1}{4}}
\def\what{\widehat}
\def\cE{\mathcal{E}}
\def\cG{\mathcal{G}}
\def\Dom{{\rm Dom}}
\def\i{{\rm i}}
\def\h{\langle h \rangle}
\def\epsi{\langle \epsilon\rangle}
\def\cN{{\mathcal N}}
\def\jh{\langle h\rangle}
\def\sh{|h|}
\def\jr{\langle x\rangle}
\def\t{{\scriptscriptstyle\#}}
\def\hrr{\hat{\rr}}

\title[Resolvent and propagation estimates for Klein-Gordon equations]{Resolvent and propagation estimates \\ for Klein-Gordon equations \\ with non-positive energy}
\author{V. Georgescu}
\address{D\'epartement de Math\'ematiques, UMR 8088,
Universit\'e de Cergy-Pontoise, 
95000 Cergy-Pontoise Cedex, France }
\email{Vladimir.Georgescu@math.cnrs.fr}
\author{C. G\'erard}
\address{D\'epartement de Math\'ematiques, Universit\'e de Paris XI, 91405 Orsay Cedex France}
\email{christian.gerard@math.u-psud.fr}
\author{D. H\"{a}fner}
\address{Universit\'e de Grenoble 1, Institut Fourier, UMR 5582 CNRS, BP 74 38402 Saint-Martin d'H\`eres France}
\email{Dietrich.Hafner@ujf-grenoble.fr}
\date{March 2013}
\begin{abstract}
 We study  in this paper an abstract class of Klein-Gordon equations:
 \[
\p_{t}^{2}\phi(t)- 2\i k \p_{t}\phi(t)+ h \phi(t)=0,
\]
where $\phi: \rr\to \cH$, $\cH$ is a (complex) Hilbert space, and $h$, $k$ are  self-adjoint, resp. symmetric operators on $\cH$.

We consider their generators $H$ (resp. $K$) in the two natural spaces of Cauchy data, the {\em energy} (resp. {\em charge}) {\em spaces}.
We do not assume that the dynamics generated by $H$ or $K$ has any positive conserved quantity, in particular these operators may have complex spectrum.  Assuming conditions on $h$ and $k$ which allow to use the theory of selfadjoint operators on {\em Krein spaces}, we prove weighted  estimates on the boundary values of the resolvents of $H$, $K$ on the real axis. From these resolvent estimates we obtain corresponding propagation estimates on the behavior of the dynamics for large times. 

Examples include wave or Klein-Gordon equations on asymptotically euclidean or asymptotically hyperbolic manifolds, minimally coupled with an external electro-magnetic field decaying at infinity.
\end{abstract}
\maketitle
\section{Introduction}\label{sec-2}\init
This paper  is devoted to  the proof of resolvent and propagation estimates for the generators of a class of abstract Klein-Gordon equations
\begin{equation}
\label{ei.1}
\p_{t}^{2}\phi(t)- 2\i k \p_{t}\phi(t)+ h \phi(t)=0,
\end{equation}
where $\phi: \rr\to \cH$, $\cH$ is a (complex) Hilbert space, and $h$, $k$ are  self-adjoint, resp. symmetric operators on $\cH$.

There are many natural examples of such abstract class of equations: one class is obtained by considering Klein-Gordon equations
\[
-\nabla^{a}\nabla_{a}\phi+ m^{2}\phi=0,
\]
on a Lorentzian manifold having a global Killing vector field, 
corresponding in (\ref{ei.1}) to $\p_{t}$.  
A related class is obtained  by perturbing a static Klein-Gordon equation:
\[
\p_{t}^{2}\phi-\nabla^{j}\nabla_{j}\phi+ m^{2}\phi=0
\]
on $\rr_{t}\times N$ ($N$ is a Riemannian manifold) by 
minimal coupling this equation to an external electro-magnetic field $A_{a}$
 independent on $t$. We obtain then the equation
 \[
(\p_{t}- \i v(x))^{2}\phi-(\nabla^{j}-\i A^{j}(x))(\nabla_{j}- \i A_{j}(x))\phi+ m^{2}\phi=0,
\]
which can be put in the form (\ref{ei.1}).

An example to keep in mind is the Klein-Gordon equation on Minkowski space minimally coupled with an external electric field:
\beq\label{ei.2}
(\p_{t}- \i v(x))^{2}\phi(t,x)- \Delta_{x}\phi(t,x)+ m^{2}\phi(t,x)=0,
\eeq
for which $\cH= L^{2}(\rr^{d}, dx)$, $h=-\Delta_{x}+ m^{2}- v^{2}(x)$, $k= v(x)$ is a (real) electric potential and  $m\geq 0$ is the mass of the Klein-Gordon field. We will use this example to describe the results and methods of the present work.
\subsection{Description of the main results}
The equation (\ref{ei.2}) has two natural conserved quantities, the {\em charge}:
\[
\int_{\rr^{d}} \Big(
\i \p_{t}\overline{\phi}(t, x)\phi(t,x)- \i \overline{\phi}(t,x)\p_{t}\phi(t,x)- 2 v(x)|\phi(t,x)|^{2}
\Big)dx,
\]
and the {\em energy}:
\[
\int_{\rr^{d}} \Big(
|\p_{t}\phi(t,x)|^{2} + |\nabla_{x}\phi(t,x)|^{2}+ (m^{2}- v^{2}(x))|\phi(t,x)|^{2}
\Big)dx,
\]
both related to the symplectic nature of (\ref{ei.2}). In order to associate a generator to (\ref{ei.2}), one has to consider a {\em Cauchy problem}. There are two natural ways to define Cauchy data at time $t$. On can set 
\begin{equation}
\label{ei.3}
f(t)= \lin{\phi(t)}{\i^{-1}\p_{t}\phi(t)- v\phi(t)},
\end{equation}
so that 
\[
f(t)= \e^{\i t K }f(0), \ K= \mat{v}{\one}{-\Delta_{x}+ m^{2}}{v}.
\]
This choice is natural when one emphasizes the conservation of the charge, which takes the simple form:
\[
q(f, f)= \int_{\rr^{d}}\overline{f_1}(x)f_{0}(x)+ \overline{f_0}(x)f_{1}(x)dx, \ f= \lin{f_{0}}{f_{1}}.
\]
Another choice, more common in the PDE literature is:
\begin{equation}
\label{ei.4}
f(t)= \lin{\phi(t)}{\i^{-1}\p_{t}\phi(t)},
\end{equation}
so that
\[
f(t)= \e^{\i tH}f(0), \ H= \mat{0}{\one}{-\Delta_{x}+m^{2}- v^{2}}{2v}.
\]
With this choice the energy takes the simple form:
\[
E(f,f)= \int_{\rr^{d}}|f_{1}|^{2}(x)+ |\nabla f_{0}|^{2}(x)+ (m^{2}- v^{2}(x))|f_{0}|^{2}(x) dx,  \ f= \lin{f_{0}}{f_{1}}.
\]
Note that the two operators $K$ and $H$ are obviously related by similarity, see e.g. Subsect. \ref{sec0b.3}.

The main problem one faces when studying Klein-Gordon equations (\ref{ei.2}) is the lack of a {\em positive} conserved quantity. For example $q$ is clearly never positive definite, while $E$ is not positive definite if the electric potential $v$ becomes too large, so that $-\Delta_{x}+ m^{2}- v^{2}(x)$ acquires some negative spectrum.  In other words it is generally not possible to equip the space of Cauchy data with a Hilbert space structure such that $K$ or $H$ are self-adjoint.

There are two manifestations of this problem with some physical significance. The first one, discovered long ago by physicists \cite{SSW}, is the fact that if $v$ is too large, $K$ and $H$ acquire complex eigenvalues, appearing in complex conjugate pairs. This phenomenon is sometimes called the {\em Klein paradox}. It implies the existence of exponentially growing solutions, and causes difficulties with the quantization of (\ref{ei.2}).

The second manifestation is known as {\em superradiance}. It appears for
example for the Klein-Gordon equation (\ref{ei.2}) in $1$ dimension, when the
electric potential $v$ has two limits $v_{\pm}$ at $\pm\infty$ with 
$|v_{+}- v_{-}|>m$, see \cite{Ba} for a mathematical analysis. It also appears
in more complicated models, like the Klein-Gordon equation on the 
{\em Kerr space-time}, which can be reduced to the abstract form 
(\ref{ei.1}) after some separation of variables.

Superradiance appears  when there exist infinite dimensional subspaces of Cauchy data,  asymptotic invariant under $H$, on which the energy is positive (resp. negative). If this happens a wave coming from $+\infty$ may, after scattering by the potential, return to $+\infty$ with more energy than it initially had.

Another more mathematical issue with (\ref{ei.2}) is that there are many possible topologies to put on the space of Cauchy data. If we use (\ref{ei.3}) it is natural to require that the charge should be bounded for the chosen topology. This of course does not fix the topology, but by considering the simple case $v(x)\equiv 0$ it is easy to see (see Subsect. \ref{sec0b.4})  that the natural space of Cauchy data is the {\em charge space}:
\[
\cF= H^{\12}(\rr^{d})\oplus H^{-\12}(\rr^{d}),
\]
where $H^{s}(\rr^{d})$ denotes  the usual Sobolev space of order $s$.

If we use (\ref{ei.4}), then we should require that the energy be bounded, which leads to the essentially unique choice of the {\em energy space}:
\[
\cE= H^{1}(\rr^{d})\oplus L^{2}(\rr^{d}).
\]
Note that if $m=0$ the {\em homogeneous energy space} 
\[
\dot\cE= (-\Delta_{x}- v^{2})^{-\12}L^{2}(\rr^{d})\oplus L^{2}(\rr^{d})
\]
(see Subsect. \ref{sec0.1} for this notation) could also be considered, and will actually play an important role in our work.

Let us now illustrate the results of our paper on the example (\ref{ei.2}), assuming for simplicity that $v\in \coinf(\rr^{d})$. 
Using general results on self-adjoint operators on Krein spaces, one can first show that 
\[
\begin{array}{rl}
&\sigma(H)= \sigma(K),\\[2mm]
&\sigma_{\rm ess}(H)= \sigma_{\rm ess}(K)= ]-\infty, -m] \cup [m, +\infty[\\[2mm]
&\sigma(H)\backslash \rr= \sigma(K)\backslash\rr= \cup_{1\leq j\leq n}\{\lambda_{j}, \overline{\lambda_{j}}\},
\end{array}
\]
where $\lambda_{j}$, $\overline{\lambda_{j}}$ are eigenvalues of finite Riesz index. 

The main result of this work are {\em weighted resolvent estimates}, valid near the essential spectrum of $K,H$:
\begin{equation}
\label{ei.5}
\sup_{\Re z\in I, 0<\\Im z|\leq \delta}\| \langle x\rangle_{\rm diag}^{-\delta}(H-z)^{-1}\langle x\rangle_{\rm diag }^{-\delta}\|_{B(\cE)}<\infty, \ \forall \ \12 <\delta,
\end{equation} 
Here $\langle x\rangle_{\rm diag}$  denotes the diagonal operator on the space of Cauchy data $H^{1}(\rr^{d})\oplus L^{2}(\rr^{d})$ with entries $\langle x\rangle$ (see Subsect. \ref{sec0.1}), $I\subset \rr$ is a compact interval disjoint from $\pm m$, containing no real eigenvalues of $H$, nor so called {\em critical points} of $H$ (see Sect. \ref{seckrein} for the definition of critical points). 

Similar results hold for $K$, replacing $\cE$ by $\cF$. By the usual argument based on Fourier transformation, we deduce from (\ref{ei.5}) {\em propagation estimates} on the $C_{0}-$groups $\e^{\i tH}$, $\e^{\i tK}$:
\beq
\label{ei.6}
\begin{array}{l}
\int_{\rr}\Vert
\x_{\rm diag}^{-\delta}\e^{\i tH}\chi(H)\x^{-\delta}f\Vert^2_{\cE}dt\leq C\| f\|^{2}_{\cE},\\[2mm]
\int_{\rr}\Vert
\x_{\rm diag}^{-\delta}\e^{\i tK}\chi(K)\x^{-\delta}f\Vert^2_{\cF}dt\leq C \| f\|^{2}_{\cF},
\end{array}
\eeq
where $\chi\in \coinf(\rr)$ is a cutoff function supported away from real eigenvalues and critical points of $H$ and $K$. 

From (\ref{ei.6}) it is easy to construct the short-range scattering theory for the dynamics $\e^{\i tH}$, $\e^{\i tK}$. With a little more effort, the long-range scattering theory can also be constructed. In this way the results of \cite{G}, dealing with the scattering theory of massive Klein-Gordon equations in energy spaces, can certainly be extended to  the massless case (ie to wave equations), both in the energy and charge spaces. 
\subsection{Methods}
In the usual Hilbert space setting, where $H$ is self-adjoint for some Hilbert space scalar product, the most powerful way to prove estimates (\ref{ei.5}), (\ref{ei.6}) relies on the {\em Mourre method},  i.e. on the construction of another self-adjoint operator $A$ such that
\begin{equation}
\label{ei.7}
\one_{I}(H)[H, \i A]\one_{I}(H)\geq c \one_{I}(H)+R,
\end{equation}
where $R$ is compact and $c>0$. This method can be directly applied to (\ref{ei.2}) if the energy $E(f,f)$ is positive definite, so that it defines a compatible scalar product on $\cE$. Numerous papers rely on this observation, see among many others the papers \cite{E, Lu, N,S,V}, and for more recent results based on the Mourre method \cite{Ha1, Ha2}.

If the energy is not positive, one can consider the energy space $\cE$ equipped with $E$ as a {\em Krein space}, i.e. a Hilbertizable vector space equipped with a bounded, non-degenerate hermitian form. Orthogonal and adjoints on a Krein space are defined w.r.t. the Krein scalar product, and conservation of energy is formally equivalent to the fact that the generator $H$ is self-adjoint in the Krein sense. 

There exists a class of self-adjoint operators on Krein spaces, the so-called {\em definitizable operators}, (see Subsect. \ref{seckrein.2}) which admit a  continuous and Borel functional calculus quite similar to the one of usual self-adjoint operators. A finite set of their real spectrum, called the {\em critical points}, plays the role of {\em spectral singularities} for the functional calculus. Spectral projections on intervals whose endpoints are not critical points can be defined, and they have the important property that if they do not contain critical points, then the Krein scalar product is {\em definite} (positive or negative) on their range.

In \cite{GGH1}, we exploited these properties of definitizable operators on Krein spaces to extend the Mourre method to this setting, obtaining  weighted resolvent estimates in an abstract setting.

In this work we apply the general results of \cite{GGH1} to one of the main examples of self-adjoint operators on Krein spaces, namely the generators of one of the $C_{0}-$groups associated to  the abstract Klein-Gordon equation (\ref{ei.1}). 

Note that several papers were devoted to Klein-Gordon or wave equations from the Krein space point of view, see e.g. \cite{J2, LNT1, LNT2}. However resolvent estimates near the real spectrum were never considered in the above papers. 

We obtain resolvent and propagation estimates which are generalizations of (\ref{ei.5}), (\ref{ei.6}). Examples of our abstract framework are minimally coupled Klein-Gordon or wave equations on {\em scattering} or {\em asymptotically hyperbolic} manifolds. 

Note that the typical assumption of the electric potential $v$ is that it should decay to $0$ at $\infty$. This assumption is necessary to ensure that $H$ is definitizable. Therefore the models considered in this paper, while possibly exhibiting the Klein paradox, do not give rise to superradiance. In a subsequent paper \cite{GGH2} we will prove similar results for a model exhibiting superradiance, namely the Klein-Gordon equation on {\em Kerr-de Sitter space-times}. Using the results of this paper, it is possible to prove resolvent estimates and to study scattering theory also for such superradiant Klein-Gordon equations.

\subsection{Plan of the paper}
Sect.\ \ref{sec0} contains some preparatory material, the most important dealing with quadratic operator pencils. 

In Sect.\ \ref{seckrein} we recall the theory of {\em definitizable operators} on Krein spaces. In particular we devote some effort to present a self-contained exposition of their functional calculus, which is a rather delicate but interesting topic.  Among previous contributions to this question, we mention the  works of Langer \cite{La} and Jonas \cite{J}. 

In \cite{GGH1}  we constructed the natural version of the {\em continuous} functional calculus for a definitizable operator $H$, associated to an algebra of continuous functions having asymptotic expansions of a specific order at each {\em critical point} of $H$. Although we will not need its full generality in the rest of the paper, we found it worthwhile to develop the corresponding {\em Borel}  functional calculus. An interesting feature of this calculus is that the natural algebra is not an algebra of functions on $\rr$ anymore, but has to be augmented by adding a copy of $\cc$ at each critical point.

In Sect.\ \ref{abstract-kg} we discuss in some detail the various setups for abstract Klein-Gordon equations and the possible choices of topologies on the space of initial data.

Sects.\ \ref{sec1} and \ref{sec5} are devoted to basic facts on the generators of Klein-Gordon equations in energy and charge spaces respectively. Related results can be found in \cite{LNT1, LNT2}. In particular  {\em quadratic pencils} play an important role here.

 In Sect.\ \ref{sec1b} we introduce a class of definitizable Klein-Gordon operators. We also construct an approximate diagonalization of these operators which will be needed later.

Sect.\ \ref{sec2} is devoted to the proof of a positive (in the Krein sense) commutator estimate for the operators considered in Sect. \ref{sec1b}. It relies on abstract conditions on the scalar operators $h$, $k$ appearing in (\ref{ei.1}).

In Sect.\ \ref{sec4} resolvent estimates are proved for the generators $H$ on energy spaces. From them we deduce similar estimates for the quadratic pencils considered in Sect. \ref{sec1}, which in turn imply resolvent estimates for the generators $K$ on charge spaces.

Sect.\ \ref{secpropest} is devoted to the proof of propagation estimates for the groups $\e^{\i tH}$ and $\e^{\i tK}$. They follow from resolvent estimates by the standard arguments, usually applied in the Hilbert space setting. 
\def\KG{ Klein-Gordon }

In Sect.\ \ref{sec3} we give various examples of our abstract class of Klein-Gordon equations. The first examples are \KG equations  on {\em scattering manifolds}, minimally coupled to external electro-magnetic fields. The massive and massless cases are discussed separately, a {\em Hardy inequality} playing an important role in the massless case. The second examples are \KG equations on {\em asymptotically hyperbolic manifolds}, again with minimal coupling.

Various technical proofs are collected in Appendices \ref{secapp.2}, \ref{secapp.3}.

\section{Some preparations}\label{sec0}\init
In this section we collect some notation and preparatory material which will be used in later sections.
\subsection{Notations}\label{sec0.1}

{\em Sets}

- if $X,Y$ are sets and $f:X\to Y$ we write $f:X\tilde\to Y$ if $f$ is
bijective.  If $X, Y$ are equipped with topologies, we use the same notation if $f:X\to Y$ is a homeomorphism.
 
 - if $I\subset \rr$  and $f$ is  a real valued function defined on $I$ then $f(I)$ denotes the image of $I$ under $f$.

 Examples of this notation used in Subsect. \ref{sec2.3} are $\sqrt{I}$, $I^{2}$ and $|I|$.

- we set $\langle  \lambda\rangle:= (\lambda^{2}+1)^{\12}$, $\lambda\in \rr$.

{\em Linear operators}

- if $E\subset F$ are Banach spaces, we denote by $[E, F]_{\theta}$, $0\leq \theta \leq 1$ the complex interpolation space of order $\theta$.

- if $A$ is a closed, densely defined operator, we denote by $\rho(A)\subset \cc$ its resolvent set and by $\Dom A$ its domain.

- let $X,Y,Z$ be Banach spaces such that $X\subset Y\subset Z$ with
continuous and dense embeddings.  Then to each continuous operator
$\what S:X\tilde \to Z$ one may associate a densely defined operator $S$
acting in $Y$ defined as the restriction of $\what S$ to the domain
$\Dom S=(\what S)^{-1}(Y)$. 

- if $\cH_{1}$, $\cH_{2}$ are Hilbert spaces, we denote by $B_{\infty}(\cH_{1}, \cH_{2})$ the ideal of compact operators from $\cH_{1}$ to $
\cH_{2}$ and set $B_{\infty}(\cH)= B_{\infty}(\cH, \cH)$.

- If $a, b$ are linear operators, then we set ${\rm ad}_{a}(b):= [a, b]$. Usually in this paper commutators are defined in the operator sense, i.e.\ 
$[a, b]$ has domain $\Dom(ab)\cap \Dom(ba)$.

 - if $A, B$ are two positive self-adjoint operators on a Hilbert space $\cH$, we write $A\sim B$ if
\[
 \Dom A^{\12}=:\Dom B^{\12}\hbox{ and } c^{-1}A\leq B\leq c A\hbox{ on }\Dom A^{\12}, \ c>0.
\]

 {\em Dual pairs}
 
-Let $\cG,\cH$ be reflexive Banach spaces and $\cE=\cG\oplus\cH$.
The usual realization $(\cG\oplus\cH)^{*}=\cG^{*}\oplus\cH^{*}$ of
the adjoint space will  not be convenient in the sequel, we shall
rather set $\cE^{*}:=\cH^{*}\oplus\cG^{*}$ so that 
\[
\langle w | f\rangle= \langle w_{0}| f_{1}\rangle + \langle w_{1}| f_{0}\rangle, \hbox{ for }
f=(f_{0}, f_{1})\in \cE,  \ w= (w_{0}, w_{1})\in\cE^{*}.
\]

 For example, if $\cH=\cG^{*}$, so $\cH^*=\cG$, the
adjoint space of $\cE=\cG\oplus\cG^{*}$ is identified with itself
$\cE^{*}=\cE$.

{\em Scale of Sobolev  spaces}

Let  $\cH$  be a Hilbert space with norm $\|\cdot\|$ and scalar
product $(\cdot| \cdot)$. We identify $\cH$ with its adjoint
space $\cH^{*}=\cH$ via the Riesz isomorphism.   Let $h$ be a self-adjoint operator on $\cH$. 

We can associate to it the 
{\em non-homogeneous Sobolev spaces}
\[
 \langle h\rangle^{-s}\cH:= \Dom |h|^{s}, \   \langle h\rangle^{s}\cH:= ( \langle h\rangle^{-s}\cH)^{*}, \ s\geq 0.
\]
The spaces $ \langle h\rangle^{-s}\cH$ are equipped with the graph norm $\| \langle h\rangle^{s}u\|$.

If moreover ${\rm Ker}h=\{0\}$, then we can also define the   {\em  homogeneous Sobolev spaces} $|h|^{s}\cH$ 
equal to the completion of $\Dom |h|^{-s}$ for the norm $\||h|^{-s}u\|$.

The notation $\langle h\rangle^{s}\cH$ or $|h|^{s}\cH$ is very convenient but somewhat ambiguous because usually $a\cH$ denotes the image 
of $\cH$ under the linear operator $a$. Let us explain how to reconcile these two meanings:

let $\cH_{\rm c}$ be the space of $u\in \cH$ such that $u= \one_{I}(h)u$, for some compact $I\subset \rr\backslash \{0\}$. We equip $\cH_{\rm c}$ with its natural topology by saying  that $u_{n}\to u$ in $\cH_{\rm c}$ if there exists $I\subset \rr\backslash \{0\}$ compact such that $u_{n}= \one_{I}(h)u_{n}$ for all $n$ and  $u_{n}\to u$ in $\cH$. We set
$\cH_{\rm loc}:= (\cH_{\rm c})^{*}$. Then  $|h|^{s}$ and $\langle h\rangle^{s}$ preserve  $
\cH_{\rm c}$ and $\cH_{\rm loc}$, and  $\langle h\rangle^{s}\cH$, resp. $|h|^{s}\cH$ are the images in $\cH_{\rm loc}$ of $\cH$ under $\langle 
h\rangle^{s}$, resp. $|h|^{s}$. It follows that these
spaces are  subspaces (equipped with finer topologies) of $\cH_{\rm loc}$,  in particular they are pairwise compatible. 
Let us mention some properties of these spaces:
\[
\begin{array}{l}
 \langle h\rangle^{-s}\cH \subset \langle h\rangle^{-t}\cH, \hbox{  if }t\leq s, \  \langle h\rangle^{-s}\cH\subset |h|^{-s}\cH, |h|^{s}\cH\subset \h^{s}\cH
\hbox{ if }s\geq 0,\\[2mm]
\h^{0}\cH= |h|^{0}\cH= \cH, \ \h^{s}\cH= (\h^{-s}\cH)^{*}, \ |h|^{s}\cH= (|h|^{-s}\cH)^{*},\\[2mm]
0\in \rho(h)\Leftrightarrow \ \h^{s}\cH= |h|^{s}\cH\hbox{ for some }s\neq 0 \Leftrightarrow \h^{s}\cH= |h|^{s}\cH\hbox{ for all }s.
\end{array}
\]
Moreover the operator $|h|^{s}$ is unitary from $|h|^{-t}\cH$ to $|h|^{s-t}\cH$ for all $s,t\in \rr$.

The following fact is a rephrasing of the Kato-Heinz theorem:

- if $a\sim b$ then $a^{s}\cH= b^{s}\cH$ for all $|s|\leq \12$.
\label{p:KH}

{\em Smoothness of operators}

Let $\cH_{1}$, $\cH_{2}$ be two Banach spaces such that $\cH_{1}\subset \cH_{2}$ continuously and densely. Let $\{T_{t}\}_{t\in \rr}$ be a $C_{0}-$group on $\cH_{2}$, preserving $\cH_{1}$. It follows by \cite[Prop. 3.2.5]{ABG} that $T_{t}$ defines a $C_{0}-$group on $\cH_{1}$. If $a$ is the generator of $T_{t}$ on $\cH_{2}$, so that $T_{t}= \e^{\i t a}$ on $\cH_{2}$, then the generator of $T_{t}$ on $\cH_{1}$ is $a_{\mid \cH_{1}}$ with domain $\{u\in \cH_{1}\cap \Dom a  : au\in \cH_{1}\}$. 

We denote by $C^{k}(a; \cH_{1}, \cH_{2})$ (resp. $C^{k}_{\rm u}(a; \cH_{1}, \cH_{2})$) for $k\in \nn$ the space of operators $b\in B(\cH_{1}, \cH_{2})$ such that $\rr\ni t\mapsto \e^{\i t a}b\e^{-\i ta}$ is $C^{k}$ for the strong (resp. operator) topology of $B(\cH_{1}, \cH_{2})$. 

One defines similarly $C^{s}_{({\rm u})}(a; \cH_{1}, \cH_{2})$ first for $0<s<1$, then for all non  integers $s\in \rr^{+}$ by requiring the H\"{o}lder continuity of the above map. Note that by the uniform boundedness principle, the spaces $C^{s}(a; \cH_{1}, \cH_{2})$ and $C^{s}_{\rm u}(a; \cH_{1}, \cH_{2})$ coincide for non integer $s$.
It follows also from the same argument that $C^{k}(a; \cH_{1}, \cH_{2})\subset C^{s}_{\rm u}(a; \cH_{1}, \cH_{2})$ for $0<s<k$.

If $b\in C^{1}(a; \cH_{1}, \cH_{2})$ then $b$ maps $\Dom a_{\mid \cH_{1}}$ into $\Dom a$ and ${\rm ad}_{a}b:= ab- ba\in B(\cH_{1}, \cH_{2})$.

If $\cH_{1}= \cH_{2}= \cH$, the above spaces are simply denoted by $C^{s}_{({\rm u})}(a;\cH)$ or even $C^{s}_{({\rm u})}(a)$ if $\cH$ is fixed from the context.

\subsection{Quadratic pencils}\label{sec0.2}

In this subsection we prove some basic results about a quadratic operator pencil related to the abstract Klein-Gordon operator.

Let $\cH$ be a Hilbert space, $h$ be a self-adjoint operator on $\cH$
  and $\h^{-s}\cH$ the Sobolev spaces  introduced in Subsect. \ref{sec0.1}.  Let
  $k:\h^{-\12}\cH\to\cH$ be a continuous symmetric operator and denote
  also $k$ its unique extension to a continuous map
  $\cH\to\h^{\12}\cH$. Denote 
  \[
  h_0=h+k^2:\h^{-\12}\cH\to\h^{\12}\cH
  \]
   and
  \[
  p(z)= h+z(2k-z)=h_0-(k-z)^2\in B(\h^{-\12}\cH, \h^{\12}\cH), \ z\in \cc.
  \]
  
\begin{definition}\label{defderho}
  We denote by $\rho(h,k)$ the set of $z\in\cc$ such
  that \[
  p(z):\h^{-\12}\cH\tilde\to\h^{\12}\cH.
  \] 
\end{definition}

Observe that the domain in $\cH$ of the operator
$p(z):\h^{-\12}\cH\to\h^{\12}\cH$ is equal to $\h^{-1}\cH$, i.e.
$\h^{-1}\cH=p(z)^{-1}\cH$. Indeed, for $f\in\h^{-\12}\cH$ we have
$p(z)f=hf+z(2k-z)f$ and the last term belongs to
$\cH$, hence $p(z)f\in\cH$ if and only if $hf\in\cH$.  Note
also that the relation $p(z)^{*}=p(\bar z)$ in
$B(\h^{-\12}\cH,\h^{\12}\cH)$ is obvious. It follows that the map
$p(z):\h^{-\12}\cH\to\h^{\12}\cH$ naturally induces operators in
$B(\h^{-1}\cH,\cH)$ and $B(\cH,\jh\cH)$.
  
The following two results  are proved in \cite[Lemmas 8.1, 8.2]{GGH1}.
\begin{lemma}\label{lm:kg}
  The operator induced by $p(z)$ in $\cH$ is a closed operator
  and its Hilbert space adjoint is the operator induced by
  $p(\bar z)$ in $\cH$. In other terms, the relation
  $p( z)^*=p(\bar{ z})$ also holds in the sense of closed
  operators in $\cH$. The following conditions are equivalent:
  \[
  \begin{array}{l}
(1)\ p( z):\h^{-1}\cH\tilde\to\cH, \
(2)\  p(\bar{ z}):\h^{-1}\cH\tilde\to\cH,\\[2mm]
(3) \ p( z):\h^{-\12}\cH\tilde\to\h^{\12}\cH, \ 
(4) \ p(\bar{ z}):\h^{-\12}\cH\tilde\to\h^{\12}\cH,\\[2mm]
(5) \ p( z):\cH\tilde\to\jh\cH, \
(6) \ p(\bar{ z}):\cH\tilde\to\jh\cH.
\end{array}
\]
In particular, the set 
\begin{equation}\label{eq:rhohk}
\rho(h,k):=\{z\in\cc\mid p(z):\jh^{-\12}\cH\tilde\to\jh^{\12}\cH\}
=\{z\in\cc\mid p(z):\jh^{-1}\cH\tilde\to\cH\}
\end{equation}
is invariant under conjugation.
\end{lemma}

\begin{proposition}\label{pr:idiot}
 Assume that $h$ is bounded below. Then there exists $c_{0}>0$ such that
 \[
\{ z\ : |{\rm Im} z| > |{\rm Re}z| + c_{0}\}\subset \rho(h,k).
\]
\end{proposition}
\section{Operators on Krein spaces}\label{seckrein}
In this section we review some basic facts about Krein spaces and self-adjoint operators on Krein spaces. We refer the reader for more details to  the survey paper \cite{La}, or to \cite{G}, \cite{GGH1}. We also describe the natural extension of the continuous functional calculus constructed in \cite{GGH1} to the Borel case.
\subsection{Krein spaces}\label{seckrein.1}
 If $\cH$ is a topological complex vector space, we denote by $\cH^{\t}$ the space of continuous linear forms on $\cH$ and by $\langle w, u\rangle$, for $u\in \cH$, $w\in \cH^{\t}$ the duality bracket between $\cH$ and $\cH^{\t}$.

\begin{definition}
 A {\em Krein space}  $\cK$ is a hilbertizable vector  space  equipped with a bounded hermitian form $\langle u| v\rangle$ non-degenerate in the sense that if $w\in \cK^{\t}$ there exists a unique $u\in \cK$ such that 
\[
\langle u| v\rangle= \langle w, v\rangle, \ v\in \cK.
\]
\end{definition}
If $\cK_{1}$ is a subspace of $\cK$, we denote by $\cK_{1}^{\perp}$ the orthogonal of $\cK_{1}$ for $\langle \cdot| \cdot\rangle$.

If we fix a scalar product $(\cdot |\cdot)$ on $\cK$ endowing $\cK$ with its hilbertizable topology, then by  the Riesz theorem there exists a bounded, invertible self-adjoint operator $M$ such that
\[
\langle u| v\rangle= (u| Mv), \ u, v\in \cK.
\]
Using the polar decomposition of $M$, $M= J|M|$ where $J=J^{*}$, $J^{2}=\one$, one can equip $\cK$ with the equivalent scalar product 
\beq\label{e0.00}
(u| v)_{M}:=(u||M| v),
\eeq so that
\begin{equation}
\label{e0.01}
\langle u| v\rangle= (u|J v)_{M}, \ u, v\in \cK.
\end{equation}

\begin{definition}\label{pontryag}
 A Krein space $(\cK,\langle \cdot| \cdot\rangle)$ is a {\em Pontryagin space} if either $\one_{\rr^{-}}(J)$ or $\one_{\rr^{+}}(J)$ has  finite rank.
\end{definition}

Clearly this definition is independent on the choice of the scalar product $(\cdot | \cdot)$.

Replacing $\langle \cdot| \cdot\rangle$ by $-\langle \cdot| \cdot\rangle$ we can assume that $\one_{\rr^{-}}(J)$ has finite rank, which is the usual convention for Pontryagin spaces.

Let $A:\Dom A\to \cK$ be a densely defined linear operator on the Krein space $\cK$. The adjoint $A^{\dag}$ of $A$ on $(\cK, \langle \cdot| \cdot\rangle)$ is defined as 
\[
\Dom A^{\dag}:=\{u\in \cK  : \exists \   f=: A^{\dag }u\hbox{ such that } \langle f|v\rangle= \langle u| Av\rangle, \ \forall \ v\in \Dom A\}.
\]
We will sometimes use the following easy fact: there is a constant $C>0$ such that
\beq\label{krein.1}
 C^{-1}\|A\|\leq \|A^{\dag}\|\leq C\|A\|, \ A\in B(\cK).
\eeq
A densely defined operator $H$ is {\em self-adjoint } on $\cK$ if $H= H^{\dag}$. The following fact is often useful.
\begin{lemma}\label{stupid}
Let $H$ be closed and densely defined on $\cK$. Assume that for some 
$z\in \rho(H)\cap \overline{\rho(H)}$ 
one has $((H- z)^{-1})^{\dag}= (H-\bar{ z})^{-1}$. Then $H= H^{\dag}$.
\end{lemma}
\subsection{Definitizable operators on Krein spaces}\label{seckrein.2}
Not much of interest can be said about self-adjoint operators on a Krein space, except for the trivial fact that $\rho(H)= \overline{\rho(H)}$. There is however a special class of self-adjoint operators, called {\em definitizable}, which admit a functional calculus close to the one of usual self-adjoint operators on a Hilbert space.

\begin{definition}
A self-adjoint operator $H$ is {\em definitizable} if
\ben
\item $\rho(H)\neq \emptyset$,
\item there exists a real polynomial $p(\lambda)$ such that
\beq\label{flamby}
\langle u| p(H)u\rangle\geq 0, \ \forall\  u\in \Dom H^{k}, \ k:={\rm deg}\,p.
\eeq
\een
An operator $H$ on a Krein space $\cK$ which is definitizable with an even definitizing polynomial will be called {\em even-definitizable}.
\end{definition}
%
The following result is well-known, see e.g. \cite[Lemma 1]{J}.
\begin{proposition}\label{frogu}
Let $H$ be definitizable. Then: 
 \ben
\item If $z\in \sigma(H)\backslash \rr$  then 
$p(z)=0$ for each definitizing polynomial $p$,

\item There is a definitizing polynomial $p$ such that
  $\sigma(H)\setminus\rr$ is exactly the set of non-real zeroes of
  $p$,

\item Moreover, this $p$ may be chosen such that if
  $\lambda\not\in\rr$ is a zero of multiplicity $k$ of $p$ then
  $\lambda$ is an eigenvalue of $H$ of Riesz index $k$,

\item The non-real spectrum of $H$ consists of a finite number of 
eigenvalues of finite Riesz index distributed symmetrically with respect to 
the real axis.
\een
\end{proposition}

The usefulness of the notion of Pontryagin spaces comes from the following theorem (see \cite{La}).
\begin{theoreme}\label{arsouille}
 A self-adjoint operator $H$ on a Pontryagin space is even-definitizable.
 \end{theoreme}

 The following result is easy (see Langer \cite{La}).  If $\lambda$ is an isolated point of $\sigma(H)$  the {\em Riesz spectral projection} on $\lambda$ is:
\[
E(\lambda, H):=\frac{1}{2\i \pi}\ointctrclockwise_{C}(z-H)^{-1}d z
\]
where $C$ is a small curve in $\rho(H)$ surrounding $\lambda$. 

\begin{proposition}\label{coco}
Let $H$ be a definitizable self-adjoint operator and
 \[
\one_{\rm pp}^{\cc}(H)= \sum_{\lambda\in \sigma(H),
 \ {\rm Im}\lambda>0}
\left( E(\lambda, H)+ E(\overline{\lambda}, H) \right)
, \quad \cK_{\pp}^{\cc}:= \one_{\rm pp}^{\cc}(H)\cK.
\]
Then  $\one_{\rm pp}^{\cc}(H)$ is a projection, $\one_{\rm pp}^{\cc}(H)= (\one_{\rm pp}^{\cc}(H))^{\dag}$, hence $\cK_{\rm pp}^{\cc}$ is a Krein space and
\[
\cK= \cK_{\rm pp}^{\cc}\oplus(\cK_{\rm pp}^{\cc})^{\perp}.
\]
\end{proposition}

\subsection{$C^{\alpha}$ functional calculus}\label{seckrein.3}

In this subsection we recall some results of \cite{GGH1}, extending earlier results of \cite{J, La} on the  continuous functional calculus for definitizable operators. It turns out that a definitizable operator $H$ admits a functional calculus associated to the algebra of bounded continuous functions on $\rr$ having an asymptotic expansion of a specific order at each {\em critical point} of $H$ (see Def.\  \ref{defdecrit}).

Let $\hat{\rr}= \rr\cup\{\infty\}$ be the one point compactification of $\rr$, so that $C(\hat\rr)$ is identified with the set of continuous functions $\rr\to \cc$ having a finite limit at $\infty$.

We equip $\hat\rr\times\nn$ with the  order relation defined by  $(\xi, s)\leq (\eta, t)$ iff $\xi= \eta$ and $s\leq t$.
 If $\omega=(\xi,s)\in\hat\rr\times\nn$ we denote by $\chi_{\omega}$  the
rational function  equal to  $(x-\xi)^{s}$ if
$\xi\in\rr$ and $x^{-s}$ if $\xi=\infty$. We set also
$\rho_{\omega}=\chi_{\omega}^{-1}$.

\begin{definition}\label{defdede}
Let $\omega=(\xi,s)\in\hat\rr\times\nn$. We denote by $C^{\omega}(\hat\rr)$ the space of  functions $\varphi\in C(\hat\rr)$ such that there is a polynomial $P$ with:
\[
\varphi(x)=\left\{\begin{array}{l}
P(x-\xi)+o(|x-\xi|^{s}),  \hbox{ if }\xi\in \rr,\\[2mm]
P(1/x)+o(|x|^{-s}), \hbox{ if }\xi=\infty.
\end{array}
\right.
\]
\end{definition}

Clearly  $C^{\mu}(\hrr)\subset C^{\omega}(\hrr)$ if $\mu\geq \omega$. If $\varphi\in C^{\omega}(\hrr)$  then the terms of degree $\leq s$ of $P$ are uniquely determined,   hence there is a unique sequence of complex numbers $\{\delta_{\mu}(\varphi)\}_{\mu\leq \omega}$ such that the rational function 
\beq\label{iac1}
T^{+}_{\omega}\varphi:= \sum_{\mu\leq \omega} \delta_{\mu}(\varphi )\chi_{\mu}
\eeq satisfies
\beq\label{asympto}
\varphi(x)= T^{+}_{\omega}\varphi(x)+ o(| \chi_{\omega}(x)|).
\eeq
Set
\beq\label{iac2}
T_{\omega}\varphi:=\sum_{\mu<\omega}\delta_{\mu}(\varphi)\chi_{\mu},  \ R_{\omega}\varphi:=\rho_{\omega}(\varphi-T_{\omega}\varphi),
\eeq
so that
\[\varphi = T_{\omega}\varphi + \chi_{\omega}R_{\omega}\varphi.
\]
Note that if $\omega= (\xi, 0)$ then $R_{\omega}\varphi = \varphi$. If follows that 
\[
\|\varphi\|_\omega:= \sum_{\mu\leq\omega}\sup|R_\mu\varphi|
\]
is a norm on $C^{\omega}(\hat\rr)$ dominating the sup norm.

An element $\omega\in\hat\rr\times\nn$ may be seen  as a
function $\hat\rr\to\nn$ with support containing at most one point.
A function  $\alpha:
\hat{\rr}\to \nn$ with finite support is called an {\em order function}. 
We write
$\omega\preceq\alpha$ if $\omega=(\xi,s)\in\hat\rr\times\nn$ and
$s\leq\alpha(\xi)$. Then $\omega\prec\alpha$ means
$\omega\preceq\alpha$ and $s< \alpha(\xi)$.

To each definitizable operator  one can associate a natural order function:
\begin{definition}\label{isap}
  Let $H$ be a definitizable operator on $\cK$. 
  \ben
\item To each definitizing polynomial $p$ for $H$ we associate an
  order function $\beta$ as follows: if $\xi\in\rr$ then $\beta(\xi)$
  is the multiplicity of $\xi$ as zero of $p$ and $\beta(\infty)=0$
  if $p$ is of even degree and $\beta(\infty)=1$ if $p$ is of odd
  degree.
\item The order function $\alpha_{H}$ of $H$ is the infimum over all
  definitizing polynomials for $H$ of the above functions $\beta$.
  \een
\end{definition}
 If $\alpha$ is an order function, we set 
\[
 C^{\alpha}(\hat\rr):= \cap_{\omega\preceq \alpha}C^{\omega}(\hat\rr),
\]
which, equipped with the norm $\|\varphi\|_{\alpha}:= \sup_{\omega\preceq \alpha}\|\varphi\|_{\omega}$, is a unital Banach $*$-algebra  for the usual  operations.

The following theorem is shown in \cite[Thm. 4.9]{GGH1}
\begin{theoreme}\label{th:main}
  Let $H$ be a self-adjoint definitizable operator on the Krein
  space $\mathcal{H}$ with $\sigma(H)\subset\rr$.Then there is a unique
  linear continuous map \[
  C^{\alpha_{H}}(\hrr)\ni \varphi\mapsto\varphi(H)\in B(\cK)
  \]  such that if
  $\varphi(\lambda)=(\lambda-z)^{-1}$ for $z\in \cc\backslash \rr$ then
  $\varphi(H)=(H-z)^{-1}$. This map is a morphism of unital
  $*$-algebras.
\end{theoreme}

Thm.\ \ref{th:main} implies  an optimal estimate of the resolvent of
a definitizable operator. We first introduce some terminology.
\begin{definition}\label{defdecrit}
 We  set $\sigma_{\cc}(H):= \sigma(H)\backslash \rr$, $c(H):= \{\omega\in \hat{\rr} : \alpha_{H}(\xi)\neq 0\}$. The set $c(H)$ is called the set of {\em critical points} of $H$. 
\end{definition}

Let $H$ be a definitizable operator.  Note that Def.\ \ref{isap} extends naturally to give an order function on $\hat{\cc}=\cc\cup \{\infty\}$, still denoted  by $\alpha_{H}$. The following result is proved in \cite[Prop.\ 4.15]{GGH1}.

\begin{proposition}\label{pr:reso}
  With the preceding notations, there exists $c>0$ such that
\begin{equation}\label{eq:reso}
c\|(H-z)^{-1}\|\leq  \sum_{\xi\in\sigma_{\cc}(H)} |z-\xi|^{-\alpha_H(\xi)}
+|\mathrm{Im} z|^{-1}
\Big(1+\sum_{\xi\in c(H)\cap \rr} |z-\xi|^{-\alpha_H(\xi)} +
|z|^{\alpha_H(\infty)} \Big)
\end{equation}
for all $z\not\in\sigma_\cc(H)\cup\rr$.  Note that $\alpha_H(\infty)$ is
either $0$ or $1$.
\end{proposition}
We will use the following corollary of Prop.\ \ref{pr:reso}, giving estimates on $(H-z)^{-1}$ in a bounded region or in  a conic neighborhood of infinity in $\cc\backslash\rr$.
\begin{corollary}\label{besson}
 Let   for $R, a, \delta>0$:
\[
\begin{array}{rl}
U_{0}(R, a)&=\{z\in \cc:   0<| {\rm Im}z|<a, \ |{\rm Re}z|\leq R\},\\[2mm]
U_{\infty}(R, \delta)&=\{z\in \cc:     0<| {\rm Im}z|\leq \delta|{\rm Re}z|, \  |{\rm Re}z|\geq R\}.
\end{array}
\]
where $R, a$ are chosen such that $\sigma_{\cc}(H)$ does not intersect $U_{0}(R, a)$ and $U_{\infty}(R, \delta)$.
Then there exists $C>0$ such that:
\[
\|(H-z)^{-1}\|\leq \left\{
\begin{array}{l}
C| {\rm Im}z|^{-m-1}, \hbox{ for }z\in U_{0}(R, a),\\[2mm]
C \langle z\rangle^{\alpha_{H}(\infty)}|{\rm Im}z|^{-1}, \hbox{ for }z\in U_{\infty}(R, \delta),
 \end{array}
\right.
 \]
 where $m= \sup_{\xi\in \rr}\alpha_{H}(\xi)$.

\end{corollary}
It is sometimes convenient to have a concrete expression of $\varphi(H)$ if $\varphi\in \coinf(\rr)$. Let
 $\tilde{\varphi}\in \coinf(\cc)$ be an {\em almost-analytic extension} of $\varphi$, satisfying:
\[
\tilde{\varphi}_{|\rr}=\varphi,\ \left| \frac{\p\tilde{\varphi}(z)}{\p \overline{z}} \right|\leq C_{N}|{\rm Im z}|^{N}, \ \forall \ N\in \nn.
\]
For $m\in \nn$ we set $\| \varphi\|_{m}:= \sum_{0\leq k \leq m}\|\p_{x}^{k}\varphi\|_{\infty}$.
Then we have:
\begin{equation}
\label{irlitu}
\varphi(H)= \frac{\i}{2\pi}\int_{\cc}\frac{\p \tilde{\varphi}}{\p\overline{z}}(z)(z-H)^{-1}d z \wedge d \overline{z},
\end{equation}
where  due to Corollary  \ref{besson} the integral is norm-convergent  and one has 
\beq\label{arsitoli}
\|\varphi(H)\|\leq C \| \varphi\|_{m},\hbox{ for some }m\in \nn.
\eeq
\subsection{Borel functional calculus}\label{borelborel}
In this subsection we extend the results of Subsect.\ \ref{seckrein.3} to cover the Borel functional calculus.  Similar results were already obtained by Jonas \cite{J}, see also \cite{Wora}, although we believe that our approach is simpler and more transparent.

The standard method to obtain a Borel functional calculus from a continuous one relies on the Riesz and monotone class theorems  (see Thm.\ \ref{arsi} and the beginning of the Appendix \ref{proof-iaca} for details). 

In our case we have to follow the same procedure, starting from the algebra $C^{\alpha}(\hrr)$ instead of $C(\hrr)$. In turns out that the resulting algebra is {\em not} an algebra of functions on $\hrr$, because after a bounded limit, the top order term in the asymptotic expansion (\ref{asympto}) is not uniquely determined.  Instead the resulting algebra is a direct sum of a sub-algebra of bounded Borel functions satisfying (\ref{asympto-bis}) below, and of a finite dimensional space.

We first introduce some definitions.
Denote by $B(\hrr)$ the space of bounded Borel functions on $\hrr$.  
\begin{definition}\label{defdeder}
Let $\omega=(\xi,s)\in\hat\rr\times\nn$. We denote by $L^{\omega}(\hat\rr)$ the space of  functions $\varphi\in B(\hat\rr)$ such that there is a polynomial $P$ with:
\[
\varphi(x)=\left\{\begin{array}{l}
P(x-\xi)+O(|x-\xi|^{s}),  \hbox{ if }\xi\in \rr,\\[2mm]
P(1/x)+O(|x|^{-s}), \hbox{ if }\xi=\infty.
\end{array}
\right.
\]
\end{definition}
Again $L^{\mu}(\hrr)\subset L^{\omega}(\hrr)$ if $\mu\geq \omega$. If $\varphi\in L^{\omega}(\hrr)$ for 
$\omega= (\xi, s)$, $s\geq 1$, the terms of degree $<s$ of $P$ are uniquely determined, hence there is a unique sequence $\{\delta_{\mu}(\varphi)\}_{\mu<\omega}$  such that the rational function $T_{\omega}\varphi$  defined in (\ref{iac2}) satisfies 
\beq\label{asympto-bis}
\varphi(x)= T_{\omega}\varphi(x)+ O(|\chi_{\omega}(x)|).
\eeq
If $\omega= (\xi, 0)$ we set  $\delta_{\omega}(\varphi):= \varphi(\xi)$.  We equip $L^{\omega}(\hat\rr)$ with the norm $\|\varphi\|_{\omega}$ as before and if $\alpha$ is an order function, we introduce the space $L^{\alpha}(\hat\rr):= \bigcap_{\omega\preceq \alpha}L^{\omega}(\hat\rr)$ equipped with the norm $\|\varphi\|_{\alpha}$.

 Clearly $L^\alpha(\hrr)$ is a unital
  Banach $*$-algebra for the usual algebraic operations.

\begin{definition}\label{iaco}
Let $\tilde{\alpha}= \{(\xi, \alpha(\xi)): \xi\in \supp \alpha\}\subset \hrr\times \nn$. We set
\[
\Lambda^{\alpha}:= L^{\alpha}(\hat\rr)\oplus \cc^{\tilde{\alpha}},
\] 
and:
\[
\begin{array}{rl}
I:& C^{\alpha}(\hrr)\to \Lambda^{\alpha}\\
&\varphi\mapsto (\varphi, (\delta_{\omega}(\varphi))_{\omega\in \tilde{\alpha}}).
\end{array}
\]
\end{definition}
For $\varphi=(\varphi^{0}, (a_{\omega})_{\omega\in \tilde{\alpha}})\in \Lambda^{\alpha}$ and $\omega\preceq \alpha$,  we define
\[
\delta_{\omega}(\varphi):= \left\{
\begin{array}{rl}
&\delta_{\omega}(\varphi^{0})\hbox{ if } \omega\not\in \tilde{\alpha},\\
&a_{\omega}\hbox{ if }\omega\in \tilde{\alpha},
\end{array}
\right.
\]
which allows to write $\varphi$ as $(\varphi^{0}, (\delta_{\omega}(\varphi))_{\omega\in \tilde{\alpha}})$. We can then equip $\Lambda^{\alpha}$ with a $*-$algebra structure by setting:
\begin{align*}
\varphi\psi :=&
(\varphi^{\circ},(\delta_{\omega}(\varphi))_{\omega\in\tilde\alpha}) \cdot
(\psi^{\circ},(\delta_{\omega}(\psi))_{\omega\in\tilde\alpha}) \\
=&
\left(\varphi^{\circ}\psi^{\circ},
\big({\textstyle\sum_{\mu+\nu=\omega}}
\delta_{\mu}(\varphi)\delta_{\nu}(\psi)
\big)_{\omega\in\tilde\alpha}
\right),\\[1mm]
(\varphi^{0},  ( \delta&_{\omega}  (\varphi))_{\omega\in \tilde{\alpha}})^{*}:=(\overline{\varphi^{0}},  (\overline{\delta_{\omega}}(\varphi))_{\omega\in \tilde{\alpha}}). 
\end{align*}
It is easy to see that $\Lambda^{\alpha}$, equipped with the norm
\begin{equation}\label{eq:lambda}
\|\varphi\|_{\Lambda_{\alpha}}=
\max_{\omega\preceq\alpha}\max\big\{\|\varphi^{\circ}\|_{\omega},
{\textstyle\sum_{\mu\leq\omega}}|\delta_{\mu}(\varphi)| \big\}
\end{equation}
is a unital Banach $*$-algebra with $(1, 0)$ as unit. The embedding 
$I:C^{\alpha}(\hrr)\to\Lambda^{\alpha}$ is isometric hence 
$C^{\alpha}(\hrr)$ is identified with a closed $*$-subalgebra of 
$\Lambda^{\alpha}$.

\begin{definition}
 A sequence $(\varphi_{n})_{n\in \nn}$ in $\Lambda^{\alpha}$ is b-convergent to $\varphi$  if 
 $\sup_{n}\| \varphi_{n}\|_{\Lambda^{\alpha}}<\infty$ and $\lim_{n}\delta_{\omega}(\varphi_{n})= \delta_{\omega}(\varphi)$ for each $\omega\preceq \alpha$. 
\end{definition}
Clearly  the b-convergence of $(\varphi_{n})$ to $\varphi$ implies the b-convergence of $(\varphi^{0}_{n})$ to $\varphi^{0}$.

The main result of this subsection is the following theorem which is the natural extension of Thm.\ \ref{th:main} to the Borel case.
\begin{theoreme}\label{iaca}
  Let $H$ be a self-adjoint definitizable operator on a Krein space
  $\cK$ with $\sigma(H)\subset\rr$ and order function $\alpha_{H}$.
  Then there is a unique linear weakly b-continuous map
  \[
  \Lambda^{\alpha_{H}}\ni \varphi\mapsto\varphi(H)\in B(\cK)
  \] 
  such that  if 
  $\varphi=Ir_{z}$, with $r_{z}(\lambda)=(\lambda-z)^{-1}$  
  and $z\in\cc\backslash \rr$, 
  then $\varphi(H)=(H-z)^{-1}$.  This
  map is a norm continuous morphism of unital $*$-algebras.
\end{theoreme}
The proof will be given in Appendix \ref{secapp.3}. 
\begin{corollary}\label{iacu}
Let $H$ a self-adjoint definitizable operator as above.
 Let $J\subset \hrr$ an open set such that $\overline{J}\cap \supp \alpha_{H}= \emptyset$ and  $B_{J}(\hrr)\subset B(\hrr)$ be the $*-$ideal of functions   supported in $\overline{J}$. Then the map
 \[
 B_{J}(\hrr)\ni \varphi\mapsto \varphi(H):= (\varphi, 0)(H)\in B(\cK)
\]
is a $*-$morphism, continuous for the norm topologies of $B_{J}(\hrr)$ and $B(\cK)$ and weakly b-continuous.  
 \end{corollary}
\proof  Let us denote by $C_{J}(\hrr)\subset C(\hrr)$ the $*-$ideal of functions supported in $\overline{J}$. Clearly $C_{J}(\hrr)\subset C^{\alpha_{H}}(\hrr)$ isometrically. Moreover $I\varphi= (\varphi, 0)$ for $\varphi\in C_{J}(\hrr)$, if $I: C^{\alpha_{H}}(\hrr)\to \Lambda^{\alpha_{H}}$ is defined in Def.\ \ref{iaco}.  Finally if $\varphi_{n}\in B_{J}(\hrr)$ and ${\rm b-}\lim_{n} \varphi_{n}= \varphi$ then ${\rm b-}\lim_{n}(\varphi_{n}, 0)=(\varphi, 0)\in \Lambda^{\alpha}$.  These facts imply the corollary. \qed

\subsection{Existence of the dynamics}\label{exidyn}
Let us mention a well-known consequence of Corollary  \ref{iacu} about the existence of the dynamics generated by an even-definitizable operator.

Let  $H$ be \emph{even-definitizable}, $f_{t}:\ x\mapsto \e^{\i t x}$ and $\chi\in \coinf(\rr)$ such  all finite critical points  of $H$ are in the support of $\chi$.  We write $f_{t}= \chi f_{t}+ (1- \chi)f_{t}$, and extend $(1- \chi)f_{t}$ arbitrarily at $\infty$. We can define $(\chi f_{t})(H)$ by 
Thm.\ \ref{th:main} and $((1- \chi)f_{t})(H)$ by Corollary \ref{iacu}. We set then  
\[
f_{t}(H):= (\chi f_{t})(H)+ ((1- \chi)f_{t})(H)\in B(\cK),
\]
 which is independent on the choice of $\chi$ with the above properties.

 The space $\cK_{\rm pp}^{\cc}= \one_{\rm pp}^{\cc}(H)\cK$ is finite dimensional and invariant under $H$, hence we can obviously define $(\e^{\i tH})_{|\cK_{\rm pp}^{\cc} }$. We then set
 \[
\e^{\i t H}:= f_{t}(H)+ (\e^{\i tH})_{|\cK_{\rm pp}^{\cc} }, \ t\in \rr.
\]
It is easy to see that $\{\e^{\i t H}\}_{t\in \rr}$ is a $C_{0}-$group on $\cK$, with $(\e^{\i t H})^{\dag}= \e^{- \i t H}$, i.e. a unitary $C_{0}-$group on $(\cK, \langle \cdot| \cdot\rangle)$. Moreover $H$ is the generator of $\{\e^{\i t H}\}_{t\in \rr}$ and there exist $C, \lambda>0$, $n\in \nn$ such that
\beq\label{azer}
\| (\e^{\i tH})_{| \cK_{\rm pp}^{\cc}}\|\leq C \e^{\lambda|t|}, \
\| (\e^{\i tH})_{| (\cK_{\rm pp}^{\cc})^{\perp}}\|\leq C \langle t\rangle^{n}, \ t\in \rr.
\eeq

\section{Abstract Klein-Gordon equations}\label{abstract-kg}
Let us discuss in more details the Klein-Gordon equation (\ref{ei.1}).
The scalar product on $\cH$ will be denoted by $(u|v)$ or sometimes by $\overline{u}\cdot v$.

To associate a generator to (\ref{ei.1}) one has to turn this equation into a first order evolution equation. 
It turns out that there are several ways to do this, leading to different generators, and different topological spaces of Cauchy data. 

In order to present the results of this paper, we  first discuss these questions in an informal way, without worrying about the problems of existence, uniqueness or even the meaning of
solutions to (\ref{ei.1}).

\subsection{Symplectic setup}\label{sec0b.1}
The most natural approach   is to consider  $\cY=\cH\oplus \cH$ whose elements are denoted by $(\varphi,
\pi)$,  and to equip it with  the complex {\em symplectic form} 
(i.e.\ sesquilinear, non-degenerate, anti-hermitian):
\[
\overline{(\varphi_{1}, \pi_{1})}\omega(\varphi_{2}, \pi_{2}):= \overline{\pi_{1}}\cdot \varphi_{2}-\overline{\varphi_{1}}\cdot\pi_{2}  .
\]
The {\em classical Hamiltonian} is:
\[
E(\varphi, \pi):= \overline{(\pi+ \i k \varphi)}\cdot (\pi+ \i k \varphi)+ \overline{\varphi}\cdot h\varphi .
\]
We consider
$\omega, E$ as maps from $\cY$ to $\cY^{*}$, where $\cY^{*}$ is the
space of anti-linear forms on $\cY$ and set:
\beq\label{e0.002}
A:=-\i \omega^{-1}E=\mat{k}{-\i}{\i h_{0}}{k},
\eeq
for $h_{0}= h+ k^{2}$. In other words $\e^{\i t A}$ is the symplectic flow obtained from the classical Hamiltonian  $E$.

If we set
\[
\lin{\varphi(t)}{\pi(t)}:= \e^{\i t A}\lin{\varphi}{\pi}
\]
then
$\phi(t):= \varphi(t)$ solves the Cauchy problem:
\[\left\{
\begin{array}{l}
\p_{t}^{2}\phi(t)- 2\i k \p_{t}\phi(t)+ h \phi(t)=0,\\[2mm]
 \phi(0)= \varphi, \ \p_{t} \phi(0)= \pi+\i k \varphi.
\end{array}\right.
\]

\subsection{Quadratic pencils and stationary solutions}\label{sec0b.0}
If we look for a solution of (\ref{ei.1}) of the form $\phi(t)= \e^{\i t  z}\phi$ (or equivalently set $\i^{-1}\p_{t}=  z$), we obtain that $\phi$ should solve
\[
p( z)\phi=0, \hbox{ for }p( z)= h_{0}- (k- z)^{2}.
\]
The map $ z\mapsto p( z)$, called a {\em quadratic pencil}, is further  discussed in Subsect.\ \ref{sec0.2}.

\subsection{Charge setup}\label{sec0b.2}
Since we work on a complex symplectic space, it is more convenient to turn the symplectic form $\omega$ into a hermitian form. In fact 
setting 
\[
f:=\left(\begin{array}{l}
\varphi\\
\i^{-1}\pi
\end{array}\right)=\left(\begin{array}{l}
f_{0}\\f_{1}
\end{array}\right)
\]
the hermitian form $q:= \i \omega$, called the {\em charge}, takes the form:
\[
 \overline{f}q f= (f_{1}| f_{0})+ (f_{0}| f_{1}),
\]
and the energy $E$ becomes:
\[
 E(f,f)= \|f_{1}+ kf_{0}\|^{2}+ (f_{0}| h f_{0}).
\]
Note that from (\ref{e0.002}) we obtain
\begin{equation}
\label{e0.003}
E(f, f)= \overline{f}qKf.
\end{equation}
If
\[
 f(t):= \e^{\i tK}f, \hbox{ for }K:= \mat{k}{\one}{h_{0}}{k},
\]
then $\phi(t)= f_{0}(t)$ solves the Cauchy problem:
\[\left\{
\begin{array}{l}
\p_{t}^{2}\phi(t)- 2\i k \p_{t}\phi(t)+ h \phi(t)=0,\\[2mm]
 \phi(0)= f_{0}, \ \i^{-1}\p_{t} \phi(0)-k \phi(0)=f_{1}.
\end{array}\right.
\]
\subsection{PDE setup}\label{sec0b.3}
Finally let us describe  the standard setup used in 
partial differential equations.  We  set:
\[
f(t)= \e^{\i tH}f, \hbox{ for }H:= \mat{0}{\one}{h}{2k},
\]
and $\phi(t)= f_{0}(t)$ solves the Cauchy problem:
\[\left\{
\begin{array}{l}
\p_{t}^{2}\phi(t)- 2\i k \p_{t}\phi(t)+ h \phi(t)=0,\\[2mm]
 \phi(0)= f_{0}, \ \i^{-1}\p_{t} \phi(0)=f_{1}.
\end{array}\right.
\]
 The charge  and energy become:
 \[\begin{array}{rl}
 \overline{f}qf=&(f_{1}| f_{0})+ (f_{0}| f_{1})- 2 (f_{0}| k f_{0}),\\[2mm]
E(f, f)=& \|f_{1}\|^{2}+ (f_{0}|hf_{0}).
 \end{array}
\]
Note that if 
\[
\Phi= \mat{\one}{0}{k}{\one},
\]
then $H\Phi= \Phi K$.

\subsection{The choice of functional spaces}\label{sec0b.4}
Let us now discuss the choice of the possible topologies to put on the spaces of Cauchy data.  We will use the abstract Sobolev spaces $\langle h\rangle^{s}\cH$ and $|h|^{s}\cH$ associated to the self-adjoint operator $h$, whose definition and properties are given in Subsect.\ \ref{sec0.1}.

The first  natural choices correspond to topologies for  which the symplectic form $\omega$ is bounded.  
Note that our choice of $\cY= \cH\oplus \cH$ as symplectic space in Subsect.\ \ref{sec0b.1} was quite arbitrary. In fact we can choose a reflexive Banach space $\cG$ and set $\cY= \cG\oplus \cG^{*}$ equipped with
\[
\overline{g}\omega f:= \langle g_{1}| f_{0}\rangle -\langle g_{0}| f_{1}\rangle,
\]
where $\langle g_{0}| f_{1}\rangle= f_{1}(g_{0})$ and $\langle g_{1}| f_{0}\rangle = \overline{\langle f_{0}| g_{1}\rangle}$. Clearly $\omega$ is sesquilinear, anti-hermitian, non degenerate and bounded on $\cY$.

Examples of such symplectic spaces are  the {\em charge spaces}:
\[
\cK_{\theta}= \jh^{-\theta}\cH\oplus \jh^{\theta}\cH, \ \dot\cK_{\theta}=\sh^{-\theta}\cH\oplus \sh^{\theta}\cH, \ \theta\geq 0.
\]
In this case it is convenient to use the charge setup. An additional requirement is of course that $K$ should be well defined as a closed operator on $\cK_{\theta}$ or $\dot \cK_{\theta}$, possibly with non-empty resolvent set, and that $K$ be the generator of a strongly continuous group $\e^{\i tK}$.

Another possibility  often used in partial differential equations is to forget about the symplectic form and consider instead spaces on which the energy $E$ is bounded.   It is then more convenient  to use the PDE setup, and to work with the generator $H$. Reasonable  choices are then the {\em energy spaces}: 
\[
\cE= \jh^{-\12}\cH\oplus \cH, \ \dot \cE= \sh^{-\12}\cH\oplus \cH.
\]
To select  convenient spaces among all these, it suffices to consider the 'static' Klein-Gordon equation:
\beq\label{e0b.1b}
\p_{t}^{2}\phi(t) + \epsilon^{2}\phi(t)=0,
\eeq
corresponding to $h= \epsilon^{2}$, $k=0$ (we assume of course that $\epsilon\geq 0$ is unbounded). In this case we have
\[
H= K= \mat{0}{\one}{\epsilon^{2}}{0}=: H_{0}.
\]
On any space of Cauchy data, the group $\e^{\i tH_{0}}$ will be formally given by:
\[
 \e^{\i tH_{0}}= \mat{\cos( \epsilon t)}{\i \epsilon^{-1}\sin(\epsilon t)}{ \i \epsilon \sin(\epsilon t)}{\cos{\epsilon t}}.
\]
We see that  among these spaces the only ones on which $\e^{\i tH_{0}}$ is bounded are $\dot \cE$, $\cE$, $\dot\cK_{\4}$ and $\cK_{\4}$. The first two are the
usual homogeneous and non-homogeneous energy spaces. The last two are called the homogeneous and non-homogeneous {\em charge spaces}. Note that the space $\dot\cK_{\4}$ appears naturally as the {\em one-particle space} in the Fock quantization of (\ref{e0b.1b}).

In this paper we consider the two operators $H$ acting on $\cE$ and $K$ 
acting on $\cK_{\4}$.

\section{Klein-Gordon operators in energy spaces}\label{sec1}\init

In this section we discuss the properties of the operator $H$ in Subsect.\ \ref{sec0b.3} considered as acting on the energy spaces $\cE$ or $\dot \cE$.
\subsection{Non-homogeneous energy space}\label{sec1.1}

Let us fix  a self-adjoint operator $h$ on $\cH$ and  a bounded, symmetric operator $k:\h^{-\12}\cH\to \cH$ as in Subsect. \ref{sec0.2}.

The \emph{ energy space} $\cE$ and its adjoint space $\cE^{*}$ 
are defined by
\begin{equation}\label{eq:energy}
\cE:=\h^{-\12}\cH\oplus\cH \quad\text{and}\quad
\cE^{*}:=\cH\oplus\h^{\12}\cH,
\end{equation}
where we used the convention explained in Subsect. \ref{sec0.1}.
We have a continuous and dense embedding $\cE\subset\cE^{*}$.

\begin{lemma}\label{1.1}
\ben
\item $h: \h^{-\12}\cH\tilde\to\h^{\12}\cH$ iff $0\in \rho(h)$ iff $0\in \rho(h, k)$.
\item If $0\in \rho(h)$ then $\cE$ equipped with the hermitian sesquilinear form:
\[
\langle f| f\rangle_{\cE}:= (f_{0}| h f_{0})+ (f_{1}| f_{1})
\]
is a Krein space.
\item if in addition ${\rm Tr}\one_{]-\infty, 0]}(h)<\infty$,  then 
$(\cE, \ \langle\cdot|\cdot\rangle_{\cE})$ is Pontryagin.
\een
\end{lemma}  
\proof (1) follows from Lemma \ref{lm:kg}. (2) and (3) are immediate.  \qed

\subsection{Klein-Gordon operators on energy space}\label{sec1.2}

We set   \begin{equation}\label{eq:kg0}
\what H:= \mat{0}{\one}{h}{2k}\in B(\cE, \cE^{*}).
\end{equation}
\begin{definition}\label{def:energy}
The {\em energy Klein-Gordon operator}  is the operator $H$ induced by $\what H$ in $\cE$. Its domain is  
 given by 
\begin{equation}\label{eq:DH}
\Dom H:=\cD=  \h^{-1}\cH\oplus\h^{-\12}\cH= (\what H - z)^{-1}\cE, \  z\in \rho(h,k).
\end{equation}
We have 
\[
H=  \mat{0}{\one}{h}{2k}.
\]
\end{definition}

\begin{proposition}\label{pr:kge}
\ben
  \item one has $\rho(H)= \rho(h, k)$.
  \item  In particular, if
  $\rho(h,k)\neq\emptyset$ then $H$ is a closed densely defined
  operator in $\cE$ and its spectrum is invariant under complex
  conjugation.  
  \item If $ z\in\rho(h,k)$ then
\begin{equation}\label{eq:inver}
(H- z)^{-1}=p( z)^{-1}\mat{ z-2k}{1}{h}{ z}.
\end{equation}
\een
\end{proposition}
\proof We will prove (1) and (3). Note that (2) will follow then from Lemma \ref{lm:kg}.

 Let $ z\in \rho(H)$.  If $f_{0}\in\h^{-1}\cH$ with $p( z)f_{0}=0$, then $f=(f_{0},  z f_{0})\in {\rm Ker}(H- z)$ hence $f_{0}=0$.
 If $g_{1}\in \cH$ then  $g=(0, g_{1})\in \cE$ and if $f= (H- z)^{-1}g$ then $p( z)f_{0}=g_{1}$, hence $p( z):\h^{-1}\cH\tilde\to \cH$ and $
 z\in \rho(h, k)$. Therefore $\rho(H)\subset \rho(h,k)$.

Conversely let $ z\in \rho(h, k)$ so that $p( z):\h^{-1}\cH\tilde\to\cH$. We shall
show that $ z\in \rho(H)$ and 
\begin{equation}\label{eq:inverse}
(H- z)^{-1}=\mat{\ell w}{\ell}{\ell h}{ z \ell}, \ \ell=p( z)^{-1}, \ w=  z- 2k,
\end{equation}
completing the proof of (1) and (3).
One must interpret carefully the operators appearing in the matrix
above because $(H- z)^{-1}$ must send $\cE$ into $\cD$. More
precisely, since $hf_{0}\in\h^{\12}\cH$ if $f_{0}\in\h^{-\12}\cH$, the
factor $\ell$ in the product $\ell h$ is not the inverse of
$p( z):\h^{-1}\cH\tilde\to\cH$ but of its extension
$p( z):\h^{-\12}\cH\tilde\to\h^{\12}\cH$. We can do this thanks to
Lemma \ref{lm:kg}. Now a mechanical computation implies
\[
\begin{array}{rl}
&\mat{\ell w}{\ell}{\ell h}{ z \ell}
\mat{- z}{1}{h}{2k- z}\lin{f_{0}}{f_{1}}\\[4mm]
=&\mat{\ell w}{\ell}{\ell h}{ z \ell }
\lin{- z f_{0}+f_{1}}{hf_{0}-wf_{1}}
=\lin{f_{0}}{f_{1}},
\end{array}
\]
for all $f=(f_{0}, f_{1})\in \cD$. Similarly for  
 $g=(g_{0}, g_{1})\in\cE$ we compute
\[
\begin{array}{rl}
&\mat{- z}{1}{h}{2k- z}
\mat{\ell w}{\ell }{\ell h}{ z \ell }
\lin{g_{0}}{g_{1}}\\[4mm]
=&\mat{- z}{1}{h}{-w}
\lin{\ell wg_{0}+\ell g_{1}}{\ell hg_{0}+ z \ell g_{1}}
=\lin{g_{0}}{g_{1}}
\end{array}
\]
which holds because $h\ell w=w\ell h$ on $\h^{-\12}\cH$, where $\ell $ is the inverse of
$p( z):\h^{-\12}\cH\to\h^{\12}\cH$. Thus $ z\in \rho(H)$ and 
$(H- z)^{-1}$ is given by (\ref{eq:inverse}).
 \qed

\medskip

%
%
%
%
%
%
%

\begin{theoreme}\label{pr:sad}
Assume that $0\in \rho(h)$.
\ben
 \item  Then  $H$ is a self-adjoint operator on the Krein space $(\cE, \langle\cdot|\cdot\rangle_{\cE})$ with $\rho(H)\neq \emptyset$.
 \item If in addition ${\rm Tr}\one_{]-\infty, 0]}(h)<\infty$, then $H$ is even-definitizable.
 \een
\end{theoreme}

\proof If $0\in \rho(h)$ then $0\in \rho(h, k)= \rho(H)$ and from
\eqref{eq:inverse} we get
\begin{equation}\label{eq:inv}
H^{-1}=\mat{-2h^{-1}k}{h^{-1}}{\one}{0}.
\end{equation}
By Lemma \ref{stupid} it suffices to show that $(H^{-1})^{\dag}= H^{-1}$, which is a  simple computation.
This proves (1). Since any self-adjoint operator with non-empty resolvent set on a Pontryagin space is even-definitizable, (2) follows from Lemma  \ref{1.1}.
\qed

\subsection{Homogeneous energy space}\label{sec1.3}
Assume that ${\rm Ker}\,h=\{0\}$. Then we can introduce the 
 \emph{homogeneous energy space} \begin{equation}\label{eq:hon}
\dot\cE:=|h|^{-\12}\cH\oplus\cH,
\end{equation}
equipped with his canonical Hilbert space structure. Note that $\cE\subset \dot \cE$ continuously and densely. Of course $\cE= \dot \cE$ iff $0\in \rho(h)$, so the typical situation considered in the sequel is  $0\in \sigma(h)$.
 
The following analog of Lemma \ref{1.1} is obvious.
\begin{lemma}\label{1.2}
 Assume that ${\rm Ker}\,h=\{0\}$. Then $\dot\cE$ equipped with $\langle\cdot|\cdot\rangle_{\cE}$ is a Krein space. If in addition ${\rm Tr}\one_{]-\infty, 0]}
(h)<\infty$, then $\dot \cE$ is Pontryagin.
\end{lemma}
\subsection{Klein-Gordon operators on homogeneous energy space}\label{sec1.4}
\begin{definition}\label{def:hom-energy}
The {\em (homogeneous) energy Klein-Gordon operator}  is the operator $\dot H$ induced by $\what H$ in $\dot \cE$. Its domain is  
 given by 
\begin{equation}
\dot\cD=\left(|h|^{-\12}\cH\cap|h|^{-1}\cH\right)\oplus\h^{-\12}\cH= \{f\in \dot\cE : \what H f\in \dot\cE\}.
\end{equation}
which is continuously and densely embedded in $\dot\cE$. We have
\beq\label{defdeh}\dot{H}=\mat{0}{\one}{h}{2k}.
\eeq
\end{definition}
Since $\cE\subset \dot\cE$  and  $\cD\subset\dot\cD$ continuously  and densely,  $H$ may also be considered as an operator acting in
$\dot\cE$. We shall prove below that $\dot{H}$ is its closure in $\dot\cE$.

\begin{proposition}\label{ignobel}
\ben
\item  $\rho(\dot{H})=\rho(h,k)$.
\item In particular, if $\rho(h, k)\neq \emptyset$ then $\dot H$ is a closed densely defined operator in $\dot\cE$ and its spectrum if invariant under 
complex conjugation.
\item  For
$ z\in \rho(h, k)$, $ z\neq 0$ we have:
\begin{equation}\label{eq:invers}
(\dot{H}- z)^{-1}=\mat{ z^{-1}p( z)^{-1}h- z^{-1}}{p( z)^{-1}}
{p( z)^{-1}h}{ z p( z)^{-1}}.
\end{equation} 
\een
\end{proposition}

\begin{remark}
 {\rm It would be tempting to take the expression in (\ref{eq:inver}) for $(\dot{H}- z)^{-1}$. The trouble is that $kf_{0}$ does not have an obvious 
meaning under our assumptions on $k$ if $f_{0}\in |h|^{-\12}\cH$.  We obtain a meaningful formula for $(\dot{H}- z)^{-1}$ by noting that $(2k-
 z)=  z^{-1}(p( z)-h)$ for $ z\neq 0$.}
\end{remark}
\proof Let us first prove that $\rho(\dot H)\subset \rho(h,k)$. Let $ z\in \rho(\dot H)$.  Assume first that $ z\neq 0$.  Then for $g_{1}\in \cH$ 
and $g=(0, g_{1})\in \dot \cE$ there exists a unique $f=(f_{0}, f_{1})\in \dot\cD$ such that $(\dot H - z)f=g$ i.e.  $f_{1}=  z f_{0}$ and 
$p( z)f_{0}= g_{1}$. Since $f_{1}=  z f_{0}\in\cH$ and $ z\neq 0$ it follows that $f_{0}\in \h^{-1}\cH$ hence $p( z): \h^{-1}\cH\tilde\to 
\cH$  and $z\in \rho(h,k)$.

If $0\in \rho(\dot H)$, then for all $(g_{0}, g_{1})\in \dot \cE$ there exists a unique $(f_{0}, f_{1})\in|h|^{-1}\cH\cap |h|^{-\12}\cH
\oplus\langle h\rangle^{-\12} \cH$ with $f_{1}= g_{0}$ and $ hf_{0}+ 2kf_{1}=g_{1}$. This implies that $ |h|^{-\12}\cH= \h^{-\12}\cH$, hence $0\in \rho(h)$, hence $0\in 
\rho(h,k)$.

We now prove that $ \rho(h, k)\subset \rho(\dot H)$ and that (\ref{eq:invers}) holds for $ z\in \rho(h, k)$, $ z\neq 0$.

First, let  $ z\in \rho(h,k)$ with $ z\neq 0$, $g=(g_{0}, g_{1})\in \dot\cE$, and $(f_{0}, f_{1})$ given by the r.h.s. of (\ref{eq:invers}) applied to $g$.  
We begin by proving that $f\in \dot \cD$.

Note that $p(z)^{-1}g_{1}\in \h^{-1}\cH\subset |h|^{-1}\cH\cap |h|^{-\12}\cH$, and $hg_{0}\in |h|^{\12}\cH\subset \h^{\12}\cH$ hence 
$p( z)^{-1}hg_{0}\in \h^{-\12}\cH$. It follows that $f_{1}= p( z)^{-1}hg_{0}+  z p( z)^{-1}g_{1}\in\h^{-\12}\cH$.   The same argument 
shows that $f_{0}=  z^{-1}p( z)^{-1}h g_{0}-  z^{-1}g_{0}+ p( z)^{-1}g_{1}\in|h|^{-\12}\cH$. It remains to prove that $f_{0}\in |h|
^{-1}\cH$ i.e. that $ hf_{0}\in \cH$. Since $p( z)^{-1}g_{1}\in \h^{-1}\cH$ it suffices to prove that $ z^{-1}h(p( z)^{-1}h-\one)g_{0}\in \cH
$.
Note that
\[
 z^{-1}h(p( z)^{-1}h-\one)g_{0}=  z^{-1}(hp( z)^{-1}- \one)hg_{0}
= ( z- 2k)p( z)^{-1}hg_{0}.
\]
Since $g_{0}\in |h|^{-\12}\cH$, $hg_{0}\in \dot\h^{\12}\cH\subset \h^{\12}\cH$, we obtain that $p( z)^{-1}hg_{0}\in \h^{-\12}\cH$ hence $
( z- 2k)p( z)^{-1}hg_{0}\in \cH$. This completes the proof of the fact that $f\in \dot\cD$. 

It remains to prove that $(\dot H- z)f=g$, which is a standard computation.

Finally assume that $0\in \rho(h,k)$. Then $0\in \rho(h)$ which implies that $ \dot\cE= \cE$ and $\dot H = H$. Then by Prop. \ref{pr:kge}, 
$0\in \rho(H)$. This completes the proof of (1),  (3) and of the first statement of (2). 
 \qed

\medskip

\begin{theoreme}\label{prop:sadj}
Assume that there exists $ z\in \rho(h, k)$ with $ z\neq 0$. 
\ben
\item  Then $\dot H$ is self-adjoint on $(\dot\cE, \langle\cdot|\cdot\rangle_{\cE})$ and $\rho(H)\neq \emptyset$. 
\item If in addition ${\rm Tr}\one_{]-\infty, 0]}(h)<\infty$, then $\dot H$ is even-definitizable.
\een
\end{theoreme}
 \proof  An easy computation using (\ref{eq:invers}) shows that   $((\dot H- z)^{-1})^{\dag}= (\dot H- \bar{ z})^{-1}$. Then (1) follows from 
Lemma \ref{stupid}. (2) follows as before from Lemma \ref{1.2}. \qed

\medskip

We now describe the relationship between the two operators $H$ and $\dot H$.

\begin{proposition}\label{prop:equal}
 \ben
 \item $\dot H$ is the closure of $H$ in $\dot\cE$;
 \item for $ z\in \rho(h, k)$, $ z\neq 0$, $(\dot H- z)^{-1}$ maps $\cE$ into $\cD$  and $(H-  z)^{-1}= (\dot H-  z)^{-1}_{\mid \cE}$;
 \item there exists $C>0$ such that for all $ z\in \rho(h, k)$, $ z\neq 0$ one has:
 \[
\| (H- z)^{-1}g\|_{\cE}\leq C ((1+ | z|^{-1}) \| (\dot H-  z)^{-1}g\|_{\dot\cE}+  | z|^{-1}\| g\|_{\cE}), \ g\in \cE.
\]
 \een
\end{proposition}
\proof If $f=(f_{0}, f_{1})\in\dot\cD$, we pick a sequence $f_{0}^{n}\in \h^{-1}\cH$ with $f_{0}^{n}\to f_{0}$ in $|h|^{-1}\cH\cap |h|^{-\12}\cH$. 
Then $f^{n}= (f_{0}^{n}, f_{1})\in \cD$ and $f^{n}\to f$ in $\dot\cD$, $H f^{n}\to \dot H f$ in $\dot \cE$, which proves (1).
To prove (2) it suffices to note that $(2k-  z)=  z^{-1}(p( z)- h)$ on $\h^{-\12}\cH$, which proves that on $\cE$ the r.h.s. of 
(\ref{eq:inver}) and (\ref{eq:invers}) coincide. 
To prove (3) we use that  $\| f\|_{\cE}\sim \| f\|_{\dot \cE}+ \| f_{0}\|_{\cH}$. If $f= (H-  z)^{-1}g$ then $f_{0}=  z^{-1}(f_{1}- g_{0})$, hence \[
\|f_{0}\|_{\cH}\leq | z|^{-1}(\|f_{1}\|_{\cH}+ \|g_{0}\|_{\cH})\leq | z|^{-1}( \| (H- z)^{-1}g\|_{\dot\cE}+ \| g\|_{\cE}),
\]
which proves (3).
\qed

\medskip

Proposition \ref{prop:equal} has some direct consequences for the
estimates on the quadratic pencil that we collect below.
\begin{corollary}
\label{penest}
Assume that there exists $ z\in \rho(h, k)$, $ z\neq 0$ and that
${\rm Tr}\one_{]-\infty, 0]}(h)<\infty$. Then we have the following
estimates on the quadratic pencil:
\begin{eqnarray*}
\Vert \langle h\rangle^{\12}p( z)^{-1}\Vert_{B(\cH)}\leq \left\{\begin{array}{c}C((1+\vert z\vert^{-1})\vert {\rm
    Im} z\vert^{-m-1}+\vert  z\vert^{-1}),\,  z\in U_0(R,a),\\
C((1+\vert z\vert^{-1})\langle z\rangle^k\vert {\rm Im} z\vert^{-1}+\vert  z\vert^{-1}),\,  z\in
U_{\infty}(R,a). \end{array}\right.,\\
\Vert p( z)^{-1}\Vert_{B(\cH)}\leq \left\{\begin{array}{c}\frac{C}{\vert  z\vert}((1+\vert z\vert^{-1})\vert {\rm
    Im} z\vert^{-m-1}+\vert  z\vert^{-1}),\,  z\in U_0(R,a),\\
\frac{C}{\vert  z\vert} ((1+\vert z\vert^{-1})\langle z\rangle^k\vert {\rm
  Im} z\vert^{-1}+\vert z\vert^{-1}),\,  z\in U_{\infty}(R,a). \end{array}\right.
\end{eqnarray*}
\end{corollary}
\proof
By Corollary  \ref{besson} and Proposition \ref{prop:equal} we obtain:
\begin{eqnarray*}
\Vert (H- z)^{-1}\Vert_{B(\cE)}\leq \left\{\begin{array}{c}C((1+\vert z\vert^{-1})\vert {\rm
    Im} z\vert^{-m-1}+\vert  z\vert^{-1}),\,  z\in U_0(R,a),\\
C ((1+\vert z\vert^{-1})\langle z\rangle^k\vert {\rm
  Im} z\vert^{-1}+\vert z\vert^{-1},\,  z\in U_{\infty}(R,a). \end{array}\right.
\end{eqnarray*}
Using \eqref{eq:inver} we see that we have
\begin{eqnarray*}
\Vert (H- z)^{-1}(0,f)\Vert^2_{\cE}=\Vert\langle
h\rangle^{\12}p( z)^{-1}f\Vert^2_{\cH}+\vert  z\vert^2 \Vert p( z)^{-1}f\Vert^2_{\cH},
\end{eqnarray*}
which gives the result. 
\qed

\section{Klein-Gordon operators in charge spaces}\init\label{sec5}

In this section we discuss in a way parallel to Sect.\ \ref{sec1} 
the properties of the operator $K$ in Subsect.\ \ref{sec0b.2} considered as acting on the non-homogeneous charge space $\cK_{\4}$ introduced in 
Subsect.\ \ref{sec0b.4}.

Note that $K$ acting on the homogeneous charge space $\dot \cK_{\4}$ could also be considered, at the price of some technical complications. 

\subsection{Non-homogeneous charge spaces}\label{sec5.1}

In this subsection, we consider a pair of operators $(h, k)$ satisfying the conditions in Subsect.\ \ref{sec0.2}. Note that by duality and interpolation we see that
\begin{equation}
\label{e5.0}
k\in B(\langle h\rangle^{-\theta}\cH, \langle h\rangle^{\12-\theta}\cH), \ 0\leq \theta\leq \12.
\end{equation}
We define the {\em (non-homogeneous) charge spaces of order $\theta$}:
\begin{equation}
\label{e5.1}
\cK_{\theta}:= \langle h\rangle^{-\theta}\cH\oplus \langle h\rangle^{\theta}\cH, \ 0\leq \theta\leq \12.
\end{equation}
and observe that $\cE\subset \cK_{\theta}\subset\cE^{*}$ continuously and densely. 
Note also that if
\begin{equation}
\label{e5.3}
q(f, g):= (f_{0}|g_{1})_{\cH}+ (f_{1}| g_{0})_{\cH}
\end{equation}
then $(\cK_{\theta}, q)$ are Krein spaces.

As we saw in Subsect.\ \ref{sec0b.4}, 
the middle space
\begin{equation}
\label{e5.2}
\cF:= \cK_{\4},
\end{equation}
which equals the complex interpolation space $[\cE, \cE^{*}]_{\12}$ is natural even in the case of free Klein-Gordon equations.   We will forget the order $\4$ and call it the {\em non-homogeneous charge space}.
\subsection{Klein-Gordon operators on non-homogeneous charge space}\label{sec5.2}
We set 
\[
\hat{K}:= \mat{k}{\one}{h_{0}}{k}\in B(\cE, \cE^{*}).
\]
Note that there is a simple relation between $\hat{K}$ and  $\hat{H}$ defined in (\ref{eq:kg0}):  indeed, if
\begin{equation}\label{eq:phi}
\Phi=\Phi(k)=\mat{\one}{0}{k}{\one}
\quad\text{hence}\quad
\Phi\lin{f_{0}}{f_{1}}=\lin{f_{0}}{kf_{0}+f_{1}}
\end{equation}
then a straightforward computation using (\ref{e5.0}) gives 
\begin{lemma}\label{lm:eckg}
  The map $\Phi=\Phi(k):\cE^{*}\to\cE^{*}$ is an isomorphism with
  inverse $\Phi^{-1}=\Phi(-k)$. The subspaces $\cE$ and $\cF$ are
  stable under $\Phi$ and the restrictions of $\Phi$ to these
  subspaces are bijective. We have $\hat H\Phi=\Phi \hat K$.
\end{lemma}

\begin{definition}\label{defdekaka}
 The {\em charge Klein-Gordon operator} is the operator $K$ induced by $\hat K$ in $\cF$. Its domain is given by
 \begin{equation}
\label{e5.4}
\Dom\, K:=\{\ f\in \cF  : \hat K f\in \cF \}.
\end{equation}
We have
\[
K=\mat{k}{\one}{h_{0}}{\one}.
\]
\end{definition}

\begin{proposition}\label{pr:kgc}
\ben
\item One has   $\rho(K)=\rho(h,k)$.  
\item In particular, if
  $\rho(h,k)\neq\emptyset$ then $K$ is a closed, densely defined 
  operator in $\cF$ and its spectrum is invariant under complex conjugation.
  \item If $ z\in \rho(h, k)$ then
    \begin{equation}\label{eq:cinverse}
  (K- z)^{-1}=
  \mat{-p( z)^{-1}(k- z)}{p( z)^{-1}}{\one+(k- z)p( z)^{-1}(k- z)}{-(k- z)p( z)^{-1}}.  
\end{equation}
  \een
  \end{proposition}
\proof It suffices to prove (1) and (3).  We will set $l= p( z)^{-1}$, $u= k- z$ to simplify notation.

Let $ z\in \rho(K)$. If
$f_0\in\langle h\rangle^{-\12}\cH$ and $f_1=-uf_0\in\cH$ then
$h_{0}f_0+uf_1=(h_{0}-u^{2})f_0=p( z)f_0$ hence
$(K_\theta- z)(f_0,f_1)^{t}=(0,p( z)f_0)^{t}$. Thus if
$p( z)f_0=0$ then $(K- z)(f_0,f_1)^{t}=0$, in
particular $(f_0,f_1)^{t}\in \Dom K$, and so $f_0=0$. Hence
$p( z):\langle h\rangle^{-\12}\cH\to\langle h\rangle^{\12}\cH$ is injective. Now let
$b\in\cH$. Since $(K- z)\Dom K=\cF$ and
$(0,b)^{t}\in\cF$, there are $f_0\in\langle h\rangle^{-\12}\cH$ and
$f_1\in\cH$ such that $K_\theta(f_0,f_1)^{t}=(0,b)^{t}$, hence
$uf_0+f_1=0$ and $h_{0}f_0+uf_1=b$, or $p( z)f_0=b$. But
$p( z)=h- z^{2}+2 z k$ hence $hf_0=b+ z^{2}f_0-2 z
kf_0\in\cH$ so $f_0\in\langle h\rangle^{-1}\cH$. This proves that
$p( z)\langle h\rangle^{-1}\cH=\cH$ and so
$p( z):\langle h\rangle^{-1}\cH\tilde\to\cH$ and $ z\in \rho(h,k)$.

Conversely let $ z\in \rho(h,k)$, so that $p( z): \langle h\rangle^{-\12}\cH\tilde\to  \langle h\rangle^{\12}\cH$. Let
\[
G=\mat{-\ell u}{\ell }{\one+u\ell u}{-u\ell} 
\]
be the r.h.s. of (\ref{eq:cinverse}). Clearly $G\in B(\cE, \cE^{*})$  and a
simple computation gives $(\hat K- z)G=\one$ on $\cE^*$ and
$G(\hat K- z)=\one$ on $\cE$.  So $G$ is the inverse of
$\hat K- z:\cE\to\cE^*$.  If $a\in\langle h\rangle^{-\12}\cH$, $b\in\langle h\rangle^{\12}\cH$ and
$(f_0,f_1)^t:=G(a,b)^t$ then $uf_0+f_1=a$ and $h_0f_0+uf_1=b$ hence
$(f_0,f_1)^t\in \Dom K$. Thus $G\cF\subset
\Dom K$. Reciprocally, if $(f_0,f_1)^t\in \Dom K$ then
$(a,b)^t:=(K- z)(f_0,f_1)^t$ belongs to $\cF$ by
\eqref{e5.4} and $G(a,b)^t=(f_{0}, f_{1})^t$ by the preceding
computation. Thus $\Dom K\subset G \cF$. So 
$G\cF=\Dom K$ and
$K- z:\Dom K\tilde\to\cF$ with inverse given
by the restriction of $G$ to $\cF$.
 \qed

\medskip

We deduce from Prop. \ref{pr:kgc} the following analog of Thm. \ref{pr:sad}.
\begin{theoreme}\label{5.2}
 Assume that $\rho(h,k)\neq \emptyset$. Then $K$ is a self-adjoint operator on the Krein space $(\cF, q)$ with $\rho(K)\neq \emptyset$.
\end{theoreme}

Since we saw that $\hat H= \Phi \hat K\Phi^{-1}$ and $\Phi$ preserves $\cF$, it is instructive to describe the operator $\Phi K \Phi^{-1}$.
 Note that if we compute the image of the canonical Krein structure $q$ on $\cF$ under 
 $\Phi$ we get:
\beq\label{def-de-q-prime}
 q'(f, g):= q(\Phi^{-1}f, \Phi^{-1}g)= q(f, g)- 2 (f_{0}| k g_{0})_{\cH}.
\eeq
\begin{lemma}
 \ben
 \item  $\Phi K \Phi^{-1}$  is equal to the operator induced by $\hat H$ on $\cF$;
 \item $\Phi K \Phi^{-1}$  is equal to the restriction of $\hat H$ to the domain
 \[
\Phi\Dom K= \langle h\rangle^{-3/4}\cH\oplus \langle h \rangle^{-1/4}\cH= [\Dom H, \cE]_{\12}.
\]
 \een
\end{lemma}
\proof (1) is obvious, (2) is a routine computation. \qed

\section{Definitizable Klein-Gordon operators on energy spaces}\label{sec1b}\init
In this section, we describe some basic properties of a class of definitizable Klein-Gordon operators on the energy spaces $\cE$, $\dot\cE$. We also describe an approximate diagonalization of $\dot H$, which will be useful later on.

We will assume
\[
 \begin{array}{rl}
{\rm (E1)}& \Ker\, h=\{0\}\\[2mm]
{\rm (E2)}& {\rm Tr}\one_{]-\infty, 0]}(h)<\infty,\\[2mm]
{\rm (E3)}& k|h|^{-\12}\in B(\cH).
\end{array}
\]
 Condition (E3) implies 
 \[
{\rm  (E3')} \ k\langle h\rangle^{-\12}\in B(\cH),
 \]
  hence the results of Sect. \ref{sec1} hold. Moreover (E2) implies that $h$ is bounded below, hence $\rho(h,k)\neq \emptyset$ by Prop. \ref{pr:idiot}.

We set
\[
m^{2}:= \inf \sigma(h)\cap \rr^{+}, \ m\geq 0.
\]
The constant $m$ is called the {\em mass}, Klein-Gordon equations will be called {\em massive} resp.\ {\em massless} if $m>0$
resp. $m=0$.  A more common name for  a massless Klein-Gordon equation is of course a {\em wave equation}.

Note that (E1) and (E2) imply that $\sigma_{\rm ess}(h)\subset \rr^{+}$. Moreover if (E1), (E2) hold and $m>0$ then $0\not\in \sigma(h)$ hence $\sh \sim \jh$ hence $\dot \cE= \cE$ and $\dot H=H$. By Thm.\ \ref{prop:sadj}, we obtain that if (E) holds then $\dot\cE$ equipped with $\langle\cdot|\cdot\rangle_{\cE}$ if a Pontryagin space and $\dot H$ defined in Def.\ \ref{def:hom-energy} is even-definitizable.

\begin{proposition}\label{1.11}
Assume (E) and let $U$ be a compact set with  $U\subset\rho(\dot H)$ if
$m>0$ and $U\subset \rho(\dot H)\backslash \{0\}$ if $m=0$. Then there exists $C>0$ such that:
 \beq\label{e1.42}
\|(\dot H-  z)^{-1}\|_{B(\dot\cE^{*}, \dot \cE)}\leq C+ C\|(\dot H-  z)^{-1}\|_{B(\dot\cE)}, \  z\in U.
\eeq
\end{proposition}
\proof 
 If $m=0$  and $ z\in \rho(\dot H)$, $ z\neq 0$, then  $(\dot H - z)^{-1}$  is  given by the r.h.s. of (\ref{eq:inver}), using that $k\in B(\sh^{-\12}\cH, \cH)$. This also implies that $p( z)^{-1}h = \one +  z p( z)^{-1}( z- 2k)$.  Then an easy computation shows that $(\dot H- z)^{-1}\in B(\dot \cE^{*},\dot \cE )$.   If $m>0$  and $ z\in \rho(\dot H)$ then the same result holds 
using that $H= \dot H$, $\dot \cE= \cE$.
 
 Let us prove  the bound (\ref{e1.42}). We assume $m=0$, the proof for $m>0$ being simpler. We have:
\beq\label{sego7}
\begin{array}{rl}
\|(\dot H-  z)^{-1}\|_{B(\dot\cE^{*}, \dot \cE)}\leq &C \left(\|\sh^{\12} p( z)^{-1}( z-2k)\|_{B(\cH)}+ 
\| \sh^{\12}p( z)^{-1}\epsilon\|_{B(\cH)}\right.\\[2mm]
&+\left. \|  z p( z)^{-1}( z- 2k)\|_{B(\cH)}+ p( z)^{-1}\sh^{\12}\|_{B(\cK)}\right)\\[2mm]
\leq& C\| \jh^{\12} p( z)^{-1}\jh^{\12}\|_{B(\cK)}, \  z\in U.
\end{array}
\eeq
Next from the expression (\ref{eq:invers}) of $(\dot H - z)^{-1}$ we obtain that
\beq\label{e1.40}
\|\sh^{\12} p( z)^{-1}\|_{B(\cH)}+ \| p( z)^{-1}\|_{B(\cH)}\leq C \| (\dot H - z)^{-1}\|_{B(\dot \cE)},
\eeq
hence
\[
\|\jh^{\12} p( z)^{-1}\|_{B(\cH)}\leq C  \| (\dot H - z)^{-1}\|_{B(\dot \cE)}.
\]
Taking adjoints and using that $p( z)^{*}= p(\bar  z)$, we also get
\[
\|p( z)^{-1}\jh^{\12}\|_{B(\cH)}\leq C \|(\dot H -\bar z)^{-1}\|_{B(\dot \cE)}\leq C'\| (\dot H - z)^{-1}\|_{B(\dot \cE)},
\]
using (\ref{krein.1}).
 Since $p( z)^{-1}h= \one + p( z)^{-1}( z- 2k)$, we obtain 
 for $ z\in U$:
\[
\|p( z)^{-1}h\|_{B(\cH)}\leq C+ C\|p( z)^{-1}\jh^{\12}\|_{B(\cH)}
\leq C+ C\| (\dot H - z)^{-1}\|_{B(\dot \cE)}.
\]
By (\ref{e1.40}) we have the same bound for $\|p( z)^{-1}\langle h\rangle\|_{B(\cH)}$ and for $\|\langle h \rangle p( z)^{-1}\|_{B(\cH)}$ by taking adjoints. By interpolation we obtain for $z\in U$
\[
\|\langle h \rangle^{\12}p( z)^{-1}\langle h\rangle^{\12}\|_{B(\cH)}\leq C+ 
C\| (\dot H - z)^{-1}\|_{B(\dot \cE)}
\]
which using  (\ref{sego7}) completes the proof of (\ref{e1.42}). \qed
\subsection{Functional calculus}\label{sec-fcalc}
We saw that under conditions (E1), (E2), (E3'), the operator $\dot H$  is even-definitizable, hence admits a 
C$^{\alpha}$ and a $\Lambda^{\alpha}$ functional calculus,  see Subsects. \ref{seckrein.3}, \ref{borelborel}.

In this subsection, we discuss the functional calculus for $H$, in the case $m=0$, which is not completely straightforward, since in this case  
$(\cE, \ \langle\cdot|\cdot\rangle_{\cE})$
is not a Krein space.  We set
\[
\alpha_{H}:= \alpha_{\dot H}+ \one_{\{0\}},
\]
where $\alpha_{\dot H}$ is the order function of $\dot H$, (see Def. \ref{isap}).

\begin{proposition}\label{fcalc-bis}
 Assume (E1), (E2), (E3')   \ben 
    \item there exists a unique  continuous $*-$morphism
  \[
 C^{\alpha_{H}}(\hrr)\ni \varphi\mapsto \varphi(H)\in B(\cE),
 \]
 such that if $\varphi(\lambda)= (\lambda-z)^{-1}$ for $z\in \rho(H)\backslash \rr$ then $\varphi(H)= (H-z)^{-1}$.

  \item there exists a unique extension of the above map  to a weakly b-continuous map
\[
\Lambda^{\alpha_{H}}(\hrr)\ni \varphi\mapsto \varphi(H)\in B(\cE),
\]
which is a norm continuous $*-$morphism.
\een
\end{proposition}
\proof  By Prop. \ref{prop:equal}  $(H-z)^{-1}= (\dot H-z)^{-1}_{|\cE}$ for $z\in \rho(\dot H)$, $z\neq 0$.  This implies (see Prop. \ref{coco}) that $\one_{\rm pp}^{\cc}(\dot H)$ maps $\cE$ into itself and defines a bounded projection on $\cE$, naturally denoted by $\one_{\rm pp}^{\cc}(H)$, which commutes with $H$. Let us set $\cE_{1}:= (\one - \one_{\rm pp}^{\cc}(H))\cE$, which is a closed subspace of $\cE$, invariant under $(H-z)^{-1}$ for $z\in \rho(H)$.  Replacing $H$ by $H_{|\cE_{1}}$, we see that without loss of generality we can assume that $\cE_{1}= \cE$. 

Let us set $\alpha= \alpha_{\dot H}$, $\beta= \alpha_{H}= \alpha + \one_{\{0\}}$. 
For $\varphi\in C^{\beta}(\hrr)$ with $\varphi(0)=0$ we set $\tilde{\varphi}(x)= x^{-1}\varphi(x)$. Clearly $\tilde{\varphi}\in C^{\alpha}(\hrr)$ and there exists $C>0$ such that
\beq\label{iacub}
\| \tilde{\varphi}\|_{\alpha}\leq C \| \varphi\|_{\beta}, \ \forall \ \varphi\in C^{\beta}(\hrr) \hbox{ with }\varphi(0)=0.
\eeq
We claim that $\varphi(\dot H)$ is bounded from $\cE$ into itself. In fact if $g\in \cE$ we have:
\[
\begin{array}{rl}
\| \varphi(\dot H)g\|_{\cE}\leq &\|\varphi(\dot H)g\|_{\dot\cE}+ \| (\varphi(\dot H)g)_{0}\|_{\cH}=\|\varphi(\dot H)g\|_{\dot\cE}+ \| (\dot H\tilde\varphi(\dot H)g)_{1}\|_{\cH}\\[2mm]
=&\|\varphi(\dot H)g\|_{\dot\cE}+ \| (\tilde\varphi(\dot H)g)_{1}\|_{\cH}\leq \|\varphi(\dot H)g\|_{\dot\cE}+ \| \tilde\varphi(\dot H)g\|_{\dot\cE}.
\end{array}
\]
Moreover from the above inequality, Thm. \ref{th:main} applied to $\dot H$ and  (\ref{iacub}) we obtain that
\[
\|\varphi(\dot H)\|_{B(\cE)}\leq C (\| \varphi\|_{\alpha}+ \| \tilde{\varphi}\|_{\alpha})\leq C\| \varphi\|_{\beta}.
\]
Now for $\varphi\in C^{\beta}(\hrr)$ arbitrary we set $\psi(x)= \varphi(x)- \varphi(0)$ and: 
\[
\varphi(H):= \varphi(0)\one_{\cE} + \psi(\dot H)_{|\cE}.
\]
From the fact that $(H-z)^{-1}= (H-z)^{-1}_{|\cE}$, we see that $\varphi(H)= (H-z)^{-1}$ if $\varphi(x)= (x-z)^{-1}$. This yields the existence of the $*-$morphism in (1). The uniqueness follows from the density of the space of bounded rational functions in $C^{\beta}(\hrr)$, see \cite[Lemma 4.7]{GGH1}.  We deduce (2) from (1) by the argument explained in Appendix \ref{secapp.3}. \qed
 \medskip

\begin{remark}\label{fcalc-for-K}
 {\rm It is easy to construct a  similar functional calculus for the operator $K$ considered in Subsect. \ref{sec5.2}. In fact if $\varphi$ belongs to one of the algebras in Prop. \ref{fcalc-bis}, then $\varphi(H)$ is bounded on $\cE$ and thus on $\cE^*$ by duality. Recalling that $\cF=[\cE,\cE^*]_{1/2}$ we see by complex interpolation that $\varphi(H)$ defines a bounded operator on $\cF$ with similar estimates. We then define 
\[\varphi(K)=\Phi^{-1}\varphi(H)\Phi,\]
which is well defined because $\Phi$ and $\Phi^{-1}$are bounded on $\cF$.}
\end{remark}

\subsection{Essential spectrum of Klein-Gordon operators}\label{sec1b.7}
We now investigate the essential spectrum of the operators $H$ and $\dot H$.  We set 
\[
H_{0}= \mat{0}{\one}{h}{0},
\]
defined as in Def.\ \ref{def:energy} for $k=0$, so that $\Dom H_{0}= \jh^{-1}\cH \oplus \jh^{-\12}\cH$.

Similarly  we set
\[
\dot H_{0}= \mat{0}{\one}{h}{0},
\]
defined as in Def.\ \ref{def:hom-energy}, with domain $\dot\cD_{0}= \sh^{-1}\cH\cap \sh^{-\12}\cH\oplus \jh^{-\12}\cH$. Clearly
\[
\sigma_{\rm ess}(H_{0})= \sigma_{\rm ess}(\dot H_{0})= \sqrt{\sigma_{\rm ess}(h)}\cup-\sqrt{\sigma_{\rm ess}(h)} .
\]
(Recall that from (E) we saw that $\sigma_{\rm ess}(h)\subset \rr^{+}$).

We introduce the  condition:
\[
\hbox{ (A4)}\ k \jh^{-\12}\in B_{\infty}(\cH).
\]
\begin{proposition}\label{1.4old}
 Assume (E), (A4). Then:
\[
\begin{array}{rl}
(1)&(H-  z)^{-1}- (H_{0}-  z)^{-1}\in B_{\infty}(\cE^{*}, \cE), \  z\in \rho(H)\cap \rho(H_{0}),\\[2mm]
(2)&\sigma_{\rm ess}(H)=  \sigma_{\rm ess}(\dot H)=\sqrt{\sigma_{\rm ess}(h)}\cup -\sqrt{\sigma_{\rm ess}(h)} .
\end{array} 
\]
\end{proposition}

\proof 
By (A4)  we obtain that $H-H_{0}\in B_{\infty}(\cE, \cE^{*})$ which by the resolvent formula implies that  $(H- z)^{-1}- (H_{0}-  z)^{-1}\in B_{\infty}( \cE^{*}, \cE)\subset B_{\infty}(\cE)$. This implies that $\sigma_{\rm ess}(H)= \sigma_{\rm ess}(H_{0})$. 
Since by (E2) $h\one_{\rr^{-}}(h)\in B_{\infty}(\cH)$ we see by the same argument that $\sigma_{\rm ess}(H_{0})= \sigma_{\rm ess}(H_{1})$ for $H_{1}=\mat{0}{\one}{h\one_{\rr^{+}}(h)}{0}$. Using the arguments at the beginning of Subsect.\ref{sec1b.8}, we obtain that $\sigma_{\rm ess}(H_{1})= \sqrt{\sigma_{\rm ess}(h)}\cup-\sqrt{\sigma_{\rm ess}(h)}$, which proves (2) for $H$.

To prove (2) for $\dot H$ we use again the second resolvent formula, hypothesis (A4) and the fact that $(\dot H_{0}-  z)^{-1}$ maps $\dot \cE$ into $\Dom\, \dot H_{0}$. We obtain that $\sigma_{\rm ess}(\dot H)= \sigma_{\rm ess}(\dot H_{0})$. We conclude as in the case of $H$.
\qed

\medskip

For completeness we state the following proposition. 
 
\begin{proposition}\label{1.20}
 Assume  (E). Then:
 \[
\begin{array}{rl}
(1)&\sigma(H)= \sigma(\dot H),\\[2mm]
(2)&\sigma_{\rm p}(H)= \sigma_{\rm p}(\dot H).
\end{array}
\]
\end{proposition}
\proof 
By Prop. \ref{pr:kge} (1) and Prop. \ref{ignobel}, we see that $\rho(H)= \rho(\dot H)= \rho(h,k)$, which proves (1). To prove (2) note that since $H\subset \dot H$ we have $\sigma_{\rm p}(H)\subset \sigma_{\rm p}(\dot H)$. Moreover we easily see that $0\in \sigma_{\rm p}(H)\Leftrightarrow 0\in \sigma_{\rm p}(h)\Leftrightarrow 0\in \sigma_{\rm p}(\dot H)$. Since by (E1) ${\rm Ker}\,h=\{0\}$ we obtain that $0\not\in \sigma_{\rm p}(H)\cup \sigma_{\rm p}(\dot H)$.
Finally if $f\in {\rm Ker}\,(\dot H - z)$, $ z\neq 0$, we see that $f\in \cE$,  hence $\dot H f= \hat H f\in \cE$ and $f\in {\rm Ker}\,(H- z)$. 
Hence $\sigma_{\rm p}(\dot H)\subset \sigma_{\rm p}(H)$, which  completes the proof of (2). \qed

\subsection{Approximate diagonalization}\label{sec1b.8}

It will be convenient later  to diagonalize as much as possible the operator $\dot H$.  This can be done by extracting a convenient positive part from $h$.

We assume that 
\[
h= b^{2}-r \hbox{ with }
\]
\[
\begin{array}{rl}
{\rm (A1)}& b\geq 0, \ \hbox{ self-adjoint on }\cH, \ b^{2}\sim \sh,\\[2mm]
{\rm (A2)}& r\hbox{ symmetric on }\jh^{-\12}\cH, \ b^{-1}rb^{-1}\in B(\cH).
\end{array}
\]
From (A1), (E1) and the Kato-Heinz inequality (see page \pageref{p:KH}) 
we see that:
\begin{equation}
\label{e1.30}
\Ker b=\{0\}, \ \langle b\rangle^{s}\cH=\jh^{s/2}\cH, \ b^{s}\cH=\sh^{s/2}\cH, \ |s|\leq 1.
\end{equation}
Set
\[
U:= \frac{1}{\sqrt{2}}\mat{b}{\one}{b}{-\one}, \ U^{-1}=  \frac{1}{\sqrt{2}}\mat{b^{-1}}{b^{-1}}{\one}{-\one}.
\]
We see using (\ref{e1.30}) that 
\beq\label{e1.31}
U: \dot\cE\tilde\to \cH\oplus \cH=: \cK, \ U: \dot\cE^{*}\tilde\to b\cH\oplus b \cH= |L_{0}|\cK,
\eeq
for
\[
L_{0}:=  \mat{b}{0}{0}{-b}= U\mat{0}{\one}{b^{2}}{0}U^{-1}.
\] 
We will also use the space:
\[
\langle L_{0}\rangle \cK= \langle b\rangle\cH\oplus  \langle b\rangle\cH= \jh^{\12}\cH\oplus \jh^{\12}\cH.
\]
Note that  $\langle L_{0}\rangle \cK= U \cE^{*}$ iff $m>0$.

We have
\beq\label{e1.25}
\begin{array}{rl}
L:=& U\dot H U^{-1}= L_{0}+ V_{1}+ V_{2},\\[2mm]
V_{1}:=&k\mat{\one}{-\one}{-\one}{\one}, \
V_{2}:= \12 rb^{-1}\mat{-\one}{-\one}{\one}{\one}.
\end{array}
\eeq
The canonical Hilbertian  scalar product on $\cK= \cH\oplus \cH$ will be denoted by $\langle\cdot | \cdot \rangle_{0}$.
Then the Krein structure 
$\langle\cdot|\cdot\rangle_{\cE}$ is mapped by $U$ on
\beq\label{e1.24}
\langle u| u\rangle= \langle u |(\one + K)u\rangle_{0}, \ K:= -\12 b^{-1}rb^{-1}\mat{\one}{\one}{\one}{\one}.
\eeq
Clearly if $(E)$, $(A1)$, $(A2)$ hold then $L$ is even-definitizable on the Krein space $(\cK, \langle \cdot | \cdot\rangle)$.

\begin{lemma}\label{2.2bis}
 Assume (E),  (A1), (A2). Then:
 \ben
\item $L-  z: \cK\tilde\to \langle L_{0}\rangle\cK$,  for $ z\in \rho(L)= \rho(\dot H)$.
\item 
Let $U$ be a compact set with  $U\subset\rho(\dot H)$ if
$m>0$ and $U\subset \rho(\dot H)\backslash \{0\}$ if $m=0$. Then there exists $C>0$ such that:
 \[
\|(L-  z)^{-1}\|_{B(\langle L_{0}\rangle\cK,  \cK)}\leq C1+ C\|(L-  z)^{-1}\|_{B(\cK)} \quad  \forall\, z\in U.
\]
\een
\end{lemma}
Note that  Lemma \ref{2.2bis} would be immediate if $\Dom L= \Dom L_{0}$.
 \proof
If $m>0$  we know that $\jh\sim \sh$ hence  $b^{2}\sim \langle b\rangle^{2}$
by (A1). This implies that $\langle L_{0}\rangle \cK= |L_{0}|\cK$ and the lemma follows from Prop.\ \ref{1.11} and  (\ref{e1.31}). Assume that $m=0$.
Then from (A2) we see that $L- z\in B(\cK,\langle L_{0}\rangle \cK)$.
 We note then that
\beq\label{e2.3}
 \|u\|_{\langle b\rangle\cH}\sim \| \one_{[0, 1]}(b)u\|_{\cH}+ \| \one_{]1, +\infty[}(b)u\|_{b\cH}, \ u \in \langle b\rangle\cH.
\eeq
Prop.\  \ref{1.11} gives $(\dot H -  z)^{-1}: \dot\cE^{*}\to \dot\cE$, hence 
$(L- z)^{-1}: |L_{0}|\cK\to \cK$ by using again (\ref{e1.31}). Since 
$(L- z)^{-1}: \cK\to \cK$, we get  from (\ref{e2.3})
$ (L- z)^{-1}\in B(\langle L_{0}\rangle \cK, \cK)$ and
\[
\|(L- z)^{-1}\|_{B(\langle L_{0}\rangle\cK, \cK)}\leq 
C\|(L- z)^{-1}\|_{B(|L_{0}|\cK, \cK)}+ C\|(L- z)^{-1}\|_{B(\cK)} .
\]
Then we apply Prop.\ \ref{1.11}. \qed

\medskip

 We now introduce the following condition:
 \[
{\rm (A3)}\quad \ k\langle b\rangle^{-1}, \ b^{-1}rb^{-1}\in B_{\infty}(\cH).
\]
\begin{proposition}\label{5.5}
 Assume (E), (A1), (A2), (A3). Then:
 \[
\begin{array}{rl}
(1)& (L- z)^{-1}- (L_{0}- z)^{-1}\in B_{\infty}(\langle L_{0}\rangle\cK, \cK), \  z\in \rho(L)\cap \rho(L_{0}),\\[2mm]
(2)&\sigma_{\rm ess}(L)= \sqrt{\sigma_{\rm ess}(h)}\cup -\sqrt{\sigma_{\rm ess}(h)}.
\end{array}
\]
\end{proposition}
\begin{remark}
{\rm  Note that Prop.\ \ref{5.5} still holds if we replace $b^{-1}rb^{-1}\in B_{\infty}(\cH)$ by the weaker condition $\langle b\rangle^{-1}rb^{-1}\in B_{\infty}(\cH)$. If $b^{-1}rb^{-1}\in B_{\infty}(\cH)$ then $K$ defined in (\ref{e1.24}) belongs to $B_{\infty}(\cK)$, which will be useful in Sect.\ \ref{sec2}.}
\end{remark}
\proof To prove (1) we use that $(L- z)^{-1}, (L_{0}- z)^{-1}\in B(\langle L_{0}\cK, \cK)$ by Lemma \ref{2.2bis}, that $V_{1}$, $V_{2}$ defined in (\ref{e1.25}) belong to $B_{\infty}(\cK, \langle L_{0}\rangle \cK)$ and the second resolvent formula. The relation (2) follows from the analogous 
statement for 
$\dot H$ in Prop.\ \ref{1.4old}, noting that (A3) implies (A4). \qed

\medskip

We will need later the following lemma.
\begin{lemma}\label{1.21}
 Assume  (E), (A1), (A2), (A3),. Let $\chi\in \coinf(\rr)$ with  $0\not\in \supp \chi$ if $m=0$. Then
 \[
\begin{array}{rl}
(1)&\chi(L)\in B(\langle L_{0}\rangle\cK, \cK),\\[2mm]
(2)& \chi(L)- \chi(L_{0})\in B_{\infty}(\cK).
\end{array}
\]
  \end{lemma}
\proof
We use the functional calculus formula:
\[
\chi(L)= 
\frac{\i}{2\pi}\int_{\cc}\frac{\p \tilde{\chi}}{\p\overline{z}}(z)(z-L)^{-1}d z \wedge d \overline{z}.
\]
Then (1) follows from Lemma \ref{2.2bis} and the bound in Corollary \ref{besson}. Statement (2) follows from Prop. \ref{5.5} (1) and the fact that the integrals defining $\chi(L)$ and $\chi(L_{0})$ are norm convergent. \qed

\medskip

We conclude this subsection by discussing  the situation when  $h_{0}=h+k^{2}$ is positive and by formulating conditions on $k$ which imply conditions (A).

\begin{lemma}\label{5.9}
 Assume that $h_{0}\geq 0$, $\Ker\, h_{0}=\{0\}$ and:
  
  - $h_{0}\sim \sh$,
  
  - $k=k_{1}+ k_{2}$ where $k_{i}$ are symmetric on $\jh^{-\12}\cH$ and $\|k_{1}|h_{0}|^{-\12}\|_{B(\cH)}<1$, $k_{1}\langle h_{0}\rangle^{-\12}$, $k_{2}|h_{0}|^{-\12}\in B_{\infty}(\cH)$. 
  
  Then conditions (A) are satisfied for
  \[
b= (h_{0}- k_{1}^{2})^{\12}, \ r= k^{2}- k_{1}^{2}= k_{2}^{2}+ k_{1}k_{2}+ k_{2}k_{1}.
\]    
 \end{lemma}
 \proof Since $\|k_{1}|h_{0}|^{-\12}\|<1$ we have $b^{2}\sim h_{0}\sim \sh$, hence (A1) holds. We know $k_{i}\in B(b^{-1}\cH, \cH)$ hence $k_{i}\in B(\cH, b \cH)$ by duality. This implies $r\in B( b^{-1}\cH, b\cH)$, which is (A2). Similarly we obtain that $k\langle b\rangle^{-1}$ and $b^{-1}rb^{-1}$ belong to $B_{\infty}(\cH)$. \qed 

\section{Mourre estimate for  Klein-Gordon operators on energy spaces}\label{sec2}\init

This section is devoted to the proof of a Mourre estimate for Klein-Gordon operators on energy spaces. We will use the approximate diagonalization in Subsect.\ \ref{sec1b.8}, and consider the operator $L$.

\subsection{Scalar conjugate operators}\label{sec2.1}

We start with some preparations with scalar operators, i.e. operators acting on $\cH$.

Let us fix as in Subsect.\ \ref{sec1b.8} two operators $b,r$ such that (A1), (A2) hold. Let $a$ be a self-adjoint operator on $\cH$ such that:
\[
\begin{array}{rl}
{\rm (M1)}& b^{2}\in C^{2}(a).
\end{array}
\]
Then $\chi(b^{2}): \Dom\, a \to \Dom\, a$ for $\chi\in \coinf(\rr)$ and 
(see e.g. \cite[Subsect. 2.2.2]{Ha1})
\beq\label{e2.0}
 a_{\chi}:= \chi(b^{2})a\chi(b^{2})
\eeq 
is essentially self-adjoint on $\Dom\, a$. We still denote by $a_{\chi}$ 
its closure. Then $b^{2}\in C^{2}(a_{\chi})$ and 
$\ad_{a_{\chi}}^{\alpha}(b^{2})\in B(\cH)$ for $ 0\leq \alpha\leq 2$.

\begin{lemma}\label{2.1}
Assume (M1). Then:
\ben
\item $\e^{\i t a_{\chi}}:\langle b\rangle^{s}\cH\to\langle b\rangle^{s}\cH$ and defines a $C_{0}-$group on $\langle b\rangle^{s}\cH$ for $|s|\leq 2$,
\item  if $m>0$ or $m=0$ and $0\not\in \supp \chi$ then
\[
 \ad^{\alpha}_{a_{\chi}}(b)\in B(\cH) \text{ if }  
 0\leq \alpha \leq 2 \text{ and } \ 
 b, \langle b\rangle\in C^{2}(a_{\chi}; \langle b\rangle^{-1}\cH, \cH)\cap C^{2}(a_{\chi}; \cH, \langle b\rangle\cH).
\]
\een
\end{lemma}
\proof 
Since $[b^{2}, a_{\chi}]\in B(\cH)$, it follows from \cite[Appendix]{GG} that 
$ \e^{\i ta_{\chi}}$ preserves $\Dom\, b^{2}= \langle b\rangle^{-2}\cH$, hence also $\langle b\rangle^{s}\cH$ for $|s| \leq 2$ by duality and interpolation. By  \cite[Prop.\ 3.2.5]{ABG},
$\e^{\i t a_{\chi}}$ defines a $C_{0}-$group on  all these spaces.
This proves (1).  If $m>0$ or $m=0$ and $0\not\in \supp \chi$ we have
 \beq
 \label{e2.1}
\ad_{a_{\chi}}(b)=  \ad_{a_{\chi}}(f(b^{2})),
\eeq
 for   some $f\in \coinf(\rr)$. Since $b^{2}\in C^{1}(a)$ we get 
 $\ad_{a_{\chi}}(f(b^{2}))\in B(\cH)$. The same argument shows that
 \begin{equation}
\label{e2.1bis}
\ad^{\alpha}_{a_{\chi}}(b)= \chi( b^{2})M_{\alpha}\chi(b^{2}), \ M_{\alpha}\in B(\cH), \ 0\leq \alpha\leq 2.
\end{equation}
The same  hold for $\langle b\rangle$, which implies (2). \qed

\medskip

\begin{lemma}\label{segol}
 Assume (M1) and let $a_{\chi}$ be defined by (\ref{e2.0}) with $0\not \in \supp \chi$. Then 
 \[
 \langle a_{\chi}\rangle^{-\delta} (\langle b\rangle -b)\langle a_{\chi}\rangle^{\delta}\in B(\cH),  \ 0\leq \delta\leq 1.
\]
\end{lemma}
\proof  The proof is given in Subsect.\ \ref{app1.10}. \qed

\medskip

We now introduce  assumptions on $k$ and $r$.
\[
{\rm (M2)}\quad k\langle b\rangle^{-1}, \langle b\rangle ^{-1}rb^{-1}\in C^{2}(a_{\chi};  \cH), b^{-1}rb^{-1}\in C^{1}(a_{\chi}; \cH).
\]
Note that if (E), (A1), (A2) hold, then $k\langle b\rangle^{-1}$, $ \langle b\rangle^{-1}rb^{-1}$ and $b^{-1}rb^{-1}$ belong to $B(\cH)$, so assumption (M2) makes sense.

\begin{lemma}\label{7.3}
 Assume (E), (A1), (A2), (M1), (M2). Then 
 \[
k, rb^{-1}\in C^{2}(a_{\chi}; \cH, \langle b\rangle \cH).
\]
\end{lemma}
\proof Since $\langle b\rangle\in C^{2}(a_{\chi}; \cH, \langle b\rangle \cH)$ it suffices to show that $\langle b\rangle^{-1}k$, $\langle b\rangle^{-1}rb^{-1}$ belong to $C^{2}(a_{\chi}; \cH)$, which follows from (M2) and \cite[Prop.\ 5.1.7]{ABG}. \qed

\medskip

We now discuss conditions on $k$ which imply (M2), if $h_{0}:= h+ k^{2}\geq 0$ and $\sh\sim h_{0}$, similar to Lemma \ref{5.9}.

\begin{lemma}\label{7.3bis}
 Assume the hypotheses of Lemma \ref{5.9} and choose $b= (h_{0}-k_{1}^{2})^{\12}$ so that $r= k_{2}^{2}+ k_{1}k_{2}+ k_{2}k_{1}$. Assume moreover that $[k_{1},k_{2}]=0$, as an identity in $B(\jh^{-\12}\cH, \jh^{\12}\cH)$. Then if 
 \[
(M2')\quad  
k_{1}\langle b\rangle^{-1}, \ k_{2}b^{-1}\in C^{2}(a_{\chi}; \cH), \ b^{-1}k_{1}k_{2}b^{-1}\in C^{1}(a_{\chi}; \cH)
\]
condition (M2) is satisfied.
\end{lemma}
\proof Arguing as in the proof of (\ref{e2.1}) we obtain that $b\langle b\rangle^{-1}\in C^{2}(a_{\chi}; \cH)$. Since $k_{2}b^{-1}\in C^{2}(a_{\chi}; \cH)$, we obtain that $k_{2}\langle b\rangle^{-1}\in C^{2}(a_{\chi}; \cH)$, by \cite[Props.\ 5.1.7, 5.2.3]{ABG}, hence $k\langle b\rangle^{-1}\in C^{2}(a_{\chi}; \cH)$. Using that $r= k_{2}^{2}+ 2k_{1}k_{2}$ and the same argument, we also obtain the remaining conditions in (M2). \qed

\subsection{Conjugate operators for  Klein-Gordon operators}\label{sec2.2}

We first introduce some notation.
If  $c$ is  a closed densely defined operator on $\cH$,  we set 
\[
c_{\rm diag}:=\mat{c}{0}{0}{c}, \hbox{ acting on }\cK= \cH \oplus \cH.
\]
We will use the approximate diagonalization introduced in Subsect.\ \ref{sec1b.8}. Recall that $U\dot\cE= \cK$ and $U\dot\cE^{*}= |L_{0}|\cK$.

Let now $A=(a_{\chi})_{\rm diag}$ which is the generator of $(\e^{\i t a_{\chi}})_{\rm diag}$ on $\cK$.
 \begin{proposition}\label{2.3}
Assume (E), (A1), (A2), (M1), (M2). Then:
\ben 
\item $\e^{\i t A}$ is a $C_{0}-$group on $\langle L_{0}\rangle \cK$,
\item  the Krein structure $\langle\cdot | \cdot\rangle$ is of class $C^{1}(A)$,
\item $L$, $L_{0}$ belong to  $C^{2}(A; \cK, \langle L_{0}\rangle\cK)$ hence to $C^{2}(A)$.
\een
\end{proposition}
We refer to \cite[Subsect.\ 5.5]{GGH1} for the terminology in (2) above.
\proof
(1) follows from Lemma \ref{2.1} (1),  (2)  from (M2) and identity (\ref{e1.24}), and  (3)  from (M2) and identity (\ref{e1.25}). \qed

\medskip

\begin{proposition}\label{2.4}
 Assume (E), (A1), (A2),  (A3), (M1), (M2). Let $\chi\in \coinf(\rr)$ with $0\not\in \supp \chi$ if $m=0$. Then
 \[
\chi(L)[L, \i A]\chi(L)- \chi(L_{0})[L_{0},\i A]\chi(L_{0})\in B_{\infty}(\cK). 
\]
\end{proposition}
\proof From (E), (A1), (A2), (A3), we see that $L- L_{0}\in B_{\infty}(\cK, \langle L_{0}\rangle\cK)$. From Prop.\ \ref{2.3} we know that $L-L_{0}\in C^{2}(A; \cK,  \langle L_{0}\rangle\cK)\subset C^{1}_{\rm u}(A; \cK,  \langle L_{0}\rangle\cK)$.
 Therefore 
 \beq\label{e2.4}
 [L-L_{0}, \i A]\in B_{\infty}(\cK,  \langle L_{0}\rangle\cK).
 \eeq
 We write now:
\[
\begin{array}{rl}
&\chi(L)[L, \i A]\chi(L)- \chi(L_{0})[L_{0},\i A]\chi(L_{0})\\[2mm]
=&\chi(L)[L- L_{0}, \i A]\chi(L)+ \chi(L)[L_{0}, \i A] (\chi(L)- \chi(L_{0}))\\[2mm]
&+ (\chi(L)- \chi(L_{0}))[L_{0}, \i A]\chi(L_{0}).
\end{array}
\]
By  (\ref{e2.4}), the fact that $[L_{0}, \i A]\in B(\cK)$, and 
Lemma \ref{1.21} (2), this is compact.  \qed

\subsection{Mourre estimate}\label{sec2.3}

We denote by $\tau(b^{2}, a)$ the set of {\em thresholds} for $(b^{2},a)$.  If $a$ is fixed  from the context, we will often simply write $\tau(b^{2})$ for $\tau(b^{2}, a)$.
So if $\lambda\not \in \tau(b^{2},a)$ there exists an interval $I\subset \rr$, with $\lambda\in I$, a constant $c_{0}>0$ and  $R\in B_{\infty}(\cH)$ such that
\[
\one_{I}(b^{2})[b^{2}, \i a]\one_{I}(b^{2})\geq c_{0}\one_{I}(b^{2})+ R.
\]
We set
\[
\tau(b):= \sqrt{\tau(b^{2})}.
\]

In the theorem below we use the notation $c(L)$ for the set of critical points of $L$.

 Recall that  $\langle \cdot | \cdot\rangle_{0}$  denotes the Hilbertian scalar product on $\cK$. If $A\in B(\cH)$ then 
\[
A\geq_{0} 0, \hbox{ resp. }A\geq 0,  
\]
means that $A$ is self-adjoint positive for $\langle \cdot | \cdot\rangle_{0}$,
resp.  $\langle \cdot| \cdot \rangle$.

\begin{theoreme}\label{2.5}
Assume (E), (A1), (A2), (A3), (M1), (M2). 
Let $I\subset \rr^{\pm}$ a compact interval such that:
 \[
 i)\ I\cap \pm \tau(b)=\emptyset, \ ii) \ I\cap c(L)=\emptyset, \ 
 iii) \ 0\not\in I.
\]
Let $\chi\in \coinf(\rr)$ such that $\chi\equiv 1$ on $I^{2}$ and $0\not\in \supp \chi$ if $m=0$, and  $A= (a_{\chi})_{\rm diag}$, where $a_{\chi}$ is defined in (\ref{e2.0}). Then:
\ben
\item for $f\in  \coinf(\accentset{\circ}{I})$ there exists $c_{1}>0$ and $R\in B_{\infty}(\cK)$ such that
\[
\pm\Re \left(f(L)[L, \i A]f(L)\right)\geq c_{1}f^{2}(L)+ R,
\]
\item if $\lambda\in I\backslash \sigma_{p}(L)$ there exists $\delta>0, c_{2}>0$ such that
\[
\pm\Re\left(\one_{[\lambda-\delta, \lambda+ \delta]}(L)[L, \i A]\one_{[\lambda-\delta, \lambda+ \delta]}(L)\right)\geq c_{2}\one_{[\lambda-\delta, \lambda+ \delta]}(L).
\]
\een
In (1) and (2) we choose the sign $\pm$ if $I\subset \rr^{\pm}$.
\end{theoreme}
\begin{remark}
{\rm   We assume for simplicity that $0\not \in I$, even if $ m>0$. This is not a restriction since by  Prop.\ \ref{5.5} we know that $0\not\in \sigma_{\rm ess}(L)$ if $m>0$.}
\end{remark}
\proof
Since by {\it i)} $I^{2}\cap \tau(b^{2})=\emptyset$, there exists $c_{0}>0$, $R\in B_{\infty}(\cH)$ such that
\[
\one_{I^{2}}(b^{2})[b^{2}, \i a]\one_{I^{2}}(b^{2})\geq c_{0}\one_{I^{2}}(b^{2})+ R.
\]
By \cite[Thm. 2.2.4]{Ha1} this implies if $\chi\in \coinf(\rr)$ is such that $\chi\equiv 1$ on $I^{2}$, there exists $c_{1}>0$ $R_{1}\in B_{\infty}(\cH)$ such that 
\begin{equation}
\label{e2.5}
\one_{|I|}(b)[b, \i a_{\chi}]\one_{|I|}(b)\geq c_{1}\one_{|I|}(b)+ R_{1}.
\end{equation} 
This implies that if $I\subset \rr^{\pm}$ one has
\[
\pm \one_{I}(L_{0})[L_{0}, \i A]\one_{I}(L_{0})\geq_{0} c_{1}\one_{I}(L_{0})+ R_{2}, \ R_{2}\in B_{\infty}(\cK),
\]
which implies   that for $f\in \coinf(\accentset{\circ}{I})$ one has:
\begin{equation}
\label{e2.6}
\pm f(L_{0})[L_{0}, \i A]f(L_{0})\geq_{0}c_{1}f^{2}(L_{0})+ R_{3}, \ R_{3}\in B_{\infty}(\cK).
\end{equation}
Let us now set 
\[
\begin{array}{rl}
B= &f(L)[L, \i A]f(L), \ C= f^{2}(L), \\[2mm]
 B_{0}= &f(L_{0})[L_{0}, \i A]f(L_{0}), \ C_{0}= f^{2}(L_{0}),
\end{array}
\]  
and let $K$ be defined in (\ref{e1.24}). 
By (A3), we know that $K\in B_{\infty}(\cK)$.  By  Lemma \ref{1.21}, 
Prop.\ \ref{2.4}  and hypothesis (A2), we know that 
\beq\label{e2.7}
B-B_{0}, \ C-C_{0}, \ K\in B_{\infty}(\cK).
\eeq We have for $u\in \cK$:
\[
\begin{array}{rl}
\pm \Re\langle u| Bu \rangle=&\pm \Re\langle u| B_{0}u\rangle + \Re\langle u| Ru\rangle\\[2mm]
=&\pm \Re\langle u| (\one +K)B_{0}u\rangle_{0}+ \Re\langle u| Ru\rangle\\[2mm]
=&\pm \Re\langle u|B_{0}u\rangle_{0}+ \Re\langle u| Ru\rangle \\[2mm]
\geq&c_{1}\langle u| C_{0}u\rangle_{0} + \Re\langle u|Ru\rangle_{0}+  \Re\langle u| Ru\rangle\\[2mm]
=&c_{1}\langle u| Cu\rangle+ \Re\langle u| Ru\rangle,
\end{array}
\]
where $R$ denotes an element of $B_{\infty}(\cK)$ and we used  (\ref{e2.7}), (\ref{e1.24}). This proves (1).

Assume now that $\lambda\in I\backslash \sigma_{p}(L)$. Since $I$ does not contain critical points of $L$,   we know that 
$\one_{I}(L)\geq 0$  and  that the restriction of 
$\langle \cdot | \cdot \rangle$ to $\one_{I}(L)\cK$ is a Hilbertian  scalar product, equivalent to $\langle \cdot |\cdot \rangle_{0}$, and 
 the restriction of $L$ to $\one_{I}(L)\cK$ is self-adjoint in the usual sense for this scalar product.
Then (2) follows from (1) by the usual argument. \qed

\section{Limiting absorption principle}\label{sec4}\init

In this section we apply the abstract results from \cite{GGH1} to deduce weighted resolvent estimates from the positive commutator estimate proved in the previous section.   The following theorem follows directly from 
\cite[Thm.\ 7.9]{GGH1}, whose hypotheses follow from   
Thm.\ \ref{2.5} and Prop.\ \ref{2.3}.

\begin{theoreme}\label{4.1}

Assume (E), (A1), (A2), (A3), (M1), (M2).

Let $I\subset \rr$ a compact interval such that
 \[
 i)\ I\cap \pm \tau(b)=\emptyset, \ ii) \ I\cap c(L)=\emptyset, \ iii) \ 0\not\in I, \ iv) \ I\cap \sigma_{\rm p}(L)= \emptyset.
\]
Let $\chi\in \coinf(\rr)$ such that $\chi\equiv 1$ on $I^{2}$ and $0\not\in \supp \chi$ if $m=0$, and  $A= (a_{\chi})_{\rm diag}$, where $a_{\chi}$ is defined in (\ref{e2.0}). 

Then there exists $\epsilon_{0}>0$ such that for $ \delta>\12$ one has:
\[
\sup_{{\rm Re}z\in I, \ 0<|{\rm Im }z|\leq  \epsilon_{0}} \| \langle A\rangle^{-\delta}(L-z)^{-1}\langle A\rangle^{-\delta}\|_{B(\cK)}<\infty.
\]
\end{theoreme}

\subsection{Limiting absorption principle in energy space}\label{sec4.1}
After conjugation by the operator $U$ defined in Subsect.\ \ref{sec1b.8},  we immediately deduce from Thm.\ \ref{4.1} a corresponding  result on $(\dot H- z)^{-1}$ acting on the homogeneous Hilbert space $\dot \cE$.  

Clearly we have $c(L)= c(\dot H)$ since $\dot H,L$ are unitarily equivalent. Although $H$ is not necessarily definitizable on $(\cE, \langle\cdot|\cdot\rangle_{\cE})$ if $m=0$, we will still set 
\[
c(H):= c(\dot H).
\]

The weights appearing on both sides of $(\dot H -z)^{-1}$ are  not convenient for applications,  at least in the massless case, because they contain the relatively singular  operators $b$ and $b^{-1}$ (see (\ref{sego0}) below). In this subsection we consider the resolvent $(H-z)^{-1}$ on  $\cE$ and prove more useful resolvent estimates, with non singular weights.

It is convenient to formulate these estimates in terms of an additional operator on $\cH$ which dominates the conjugate operator $a_{\chi}$. Let us introduce the corresponding abstract hypothesis:

We fix  a self-adjoint operator  $\langle x\rangle \geq \one$ on $\cH$, called a {\em reference weight},  such that:
 \[
{\rm (M3)}
\left\{\begin{array}{ll}
(i)& \langle a_{\chi}\rangle \langle x \rangle^{-1}\in B(\cH), \ \forall \ \chi\in \coinf(\rr), \\[2mm]
 (ii)& 
[\langle b\rangle, \langle x\rangle^{-\delta}]\langle x\rangle^{\delta}\in B(\cH), \ 0\leq \delta\leq 1.
\end{array} 
\right.
 \]
In concrete cases (see Sect. \ref{sec3}) it is very easy to find a reference weight $\jr$.

\begin{theoreme}\label{4.2}
 Assume (E), (A1), (A2), (A3), (M1), (M2), (M3). Let $I\subset \rr$ be an interval as in Thm.\ \ref{4.1}.  Then there exists $\epsilon_{0}>0$ such that 
 for $\12<\delta\leq 1$:
 \[
\sup_{{\rm Re}z\in I, \ 0<|{\rm Im }z|\leq  \epsilon_{0}} \| (\langle x\rangle^{-\delta})_{\rm diag}(H-z)^{-1}(\langle x\rangle^{-\delta})_{\rm diag}\|_{B(\cE)}<\infty.
\]
 \end{theoreme}
\proof
Set $J= \{z\in \cc  : {\rm Re}z\in I, \  0< | {\rm Im}z|\leq \epsilon_{0}\}$. Since 
\beq\label{sego0}
U^{-1}\langle A\rangle^{-\delta}U= \mat{b^{-1}\langle a_{\chi}\rangle^{-\delta}b}{0}{0}{\langle a_{\chi}\rangle^{-\delta}},
\eeq
we obtain that if $g\in \dot \cE$ and $f= (\dot H -z)^{-1}g$ one has:
\beq\label{sego1}
\|\langle a_{\chi}\rangle^{-\delta} b f_{0}\|_{\cH}+ \| \langle a_{\chi}\rangle^{-\delta}f_{1}\|_{\cH}\leq c \left(\| \langle a_{\chi}\rangle^{\delta}bg_{0}\|_{\cH}+ \| \langle a_{\chi}\rangle^{\delta}g_{1}\|_{\cH}\right), \  z\in J.
\eeq
If $g\in \cE$, then by Prop.\ \ref{prop:equal} we know that $f\in \cE$ and $f= (H-z)^{-1}g$. Moreover since $f_{0}= z^{-1}(f_{1}- g_{0})$ and $0\not \in I$ we also obtain that:

\beq\label{sego2}
\|\langle a_{\chi}\rangle^{-\delta} f_{0}\|_{\cH}\leq c \left(\| \langle a_{\chi}\rangle^{\delta}bg_{0}\|_{\cH}+ \|\langle a_{\chi}\rangle^{-\delta}g_{0}\|_{\cH}+ \| \langle a_{\chi}\rangle^{\delta}g_{1}\|_{\cH}\right), \  z\in J.
\eeq
Writing $\langle b\rangle = b +  (\langle b \rangle -b)$ and using Lemma \ref{segol},  we obtain that:
\begin{equation}
\label{sego3}
\begin{array}{rl}
\|\langle a_{\chi}\rangle^{-\delta} \langle b\rangle f_{0}\|_{\cH}&\leq c\left(\| \langle a_{\chi}\rangle^{-\delta} b f_{0}\|_{\cH}+ \|\langle a_{\chi}\rangle^{-\delta}f_{0}\|_{\cH}\right),\\[2mm]
\|\langle a_{\chi}\rangle^{\delta} b g_{0}\|_{\cH}&\leq c\left(\|\langle a_{\chi}\rangle^{\delta} \langle b\rangle g_{0}\|_{\cH}+ \| \langle a_{\chi}\rangle^{\delta}g_{0}\|_{\cH}\right).
\end{array}
\end{equation}
From condition (M3)  {\it (i)} we obtain by interpolation that $\langle a_{\chi}\rangle^{\delta}\langle x\rangle^{-\delta}$ is bounded, hence:
\[
\begin{array}{rl}
\|\langle x\rangle^{-\delta} \langle b\rangle f_{0}\|_{\cH}&\leq c\left(\|\langle a_{\chi}\rangle^{-\delta} b f_{0}\|_{\cH}+ \|\langle x\rangle^{-\delta}f_{0}\|\right),\\[2mm]
\|\langle a_{\chi}\rangle^{\delta} b g_{0}\|_{\cH}&\leq c\left(\|\langle x\rangle^{\delta} \langle b\rangle g_{0}\|_{\cH}+ \| \langle x\rangle^{\delta}g_{0}\|_{\cH}\right),\\[2mm]
\|\langle x\rangle^{-\delta}f_{i}\|_{\cH}&\leq c \| \langle a_{\chi}\rangle^{-\delta}f_{i}\|_{\cH},\ i=0,1,\\[2mm]
\| \langle a_{\chi}\rangle^{\delta}g_{i}\|_{\cH}&\leq c\| \langle x\rangle^{\delta}g_{i}\|_{\cH}, \ i=0,1.
\end{array}
\]
Therefore we deduce from (\ref{sego1}), (\ref{sego2}) that
\beq\label{sego4}
\begin{array}{rl}
&\|\langle x \rangle^{-\delta} \langle b\rangle f_{0}\|_{\cH}+\|\langle x \rangle^{-\delta} f_{0}\|_{\cH}+\| \langle x \rangle^{-\delta}f_{1}\|_{\cH}\\[2mm]
\leq &c \left(\| \langle x\rangle^{\delta}\langle b\rangle g_{0}\|_{\cH}+ \| \langle x\rangle^{\delta}g_{0}\|_{\cH}+\| \langle x\rangle^{\delta}g_{1}\|_{\cH}\right), \  z\in J.
\end{array}
\eeq
We use now (M3) {\it( ii)} which implies that:
\[
\begin{array}{rl}
\|\langle b\rangle \langle x\rangle^{-\delta}f_{0}\|_{\cH}&\leq \|\langle x \rangle^{-\delta} \langle b\rangle f_{0}\|_{\cH}+\|\langle x \rangle^{-\delta} f_{0}\|_{\cH},\\[2mm]
\| \langle x\rangle^{\delta}\langle b\rangle g_{0}\|_{\cH}&\leq  \| \langle b\rangle\langle x\rangle^{\delta}g_{0}\|_{\cH}.
\end{array}
\]
Therefore (\ref{sego4}) yields:
\begin{equation}
\label{sego5}
\|b\langle x \rangle^{-\delta}  f_{0}\|_{\cH}+\| \langle x \rangle^{-\delta}f_{1}\|_{\cH}
\leq c \left(\| \langle b\rangle\langle x\rangle^{\delta} g_{0}\|_{\cH}+\| \langle x\rangle^{\delta}g_{1}\|_{\cH}\right), \  z\in J.
\end{equation}
Since $\langle b\rangle^{2}\simeq \langle \epsilon\rangle^{2}$, this completes the proof of the theorem. \qed

\subsection{Weighted estimates for quadratic pencils}\label{sec4.2}
In this subsection we consider weighted estimates for $p( z)^{-1}$.  It is natural to introduce the following assumption on $k$ and the reference weight $\langle x\rangle$:
\[
{\rm (M4)}\ \langle x\rangle^{\delta}k\langle x\rangle^{-\delta}\jh^{-\12}\in B(\cH), \hbox{ for }|\delta|\leq 1.
\]
Note that (M4) follows from (E2) if $k$ and $\langle x\rangle$ commute, which will be the case in the applications in Sect.\ \ref{sec3}.

\begin{proposition}\label{4.3}
 Assume (E), (M4) and let $I\subset \rr$ a compact interval with $0\not \in I$, and $0<\delta\leq 1$.
 Then the following are equivalent:
 \[
 \begin{array}{rl}
(1)& \sup_{{\rm Re}z\in I, \ 0<|{\rm Im }z|\leq  \epsilon_{0}} \| (\langle x\rangle^{-\delta})_{\rm diag}(H-z)^{-1}(\langle x\rangle^{-\delta})_{\rm diag}\|_{B(\cE)}<\infty,\\[2mm]
(2)& \sup_{{\rm Re}z\in I, \ 0<|{\rm Im }z|\leq  \epsilon_{0}} \|\jh^{\12}\langle x\rangle^{-\delta}p(z)^{-1}\langle x\rangle^{-\delta}\|_{B(\cH)}<\infty.
\end{array}
\]   
\end{proposition}
\proof The proof is an easy computation, using formula (\ref{eq:inver}), the identity $p(z)^{-1}h = \one + p(z)^{-1}z(z-2k)$ and the fact that $\langle x\rangle^{\delta}(z- 2k)\langle x\rangle^{-\delta} \jh^{-\12}$ is bounded, by (M4). The details are left to the reader. \qed

\subsection{Limiting absorption principle in charge space}\label{sec4.3}

From Prop.\ \ref{4.3} we easily get from Thm.\ \ref{4.2} similar resolvent estimates for $(K-z)^{-1}$ on the charge space $\cF$.

\begin{theoreme}\label{4.4}
Assume (E), (A1), (A2), (A3),  (M). Let $I\subset \rr$ an interval as in 
Thm.\ \ref{4.1}.  Then there exists $\epsilon_{0}>0$ such that for $\12<\delta\leq 1$ one has:
 \[
\sup_{{\rm Re}z\in I, \ 0<|{\rm Im }z|\leq  \epsilon_{0}} \| (\langle x\rangle^{-\delta})_{\rm diag}(K-z)^{-1}(\langle x\rangle^{-\delta})_{\rm diag}\|_{B(\cF)}<\infty.
\]
\end{theoreme}
\proof We use formula (\ref{eq:cinverse}) to express $(K-z)^{-1}$. We see that the estimate in the theorem holds iff 
\[
 \sup_{{\rm Re}z\in I, \ 0<|{\rm Im }z|\leq  \epsilon_{0}} \| M_{i}(z)\|_{B(\cH)}<\infty,\ i=1,\dots, 4,
\] 
for
\[
\begin{array}{rl}
M_{1}(z)=&\jh^{\4} \jr^{-\delta}p(z)^{-1} (k-z)\jr^{-\delta}\jh^{-\4}, \\[2mm]
M_{2}(z)=&\jh^{\4} \jr^{-\delta}p(z)^{-1} \jr^{-\delta}\jh^{\4}, \\[2mm]
M_{3}(z)=&\jh^{-\4} \jr^{-\delta}\left(\one+ (k-z)p(z)^{-1}(k-z)\right) \jr^{-\delta}\jh^{-\4}, \\[2mm]
M_{4}(z)=&\jh^{-\4} \jr^{-\delta}(k-z)p(z)^{-1} \jr^{-\delta}\jh^{-\4}.
\end{array}
\]
Using (M4), duality and interpolation, we see that $\jh^{-\4} \jr^{-\delta}(k-z)\jr^{\delta}\jh^{-\4}$ is bounded.  Therefore the estimate for $M_{2}(z)$ implies the others.  By Thm.\ \ref{4.2} and Prop.\ \ref{4.3} we know that
\[
 \sup_{{\rm Re}z\in I, \ 0<|{\rm Im }z|\leq  \epsilon_{0}} \| \jh^{\12}\jr^{-\delta}p(z)^{-1}\jr^{-\delta}\|_{B(\cH)}<\infty.
\]
Using duality, the fact that $p(z)^{*}= p(\overline{z})$ and
interpolation, this implies the estimate for $M_{2}(z)$. This
completes the proof of the theorem. \qed

\section{Existence of the dynamics}\label{sec7}\init

In this section we discuss the existence of the dynamics generated by
the operators $H$,  $K$ considered in Sects.\ \ref{sec1}, \ref{sec5},
\ref{sec1b}.  Note that this not a completely trivial
point, since we do not know a priori if these operators are generators
of $C_{0}-$groups. 

We will assume in this section  conditions (E).
\subsection{Existence of the dynamics for $H$}
If $m>0$ then $H= \dot H$ which is even-definitizable, hence we can define the $C_{0}-$group  $\e^{\i t\dot{H}}$, by Subsect.\ \ref{exidyn}.  

If $m=0$ we use  now Prop.\ \ref{fcalc-bis} instead of Thm.\ \ref{iaca}.
Using the bounded projection $\one_{\rm pp}^{\cc}(H)$ (see the proof of 
Prop.\ \ref{fcalc-bis}), we split $\cE$ into the direct sum
\[
\cE= \cE_{\rm pp}^{\cc}(H)\oplus \cE_{1}(H),
\]
both spaces being closed and $H$-invariant, the first one finite-dimensional. We argue as in  Subsect.\ \ref{exidyn} to construct $\e^{\i tH}_{|\cE_{1}(H)}$.
Thus there exist $C, \lambda>0$, $n\in \nn$  such that
\begin{equation}
\label{bulot}
\| (\e^{\i tH})_{|  \cE_{\rm pp}^{\cc}(H)}\|\leq C \e^{\lambda|t|}, \
\| (\e^{\i tH})_{|\cE_{1}(H) }\|\leq C \langle t\rangle^{n}, \ t\in \rr.
\end{equation}
\subsection{Existence of the dynamics for $K$}

We start with a useful observation which is further developed in \cite{GGH1}. Note that the sesquilinear form $q$ defined in (\ref{e5.3}) is defined on $\cE\times \cE^{*}$ and turns $(\cE, \cE^{*})$  into a dual pair. Since $\Phi$ defined in (\ref{eq:phi}) preserves $\cE$ and $\cE^{*}$ by Lemma \ref{e5.3}, the same is true of the sesquilinear form $q'$ defined in (\ref{def-de-q-prime}), (which is equal to $q$ transported by $\Phi$).

We check immediately that $\e^{\i tH}$ is unitary for $q'$ on $\cE$. Therefore by duality $\e^{\i tH}$ extends as a $C_{0}-$group on $\cE^{*}$, satisfying (\ref{bulot}). Since $\cF= [\cE, \cE^{*}]_{\12}$, we obtain by interpolation that $\e^{\i tH}$ and hence $\Phi^{-1} \e^{\i tH}\Phi$  extends as $C_{0}-$groups on $\cF$. Using also Lemma \ref{lm:eckg} we see that the generator of the later group is $K$, i.e.  $\Phi^{-1} \e^{\i tH}\Phi=\e^{\i tK}$.  Therefore $\e^{\i tK}$ is a $C_{0}-$group on $\cF$, satisfying (\ref{bulot}).

\section{Propagation estimates}
\label{secpropest}

In this section we will establish propagation estimates for
$\e^{\i t\dot{H}}, \e^{\i t H}$ and $\e^{\i tK}$. We will need the following assumption:

\smallskip

\noindent
(M5)  $D(\x)\cap D(b^2)$ is dense in $D(b^2)$, $\e^{\i s\x}$ sends
$D(b^2)$ into itself, and $[\x,b^2]$ extends to
a bounded operator from $D(b^2)$ to $\cH$ which we denote
$[\x,b^2]_0$.

\smallskip

Note that (M5) implies (see \cite[Prop.\ 3.2.5]{ABG}):
\begin{equation}
\label{b2expisx}
\sup_{0\le s\le 1}\Vert b^2\e^{\i s\x}u\Vert<\infty
\end{equation} 
for all $u\in D(b^2)$. 

We assume (E), (A), (M1), (M2), (M3), (M5) in the following. We also suppose that
$[k_i,\x]=0$ for $i=1,2$ which implies in particular (M4).
\begin{lemma}
\label{pdeltabound}
If $ z\in \rho(h,k)$ and $0\leq \delta\leq 1$ then 
$p^{-1}( z)$ sends $D(\x^{\delta})$ into itself. 
\end{lemma} 
\proof
We first show $p^{-1}( z)$ sends $D(\x)$ into itself. Let $u\in D(\x)$. We have to show that 
\[\sup_{\vert t \vert\leq
  1}\Vert\frac{\e^{\i t\x}-1}{t}p^{-1}(z)u\Vert<\infty.\]
We write
\begin{eqnarray*}
\frac{\e^{\i t\x}-1}{t}p^{-1}( z)u=p^{-1}( z)\frac{\e^{\i t\x}-1}{t}u+\e^{\i t\x}\frac{p^{-1}( z)-\e^{-it\x}p^{-1}( z)\e^{\i t\x}}{t}u.
\end{eqnarray*}
Clearly
\[\sup_{\vert t \vert\leq
  1}\Vert p^{-1}( z)\frac{\e^{\i t\x}-1}{t}u\Vert<\infty.\]
Let us now consider the second term. We have
\begin{eqnarray*}
\frac{p^{-1}( z)-\e^{-\i t\x}p^{-1}( z)\e^{\i t\x}}{t}u=\e^{-\i t\x}p^{-1}( z)\e^{\i t\x}\frac{\e^{-\i t\x}b^2\e^{\i t\x}-b^2}{t}p^{-1}( z)u,
\end{eqnarray*}
where we have used that $\e^{\i s\x}$ sends $D(b^2)$ into itself. 
Now note that
\[\frac{\e^{-\i t\x}b^2\e^{\i t\x}-b^2}{t}=\frac{1}{t}\int_0^t\e^{-\i\theta
\x}[\x,b^2]_0\e^{\i \theta \x}d\theta.\]
It follows using \eqref{b2expisx} that:
\[\sup_{\vert t\vert\leq 1}\Vert \e^{-\i t\x}p^{-1}( z)\e^{\i t\x}\frac{\e^{-\i t\x}b^2\e^{\i t\x}-b^2}{t}p^{-1}( z)\Vert<\infty\]
and thus the lemma for $\delta=1$. The lemma for $\delta=0$ is
obvious, the general case follows by interpolation.
\qed

\medskip

\begin{corollary}
\label{Hdeltabound}
For all $0\leq \delta\leq1$ and $\chi\in C_0^{\infty}(\rr)$ the operators
$\x^{\delta}\chi(H)\x^{-\delta}$ and $\x^{\delta}\chi(K)\x^{-\delta}$
are bounded.
\end{corollary}
\proof
By an interpolation argument it is sufficient to consider the case
$\delta=1$. Using \eqref{eq:inver} and the fact that $[\x,k]=0$ we see that for $ z\in
\rho(h,k)$ we have
\begin{eqnarray*}
\x(H- z)^{-1}\x^{-1}=\left(\begin{array}{cc} 0 & 0 \\ 1 &
    0 \end{array}
\right)+\x p^{-1}( z)\x^{-1}\left(\begin{array}{cc}
     z-2k & 1\\ - z(2k- z) &  z \end{array}\right). 
\end{eqnarray*}
By the definition of the smooth functional calculus for $H$ it is
sufficient to show that $\x p^{-1}( z)\x^{-1}$, which is
bounded on $\cH$ by Lemma \ref{pdeltabound}, fulfills suitable
resolvent estimates. Using Lemma \ref{pdeltabound} we can write the
commutator
\[[\x,p^{-1}( z)]=p^{-1}( z)\h^{\12}\h^{-\12}[b^2,\x]_0p^{-1}( z).\]
It is now sufficient to apply Corollary \ref{penest} to obtain the
required estimates. By duality $\x^{\delta}\chi(H)\x^{-\delta}$  is
bounded on $\cE^*$ and thus on $\cF$ by complex interpolation. To
obtain the result for $K$ we use that 
\[\chi(K)=\Phi^{-1}\chi(H)\Phi,\]
that $\Phi$ commutes with $\x$ and that $\Phi,\, \Phi^{-1}$ are
continuous on $\cF$.
\qed

\medskip

\begin{proposition}
Let $I\subset \rr$ an interval as in Thm.\ \ref{4.1} and $\chi\in
C_0^{\infty},\supp \chi\subset I$. Let $\12<\delta\le 1$. Then there exists $C>0$ such that:
\begin{eqnarray}
\label{propH}
\int_{\rr}\Vert
\x^{-\delta}\e^{\i tH}\chi(H)\x^{-\delta}f\Vert^2_{\cE}dt\leq C\| f\|^{2}_{\cE},\\
\label{propK}
\int_{\rr}\Vert
\x^{-\delta}\e^{\i tK}\chi(K)\x^{-\delta}f\Vert^2_{\cF}dt\leq C \| f\|^{2}_{\cF}
\end{eqnarray}
\end{proposition}
\proof
We first prove \eqref{propH}. Note that by Theorem \ref{4.2} there exists $\epsilon_{0}>0$ such that one has:
\[
\sup_{\ 0<|{\rm Im }z|\leq  \epsilon_{0}} \| (\langle x\rangle^{-\delta})_{\rm diag}(H-z)^{-1}\chi(z)(\langle x\rangle^{-\delta})_{\rm diag}\|_{B(\cE)}<\infty.
\] 
We now have to show that we can replace $\chi(z)$ by $\chi(H)$. We
choose $\tilde{\chi}\in C_0^{\infty}(I)$ with 
$\tilde{\chi}\chi=\chi$. We write~:
\begin{align}
\label{estwithchi}
\Vert \x^{-\delta}(H-z)^{-1} \chi(H) \x^{-\delta} f \Vert^2_{B({\cE})}
&\leq C \Vert \x^{-\delta}(H-z)^{-1}\tilde{\chi}(z)\chi(H) \x^{-\delta} f\Vert^2_{B({\cE})}\nonumber\\
+& C\Vert\x^{-\delta}(H-z)^{-1}(1-\tilde{\chi}(z))\chi(H)\x^{-\delta} f\Vert^2_{B({\cE})}.
\end{align}
The estimate for the first term follows from the estimate with
$\chi(z)$ and Corollary \ref{Hdeltabound}. Let us treat the second term. We claim
\begin{eqnarray*}
\Vert \x^{-\delta}(H-(\lambda+\i\epsilon))^{-1}(1-\tilde{\chi}(\lambda))\chi(H) \x^{-\delta} f\Vert^2_{B({\cE})}\leq C\langle\lambda\rangle^{-2},
\end{eqnarray*}
uniformly in $\epsilon$. Let
\begin{eqnarray*}
f^{\epsilon}_{\lambda}(x)=\langle\lambda\rangle\frac{1}{x-(\lambda+\i\epsilon)}(1-\tilde{\chi}(\lambda))\chi(x).
\end{eqnarray*}
It is sufficient to show that all the semi-norms $\Vert f^{\epsilon}_{\lambda}\Vert_m$ are uniformly bounded with respect to $\lambda,\, \epsilon$. Note that $g_{\lambda}(x)=(1-\tilde{\chi}(\lambda))\chi(x)$ vanishes to all orders at $x=\lambda$. If $\supp\chi\subset [-C,C]$ this is enough to assure that $\Vert f^{\epsilon}_{\lambda}\Vert_m$ is uniformly bounded in $\lambda\in [-2C,2C]$ and $\epsilon>0$. For $\vert\lambda\vert\ge 2C$ we observe that 
\begin{eqnarray*}
\left\vert\langle\lambda\rangle\frac{1}{x-(\lambda+\i\epsilon)}\right\vert\leq C
\end{eqnarray*}
with analogous estimates for the derivatives. Thus the second term in
\eqref{estwithchi} is also uniformly bounded in $0<\vert {\rm Im z}\vert\leq \epsilon_0$.  

We now write~:
\begin{eqnarray*}
((H-(\lambda+\i\epsilon))^{-1}-(H-(\lambda-\i\epsilon))^{-1})\chi(H)f=i\int_{-\infty}^{\infty}\e^{-\epsilon\vert t\vert}\e^{\i \lambda t}\e^{-\i Ht}\chi(H)f dt,
\end{eqnarray*}
the integral being norm convergent by (\ref{bulot}).
By Plancherel's formula this yields: 
\[
\begin{array}{rl}
&\int_{-\infty}^{\infty}\Vert
  \x^{-\delta}((H-(\lambda+\i\epsilon)^{-1}-(H-(\lambda-\i\epsilon))^{-1})\chi(H)
  \x^{-\delta}f\Vert^2_{\cE}d\lambda\\[2mm]
  =&\int_{-\infty}^{\infty}\e^{-2\epsilon \vert t\vert}\Vert  \x^{-\delta}\e^{-\i tH}\chi(H) \x^{-\delta}f\Vert^2_{\cE}dt.
\end{array}
\]
The lhs of this equation is uniformly bounded in $\epsilon$ with $\epsilon$ small enough, which implies (\ref{propH}).

Let us now prove \eqref{propK}. First note that by duality we can replace $B(\cE)$ by $B(\cE^*)$ in 
\eqref{propH}. This gives \eqref{propK} with $K$ replaced by $H$ by
complex interpolation. We then use  that 
\[\e^{\i tK}\chi(K)=\Phi^{-1}\e^{\i tH}\chi(H)\Phi,\]
that $\x^{-\delta}$ commutes with $\Phi$ and that $\Phi,\, \Phi^{-1}$
are bounded operators on $\cF$.
\qed

\section{Examples}\label{sec3}\init

In this section we describe examples of Klein-Gordon equations to which the abstract results of Sects. \ref{sec1b}, \ref{sec2} and \ref{sec4} can be applied.

 Let us first discuss how to check the abstract hypotheses (E). If $\inf \sigma_{\rm ess}(h)>0$ the only delicate condition is (E1). In fact in this case (E1) implies (E2) and also that $0\not\in \sigma(h)$. Therefore $\sh\sim \jh$. It follows that if $\epsilon\geq 0$ is a self-adjoint operator such that $\Dom\, h= \Dom\, \epsilon^{2}$ we have $\sh\sim \jh\sim \langle \epsilon\rangle^{2}$ and hence in condition (E3) we can replace $\sh^{-\12}$ by $\epsi^{-1}$. 
 
 Similarly if $b$ is an operator such that (A1) holds, we have $b^{2}\sim \epsi^{2}$ and in conditions (A2), (A3), (A4) we can replace $b^{-1}$ and $\langle b\rangle ^{-1}$ by $\epsi^{-1}$.
 
 If $\inf \sigma_{\rm ess}(h)=0$ then both (E1) and (E2) are important. Moreover it is again  important to find  a self-adjoint operator 
 $\epsilon\geq 0$ such that $\sh\sim \epsilon^{2}$. An abstract result allowing to do this is given in the following proposition.

\begin{proposition}\label{9.0}
 Let $\epsilon\geq 0$ be  a self-adjoint operator on $\cH$ and $r_{1}$, $r_{2}$ two symmetric operators on $\Dom\, \epsilon$ such that
 \[
\Ker\, \epsilon=\{0\}, \ \| r_{1}\epsilon^{-1}\|<1, \ r_{2}\epsilon^{-1}\in B_{\infty}(\cH).
\]
If $h= \epsilon^{2}-r_{1}^{2}-r_{2}^{2}$  as an identity in $B(\epsi^{-1}\cH, \epsi \cH)$ and $\Ker\, h=\{0\}$, then
\begin{equation}\label{eq:a2}
\left\{
\begin{array}{rl}
(1)& {\rm Tr}\one_{]-\infty, 0]}(h)<\infty,\\[2mm]
(2)& |h|\sim \epsilon^{2}.
\end{array}
\right.
\end{equation}
\end{proposition}

The proof  will be given in Subsect.\ \ref{app1.1}.

\subsection{Charged Klein-Gordon equations on scattering manifolds}
\label{Scattman}

Let $\cN$ be a smooth, $d-1$ dimensional  compact manifold whose elements are denoted by $\omega$. We consider a $d$ dimensional manifold:
\[
\cM\simeq  \cM_{0}\ \cup \ ]1, +\infty[\,\times \cN,
\]
where $ \cM_{0}\Subset\cM$ is relatively compact.  For $m\in \rr$ we denote by $S^m(\cM)$ the space of real valued functions $f\in C^{\infty}(\cM)$ such that 
\[
\forall \ k\in \nn,\, \alpha\in \nn^{d-1},\quad \vert\partial_s^k\partial_{\omega}^{\alpha}f(s,\omega)\vert\le C_{k,\alpha} s^{m-k}\quad\mbox{for}\quad (s,\omega)\in [1,\infty[\,\times \cN.
\]
\begin{definition}\label{def-de-conic}
A Riemannian metric $g^{0}$ on $\cM$ is called {\em conic}  if there exists $R>0$ and a Riemannian metric $h$ on $\cN$ such that 
\begin{eqnarray*}
g^0=ds^2+s^2h_{jk}(\omega)d\omega^jd\omega^k,\quad (s,\omega)\in [R,\infty[\,\times\cN.
\end{eqnarray*}
A Riemannian metric $g$ on $\cM$ is called a {\em scattering metric} if  $g=g^0+\tilde{g}$, where $g^0$ is a conic metric and $\tilde{g}$ is of the form
\begin{eqnarray*}
\tilde{g}=m^0(s,\omega)ds^2+sm^1_j(s,\omega)(dsd\omega^j+d\omega^jds)+s^2m_{jk}^2(s,\omega)d\omega^jd\omega^k
\end{eqnarray*}
with $m^l\in S^{-\mu_l}(\cM)$ for $l=0,1,2$, $\mu_{l}>0$.
\end{definition} 
We will assume in the sequel  that $g$ is a scattering metric on $\cM$ in the sense of the above definition. 
We consider a charged Klein-Gordon field $\phi$ on $\cM$ minimally coupled to an external electromagnetic field described by the electric potential $v(s,\omega)$ and the magnetic potential $A_k(s,\omega)dx^k$. It fulfills the Klein-Gordon equation:
\begin{eqnarray}
\label{KGscatt}
(\partial_t-\i v)^2\phi-(\nabla^k-\i A^k)(\nabla_k-\i A_k)\phi+m^2\phi=0.
\end{eqnarray}
Here $\nabla$ is the Levi-Civita connection associated to the metric $g$. The function $m(s, \omega)$ on $\cM$ corresponds to a variable mass term. 
The above equation writes in local coordinates:
\[
(\partial_t-\i v)^2\phi-|g|^{-1/2}(\partial_j-\i A_j) |g|^{1/2}g^{jk}(\partial_k-\i A_k)\phi+m^2(s,\omega)\phi=0,
\]
where $g^{jk}=(g_{jk})^{-1},\, |g|={\rm det}(g_{jk})$. We denote by $dv= |g|^{\12}dsd\omega$ the Riemannian volume element on $(\cM, g)$ .
Putting $\psi=|g|^{1/4}\phi$ we see that $\psi$ solves
\begin{eqnarray}
\label{KGpsi}
(\partial_t-\i v)^2\psi
-|g|^{-1/4}(\partial_j-\i A_j) |g|^{1/2}g^{jk}(\partial_k-\i A_k)|g|^{-1/4}\psi
+m^2(s,\omega)\psi=0,\nonumber
\end{eqnarray}
which is the equation we will consider. 
\begin{remark}
{\rm  The equation \eqref{KGscatt} can be seen as a Klein-Gordon equation on the lorentzian manifold $\rr\times \cM$ with metric $dt^2-g$. Our results easily generalize to the metric $c(s,\omega)dt^2-g$ 
where $0<c_1\le c(s,\omega)\le c_1^{-1}$ is a smooth function tending to $1$ at infinity. The generalization reduces to a simple change of unknown function, see \cite[Sect.\ 2.1]{GHPS} for details. }
\end{remark}
We set 
\begin{eqnarray}
\label{e3.00}
h_0:=- g^{-1/4}(\partial_j-\i A_j) g^{1/2}g^{jk}(\partial_k-\i A_k)g^{-1/4}+m^2(s,\omega)
\end{eqnarray}
acting on $\cH=L^2(\cM;dsd\omega)$, equipped with its canonical scalar product. Let also 
\[
p=-g^{-1/4}\partial_j g^{1/2}g^{jk}\partial_kg^{-1/4}.
\]
We  assume that: 
\begin{equation}
\label{e3.0}
\left\{\begin{array}{l}
 A_j(s,\omega),\, m(s,\omega)-m_{\infty}\in S^{-\mu_0}(\cM)\\
\mbox{for some}\, \mu_0>0, \ m_{\infty}:= \lim_{s\to \infty}m(s, \omega)\geq 0.
\end{array}
\right.
\end{equation}
The operator $k$ is assumed to be a multiplication operator  $k=v(s,\omega)$
with:
\begin{equation}
\label{e3.1}
\left\{\begin{array}{rl}
&v(s,\omega)= v_{\rm l}(s,\omega)+ v_{\rm s}(s,\omega),\  v_{\rm l}(s,\omega)\in S^{-\mu_{0}}(\cM), \\[2mm]
&v_{\rm s} (s,\omega)\langle p\rangle^{-1/2}\in B_{\infty}(\cH), \ \langle s\rangle^{2}v_{\rm s}(s,\omega) \langle p\rangle^{-1/2}\in B(\cH).
\end{array}
\right.
\end{equation}
It follows that $h= h_{0}-k^{2}$ is self-adjoint and bounded below with 
$\jh\sim\langle h_{0}\rangle$ and 
$\sigma_{\rm ess}(h)=[m_{\infty}^{2}, +\infty[$. 

As scalar conjugate operator we choose as usual the generator of dilations:
\[
a= \12(\eta(s)s D_{s}+ D_{s} s\eta(s)),
\]
where $\eta\in C^{\infty}(\rr, \rr^{+})$ with $\eta(s)=1$ for $s\ge 2$ and $\eta(s)=0$ for $s\le 1$. 
As reference weight we choose:
\[
\langle x\rangle= (s^{2}+1)^{\12}.
\]
\subsubsection{Massive case}
\label{sec3.1}
In this subsection we consider massive Klein-Gordon equations i.e. $m=\inf \sigma(h)\cap \rr^{+}>0$. This implies that $m_{\infty}>0$.

\begin{proposition}\label{3.1}
 Assume (\ref{e3.0}), (\ref{e3.1}), $m_{\infty}>0$ and $\Ker h=\{0\}$. Then   \ben
\item  conditions (E), (A), (M) are satisfied;
\item one has $\tau(b)=\{m_{\infty}\}$.
\een
\end{proposition}
\proof 
To check (E) we use Prop.\ \ref{9.0} with $\epsilon= h_{0}^{\12}$, $r_{1}= \one_{\{|x|\geq R\}}v$, $r_{2}= \one_{\{|x|\leq R\}}v$.  Clearly $r_{2}\epsilon^{-1}\in B_{\infty}(\cH)$, and since  $\slim_{R\to \infty} \one_{\{|x|\geq R\}}=0$, we deduce from (\ref{e3.1}) that $\|r_{1}\epsilon^{-1}\|<1$ for $R$ large enough. Moreover $|h|\sim h_{0}\sim D_{s}^{2}-\frac{1}{s^2}\Delta_{\cN}+\one$.

To check (A) we use Lemma \ref{5.9}. We fix smooth cutoff functions $F_{0}$, $F_{\infty}\in C^{\infty}(\rr)$ with
\beq\label{9.4}
\supp F_{0}\subset[-2, 2], \ F_{0}\equiv 1\hbox{ in }[-1, 1], F_{0}+ F_{\infty}=1.
\eeq
We split $k$ as $k_{1}+ k_{2}$ with
\[
k_{1}= F_{\infty}(R^{-1}s)v_{\rm l}, \ k_{2}= v_{\rm s}+ F_{0}(R^{-1}s)v_{\rm l}.
\]
Since $F_{0}(R^{-1}|x|)v_{\rm l}$ satisfies the same conditions as $v_{\rm s}$ we can assume that $k_{2}= v_{\rm s}$ in the sequel. As before for $R\gg 1$ we have $\| k_{1}\epsilon^{-1}\|<1$, $k_{1}\epsi^{-1}$, $k_{2}\epsilon^{-1}\in B_{\infty}(\cH)$. By Lemma \ref{5.9} conditions (A) are satisfied for $b= (\epsilon^{2}-k_{1}^{2})^{\12}$ and $r= k_{2}^{2}+ 2k_{1}k_{2}$.
 (M1) is clearly satisfied.  To check (M2)  we apply Lemma \ref{7.3bis} and check (M2') instead.  It is a standard fact  that $\ad^{\alpha}_{a_{\chi}}(k_{1}\langle b\rangle^{-1})\in B(\cH)$ (one may for example use pseudo-differential calculus on scattering manifolds, see e.g. \cite[Chapter 6.3]{Me95} for an overview of this calculus).  Therefore $k_{1}\langle b\rangle^{-1}\in C^{2}(a_{\chi}; \cH)$. Since  $0\not \in \sigma(b)$ we also see  that $k_{1}b^{-1}\in C^{2}(a_{\chi}; \cH)$.

 The same type of argument shows that 
\beq\label{e3.4}
a_{\chi}\langle s\rangle^{-1}\langle p\rangle^{-1/2}, \ a_{\chi}^{2}\langle s\rangle^{-2}\langle p\rangle^{-1/2}\in B(\cH).
\eeq
This implies that  $\ad^{\alpha}_{a_{\chi}}(k_{2}b^{-1})\in B(\cH)$ using (\ref{e3.1}), by undoing the commutators with $k_{2}$, and using that $0\not \in \sigma(b)$. Therefore $k_{2}b^{-1}\in C^{2}(a_{\chi}; \cH)$.   Since we saw that $k_{1}b^{-1}\in C^{2}(a_{\chi};\cH)$ this also implies by 
\cite[Prop.\ 5.2.3]{ABG} that $b^{-1}k_{1}k_{2}b^{-1}\in C^{2}(a_{\chi}; \cH)$. Hence  (M2') is satisfied. 

The fact that condition (M3) is satisfied is also a standard result
(one can either use  pseudo-differential calculus or express $\langle
b\rangle$ via almost-analytic extensions). (M4) follows from (A2)
since $k$ and $\langle s\rangle$ commute. (M5) follows from the
pseudo-differential calculus on scattering manifolds.
 
The fact that $\tau(b^{2})=m_{\infty}^{2}$ follows from \cite[Theorem 1]{Ito11}. This proves (2).
\qed

\medskip

From Prop. \ref{3.1} and Prop. \ref{1.4old} we see that
\[
 \sigma_{\rm ess}(H)= \sigma_{\rm ess}(\dot H)= ]-\infty, -m_{\infty}]\cup [m_{\infty}, +\infty[.
\]

\subsubsection{Massless case}\label{sec3.2}
We consider $h_{0}$ as in (\ref{e3.00}) satisfying  (\ref{e3.0})  but assume now that  $\inf\sigma(h_{0})=0$. This is of course equivalent to $m_{\infty}=0$. 

We assume $d\geq 3$, because the Hardy inequality on $(\cM, g)$ will play an important role.
\ Instead of (\ref{e3.1}) we assume for $k=v$ that 
\beq\label{e3.2}
\left\{\begin{array}{l}
v(s,\omega)= v_{l}(s,\omega)+ v_{s}(s,\omega),\\[2mm]
\exists \ R_{0}>1, \ 0\leq \delta <1\hbox{ such that } |v_{l}(s,\omega)|\leq \delta\frac{d-2}{2}\langle s\rangle^{-1}, \hbox{ for }s\geq R_{0},\\[2mm]
s v_{s}\langle p\rangle^{-1/2}\in B_{\infty}(\cH), \ s^{3} v_{s}\langle p\rangle^{-1/2}\in B(\cH).
\end{array}
\right.
\eeq
Note that compared to (\ref{e3.1}) we require  an extra power of $s$ in the assumptions on $v_{2}$, which is needed to control the $\sh^{-\12}$ or $b^{-1}$ term arising in (A) and (M), thanks to the Hardy inequality. As before $h= h_{0}-k^{2}$ is self-adjoint with domain $H^{2}(M)$, bounded below, $\jh\sim \langle h_{0}\rangle$ and $\sigma_{\rm ess}(h)=[0, +\infty[$.
The operators $a$ and $\langle x\rangle$ are as in the previous subsection.

We have the following analog of Prop.\ \ref{3.1}, whose proof however is more involved and relies on estimates proved in Subsect.\ \ref{secapp.2.1}.
\begin{proposition}\label{3.2}
 Assume (\ref{e3.00}), (\ref{e3.0}) with $m_{\infty}=0$, (\ref{e3.2}), 
 $\Ker\, h=\{0\}$ and $d\geq 3$.
 Then:
\ben
\item  conditions (E), (A), (M ) are satisfied;
\item one has $\tau(b)=\{0\}$.
\een
\end{proposition}
\proof 
To check (E) we use again Prop.\ \ref{9.0}, with $\epsilon$, 
$r_{1}$, $r_{2}$ as in the proof of Prop.\ \ref{3.1}.
We first claim that $v_{\rm s}\epsilon^{-1}$ is compact. We use 
  \beq\label{e3.90}
  \epsilon^{-1}= \epsi^{-1}+ \epsi^{-1}\epsilon^{-1}(\epsi - \epsilon).
  \eeq
  The term $v_{\rm s}(\epsilon+1)^{-1}$ is compact by (\ref{e3.2}), so it suffices to prove that $v_{\rm s}\epsi^{-1}\epsilon^{-1}$ is compact. We have 
 \[
v_{\rm s}\epsi^{-1}\epsilon^{-1}= \left(v_{\rm s}\s\epsi^{-1}\right)\times\left (\epsi\s^{-1}\epsi^{-1}\s\right)\times \left(\s \epsilon^{-1}\right).
\]
The first factor is compact by (\ref{e3.2}). The second is seen to be bounded  by commuting   $\s^{-1}$ through $\epsi$. The third term is bounded by Prop.\ \ref{app.1c} (1).  Therefore $v_{\rm s}\epsilon^{-1}$ is compact. 
Since $\one_{s\leq R}v_{\rm l}$ satisfies the same estimates as $v_{\rm s}$ we see that $r_{2}\epsilon^{-1}\in B_{\infty}(\cH)$. This also implies that  $\|\one_{s\geq R}v_{\rm s}\epsilon^{-1}\|\to 0$ when $R\to \infty$. By Prop.\ \ref{app.1c} (2) we obtain that $\|\one_{|x|\geq R}v_{\rm l}\epsilon^{-1}\|<1$ for $R\gg 1$. Therefore $\|r_{1}\epsilon^{-1}\|<1$ for $R\gg 1$. Applying 
Prop.\ \ref{9.0} we obtain (E1), (E2). We also get that $|h|\sim \epsilon^{2}\sim D_{s}^{2}-\frac{1}{s^2}\Delta_{\cN}$.
By what we saw above, $k\epsilon^{-1}$ is bounded, hence (E3) also holds. 

To check (A)   we use again Lemma \ref{5.9}, with the same splitting of $k$ as in the proof of Prop.\ \ref{3.1}. We already checked that the hypotheses of Lemma \ref{5.9} hold, which proves (A)

(A3) is immediate since $k_{1}\epsi^{-1}\in B_{\infty}(\cH)$. 

(M1) and the fact that $k_{1}\langle b\rangle^{-1}\in C^{2}(a_{\chi}, \cH)$ are proved  as in Prop.\ \ref{3.1}.  To prove (M2) we check the hypotheses of  Lemma \ref{7.3bis}. To prove  that $k_{2}b^{-1}\in C^{2}(a_{\chi}; \cH)$ we have to check  that $\ad^{\alpha}_{a_{\chi}}(k_{2}b^{-1})$ are bounded for 
$\alpha=1,2$.  We use that:
\[
\begin{array}{rl}
&a_{\chi}\s^{-1}\langle p\rangle^{-1/2},  \s^{-1}\langle p\rangle^{-1/2} b^{-1}a_{\chi},\ \s^{-1}b^{-1}, \\[2mm]
 &a_{\chi}a_{\chi}\s^{-2}\langle p\rangle^{-1/2}, \ \langle p\rangle^{-1/2} \s^{-2}b^{-1}a_{\chi}a_{\chi} \in B(\cH).
\end{array}
\]
The  bounds  with $a_{\chi}$ are as in (\ref{e3.4}), using that $b^{-1}a_{\chi}= \tilde{\chi}(b^{2})a\chi(b^{2})$ for $\tilde{\chi}\in \coinf(\rr)$, since $0\not \in \supp \chi$. The fact that $\x^{-1}b^{-1}$ is bounded follows from Prop. \ref{app.1c} (1), using that  $b^{2}\simeq \epsilon^{2}$.  Undoing  the commutators and  using (\ref{e3.2}) we obtain that $\ad^{\alpha}_{a_{\chi}}(k_{2}b^{-1})$ are bounded for $\alpha=1,2$.
The same argument using that $k_{1}\in O(\s^{-1})$ shows that $ad_{a_{\chi}}(b^{-1}k_{1}k_{2}b^{-1})$ is bounded. This completes the proof of (M2').

As in the massive case we prove that (M3), (M4) hold and that
$\tau(b^{2})=\{0\}$ using \cite[Theorem 1]{Ito11}.  (M5) follows from
the pseudo-differential calculus on scattering manifolds.
\qed

\medskip

As in Subsection \ref{sec3.1} one has:
\[
\sigma_{\rm ess}(H)= \sigma_{\rm ess}(\dot H)= \rr.
\]

\subsubsection{Some additional remarks in the euclidean case}
If $\cM= \rr^{d}$ and the metric $g$ is asymptotically flat, then using polar coordinates we see that $(\cM, g)$ is a scattering manifold. In this case, using results of \cite{KT}, it is possible to exclude eigenvalues and critical points embedded in the essential spectrum.

\begin{proposition}
 Assume that $\cM= \rr^{d}$ and $g$ is asymptotically flat. Assume moreover that  $v= v_{1}+ v_{2}$ where
 \[
\left\{
\begin{array}{rl}
\p_{x}^{\alpha}v_{1}\in O(\langle x\rangle^{-\mu- |\alpha|}), \ \mu>0, \ |\alpha|\leq 2,\\[2mm]
 v_{2}\hbox{ has compact support, } v_{2}\in L^{d}(\rr^{d}).
\end{array}
\right.
\]
Then  $\sigma_{\rm p}(H)\cup c(H)\subset [-m_{\infty}, m_{\infty}]$. 
If $m_{\infty}=0$ and $\Ker\, h=\{0\}$, then $\sigma_{\rm p}(H)\cup c(H)=\emptyset$. 
\end{proposition}
\proof  Since $\dot\cE$ is a Pontryagin space, we know that  critical points of $H$ are eigenvalues. From \cite[Prop.\ 3.1]{G} we know that 
$\sigma_{\rm p}(H)\subset [-m_{\infty}, m_{\infty}]$. 
Moreover $\Ker\ h=\{0\}$ implies that $\Ker\, H=\{0\}$. \qed

\subsection{Models with hyperbolic ends}\label{sec3.3}
We fix a smooth compact manifold $\cN$, whose elements will be denoted by $\omega$ and a smooth  positive density on $\cN$ denoted by $d \omega$. We set $\cM:= \rr\times \cN$, whose elements are denoted by $(s, \omega)$ and equip $\cM$ with the density 
$dsd\omega$.

In this subsection we will describe some examples of Klein-Gordon equations
\begin{equation}
\label{e3.5z}
(\p_{t}- \i k)^{2}\phi(t)+h_{0} \phi(t)=0,
\end{equation}
on the Hilbert space $\cH= L^{2}(\cM,  ds d\omega)$, to which the results of Sects.\ \ref{sec1b}, \ref{sec2} can be applied.

\begin{remark}
{\rm   All the results of this subsection extend easily to the case where  the smooth manifold $\cM$ is equal to $\cM_{0}\cup [1, +\infty[\,\times \cN$, where $\cM_{0}$ is compact. One has to assume that the restriction of $h_{0}$ to $[1, +\infty[\,\times \cN$ satisfies similar assumptions as below, and the restriction of $h_{0}$ to $\cM_{0}$ is a second order, elliptic differential operator with smooth coefficients. }
\end{remark}
We introduce the spaces of exponentially decreasing functions:
\begin{equation}
\label{e3.5}
T^{p}(\cM):=\{f\in \cinf(\cM) : \p_{s}^{\alpha}\p_{\omega}^{\beta}f\in O(\e^{ p |s|})\}, \ p\in \rr.
\end{equation}
Similarly to Subsect.\ \ref{sec3.1}, we set
\[
S^{p}(\cM):= \{f\in \cinf(\cM) : \p_{s}^{\alpha}\p_{\omega}^{\beta}f\in O(\langle s\rangle^{p-|\alpha|})\}, \ p\in \rr.
\]
As usual a function $f$ in $T^{p}(\cM)$ resp.\  in $S^{p}(\cM)$ is called {\em elliptic} if $f^{-1}\in T^{-p}(\cM)$ resp.\ $f^{-1}\in S^{-p}(\cM)$.

We fix a second order differential operator $P=P(\omega, \p_{\omega})$ on $\cN$, assumed to be self-adjoint, positive  on $L^{2}(\cN, d \omega)$ with domain $H^{2}(\cN)$.

 We consider  an operator $h_{0}$ acting   on $\coinf(\cM)$ as
 \begin{equation}
\label{e3.8}
h_{0}= -c_{0}(s,\omega)\p_{s}g_{0}(s,\omega)\p_{s}c_{0}(s,\omega)- c_{-\12}(s,\omega)P(\omega, \p_{\omega})c_{-\12}(s,\omega)+ d_{0}(s,\omega),
\end{equation}
where the coefficients $c_{0}$,  $g_{0}$, $c_{-\12}$ and $d_{0}$ satisfy:
\begin{equation}
\label{e3.9}
\left\{
\begin{array}{l}
c_{0}, g_{0}\hbox{ elliptic in }S^{0}(\cM), \ c_{0}-1, g_{0}-1\in S^{-2}(\cM),\\[2mm]
c_{-\12}\hbox{ elliptic in }T^{-\12}(\cM), \langle s\rangle^{-2}(c_{-\12}(s, \omega)- \tilde c_{-\12}(s))\in  T^{-1/2}(\cM)
\\[2mm]
 \hbox{for some } \tilde c_{-\12}(s)\hbox{ elliptic in }T^{-\12}(\cM),
\\[2mm]
d_{0}(s, \omega)\in S^{0}(\cM), \ d_{0}(s, \omega)-m^{2}_{\infty}\in S^{-2}(\cM), \ m_{\infty}\geq 0 .
\end{array}
\right. 
\end{equation}
We assume moreover that on $\coinf(\rr)$ one has:
\begin{equation}
\label{e3.10b}
h_{0}\geq  m^{2}(s, \omega)
\hbox{ for some } m\in S^{0}(\cM), \ m(s,\omega)>0, 
\ \forall \ (s, \omega)\in \cM.
\end{equation}
It is easy to see that $h_{0}$ belongs to the general class studied in \cite{FH}, \cite[Sect.\ 3.3]{Ha1}. 
Therefore $h_{0}$ is self-adjoint, bounded below  with domain 
\[
\Dom\, h_{0}= \{u\in L^{2}(\cM) : h_{0}u \in L^{2}(\cM)\}= \Dom\, h_{\rm sep}, 
\]
where
\[
h_{\rm sep}=-\p^{2}_{s}- \tilde{c}_{-\12}(s)P(\omega, \p_{\omega})\tilde{c}_{-\12}(s),
\]
is separable. Moreover the inequality (\ref{e3.10b}) still holds on 
$\Dom\, h_{0}$. One also knows that 
$\sigma_{\rm ess}(h_{0})=[m^{2}_{\infty}, +\infty[$ 
where $m_{\infty}$ is defined in (\ref{e3.9}).

Concerning the operator $k$ we assume that  $k=k(s, \omega)$ is a multiplication operator with:
\begin{equation}
\label{e3.10}
k(s, \omega)\in S^{-2}(\cM), \  k(s, \omega)m(s, \omega)^{-1}\to 0\hbox{ when }s\to \infty.
\end{equation}
From \cite{FH} we get that $h_{0}$ is self-adjoint on $\Dom\, h_{\rm sep}$.
Since (\ref{e3.10b}) implies $\Ker\, h_{0}=\{0\}$, we see that  we are dealing with a massive Klein-Gordon equation 
iff $m_{\infty}>0$.

We now describe the conjugate operator $a$, following \cite{FH} and \cite{Ha1}. 
Let us fix functions $F, \chi\in \cinf(\rr)$, with $F', \chi'\geq 0$, $F(\lambda)=0$ for $\lambda\leq -1$, $F(\lambda)=1$ for $\lambda\geq -\12$,  $\chi(s)=0$ for $s\leq 1$, $\chi(s)=1$ for $s\geq 2$. We set $F_{S}(\lambda)= F(S^{-1}\lambda)$, $\chi_{R}(s)= \chi(R^{-1}s)$ for $S, R\geq 1$ and
\[
\begin{array}{rl}
X_{S, R}(s, P)=& \chi_{R}^{2}(s)F^{2}_{S}( \sigma s-\ln (P+1))( \sigma s-\ln(P+1)+2S)\\[2mm]
&+  \chi_{R}^{2}(-s)F^{2}_{S}( - \sigma s-\ln (P+1))( \sigma s+\ln(P+1)-2S),\\[2mm]
a_{S, R}:=& \12 \left(X_{S,R}(s, P)D_{s}+ D_{s}X_{S, R}(s, P)\right).
\end{array}
\]
Let us summarize some properties of $h, a_{S, R}$, which can be proved as in  \cite{FH}, \cite{Ha1}:
\begin{proposition}\label{3.19}
Assume (\ref{e3.9}), (\ref{e3.10b}) and (\ref{e3.10}). Then:
 \ben
\item $a_{S,R}$ is essentially self-adjoint on $\Dom (h_{\rm sep}+ \langle s\rangle ^{2})$,
\item $\langle s\rangle^{-p}a_{S, R}\langle s\rangle^{p-1}\in B(\cH)$ for $p\in \rr$,
\item $h_{0}+f\in C^{2}(a_{S, R})$ for any $f\in S^{-2}(\cM)$,
\item  Let $\tau(h_{0}+f):=\bigcap_{S,R\geq 1}\tau(h_{0}+f, a_{S,R})$ (see  the beginning of Subsect. \ref{sec2.3} for notation).  Then $\tau(h_{0}+f)=\{m_{\infty}^{2}\}$.
\een
\end{proposition}
\begin{remark}
{\rm   Note that (4) means that if $\lambda\neq m_{\infty}^{2}$, then there exist  an interval $I$ with $\lambda\in I$, parameters $S, R\geq 1$, a constant $c_{0}>0$  and $K\in B_{\infty}(\cH)$ such that
\[
\one_{I}(h_{0}+f)[h_{0}+ f, \i a_{S,R}]\one_{I}(h_{0}+f)\geq c_{0}\one_{I}(h_{0}+f)+ K.
\]
In the sequel we will forget the fact that the scalar conjugate operator $a_{S,R}$ depends on parameters $S, R$, and denote it simply by $a$.}
\end{remark}

It remains to fix the  reference weight appearing in hypothesis (M3). We choose
\[
 \langle x\rangle:= (s^{2}+1)^{\12}.
\]
To prove Prop.\ \ref{3.20} below, we will need the following lemma, whose proof
may be found in Subsect.\ \ref{app1.11}.

\begin{lemma}\label{segolo}
 Let $f\in S^{-2}(\cM)$ such that $ \epsilon^{2}- f^{2}=: b^{2}\geq 0$. Then
 \[
[\langle b\rangle, \langle s\rangle^{\delta}]\in B(\cH), \ \hbox{ for }0\leq \delta\leq 1.
\]
\end{lemma} 
  
\medskip

\begin{proposition}\label{3.20}
Assume (\ref{e3.9})-(\ref{e3.10}) and $\Ker\, h=\{0\}$. Then:   
\ben
  \item conditions (E), (A), (M) are satisfied;
  \item one has $\tau(b)=\{m_{\infty}\}$.
\een
\end{proposition}
\proof 
Set $\epsilon= h_{0}^{\12}$. We first claim that 
\begin{equation}
\label{dieudo.1}
\left\{
\begin{array}{rl}
(1)& \| \one_{\{|s|\geq R\}}k\epsilon^{-1}\|\to 0\hbox{ when }R\to +\infty,\\[2mm]
(2)& \one_{\{|s|\leq R\}}k\epsilon^{-1}\in B_{\infty}(\cH).
\end{array}
\right.
\end{equation}
In fact (1) follows from the fact that $h_{0}\geq m^{2}(s, \omega)$ and $k(s, \omega)m(s, \omega)^{-1}\to 0$ when $s\to \infty$, using also Kato-Heinz inequality. To complete the proof of (\ref{dieudo.1}) it suffices to prove that $g(s)\epsilon^{-1}\in B_{\infty}(\cH)$ for $g\in \coinf(\rr)$. Note that $g(s)\epsilon^{-1}$ is bounded by (\ref{e3.10b}).

As in the proof of Prop.\ \ref{3.2} it suffices using (\ref{e3.90}) to check that $ g(s)\epsi^{-1}$ and $g(s)\epsi^{-1}\epsilon^{-1}$ are compact. 
The term $g(s)\epsi^{-1}$ is compact, by \cite[Lemma 1.2]{FH}.   We write the second term as $\epsi^{-1}g(s)\epsilon^{-1}- [\epsi^{-1},g(s)]\epsilon^{-1}$. The first term is again compact since $g(s)\epsilon^{-1}$ is bounded. We write the second term as
\[
[\epsi^{-1},g(s)]\epsilon^{-1}
=\frac{\i}{2\pi}\int_{\cc}\frac{\p \tilde{f}_{-\12}}{\p\overline{z}}(z)(z- \epsilon^{2})^{-1}[\epsilon^{2}, g]\epsilon^{-1}(z-\epsilon^{2})^{-1}d z \wedge d \overline{z},
\]
where $\tilde{f}_{-\12}$ is an almost analytic extension of $f_{-\12}(\lambda)= (\lambda^{2}+1)^{-1/4}$, satisfying
\beq\label{e3.9b}
\left\{
\begin{array}{l}
\supp \tilde{f}_{-\12}\subset \{ z\in \cc  : |{\rm Im}z|\leq C \langle {\rm Re}z\rangle\},\\[2mm]
|\frac{\p \tilde{f}_{-\12}}{\p\overline{z}}(z)|\leq C_{N}\langle z\rangle^{-3/2-N}|{\rm Im z}|^{N}, \ N\in \nn.
\end{array}
\right. 
\eeq
Since $\epsilon^{2}$ is a second order differential operator, we obtain 
$[\epsilon^{2}, g]= (\epsilon^{2}+1)B\tilde{g}(s)$ with $B$ is compact and $\tilde{g}\in \coinf(\rr)$. Therefore $(\epsilon^{2}+1)^{-1}[\epsilon^{2}, g]\epsilon^{-1}$ is compact. We use now the bounds   
\[
\|(\epsilon^{2}-z)^{-1}\|\in O(|{\rm Im }z|^{-1}), \ \|(\epsilon^{2}-z)^{-1}(\epsilon^{2}+1)\|\in O(\langle z\rangle |{\rm Im }z|^{-1}),
\]
and (\ref{e3.9b}) to obtain that $[\epsi^{-1},g(s)]\epsilon^{-1}$ is compact. This completes the proof of (\ref{dieudo.1}).  We apply then Prop.\ \ref{9.0} with $r_{1}= \one_{\{|s|\geq R\}}k$, $r_{2}= \one_{\{|s|\leq R\}}k$. The hypotheses of Prop.\ \ref{9.0} hold by (\ref{dieudo.1}), which implies (E1), (E2). Moreover $\sh\sim h_{0}$ and we can replace $\sh^{-\12}$ by $\epsilon^{-1}$ in conditions (E3) and (A). Since by (\ref{dieudo.1}) we know that $k\epsilon^{-1}\in B(\cH)$, condition (E3) holds. To check (A) we use Lemma \ref{5.9} and split $k$ as $k_{1}+ k_{2}$ with
\[
k_{1}= F_{\infty}(R^{-1}|s|)v_{\rm l}, \ k_{2}= v_{\rm s}+ F_{0}(R^{-1}|s|)v_{\rm l}
\]
for some $F_{0}$, $F_{\infty}$ as in (\ref{9.4}). By (\ref{dieudo.1}) $\|k_{1}\epsilon^{-1}\|<1$ for $R\gg 1$, $k_{2}\epsilon^{-1}$ is compact. The fact that $k\epsi^{-1}$ is compact follows once again from 
\cite[Lemma 1.2]{FH}.

Let us now check (M1), (M2'). Note first that $b^{2}= \epsilon^{2}- k_{1}^{2}$ is of the form  $\epsilon^{2}+f$ for $f\in S^{-2}(\cM)$. Moreover if $R\gg 1$ we have 
\beq\label{e3.18}
b^{2}\geq\12 m^{2}(s, \omega),
\eeq
by  (\ref{e3.10b}).  Then (M1) follows from Prop.\ \ref{3.19} (3).
Using Prop.\ \ref{3.19} (2) and the fact that $k_{1}\in S^{-2}(\cM)$, we obtain that $ k_{1}\in C^{2}(a_{\chi}; \cH)$ by undoing the commutators. Since $ \langle b\rangle^{-1}\in C^{2}(a_{\chi}; \cH)$, this proves the first  condition of (M2').

To prove the rest of (M2') we claim that $g(s)b^{-1}\in C^{2}(a_{\chi}; \cH)$ for $g\in \coinf(\rr)$. Note that this implies the last two conditions of (M2'), since $k_{2}$, $k_{1}k_{2}\in \coinf(\rr)$.

If $m_{\infty}>0$ this is  proved by the same argument as before.  If $m_{\infty}=0$, i.e. we are considering the massless case, then we argue as in the proof of Lemma \ref{3.2}: we use that   
\[
\begin{array}{rl}
&a_{\chi}\langle s\rangle^{-1}, \ \langle s\rangle b^{-1}a_{\chi}, \ \langle s\rangle^{n}g(s)b^{-1},\\[2mm]
&a_{\chi}a_{\chi}\langle s\rangle^{-2}, \ \langle s\rangle^{-2}b^{-1}a_{\chi}a_{\chi}\in B(\cH).
\end{array}
\]
The bounds with $a_{\chi}$ rely on Prop.\ \ref{3.19} (2) and the fact that $b^{-1}a_{\chi}= \tilde{\chi}(b^{2})a \chi(b^{2})$ for $\tilde{\chi}\in \coinf(\rr)$ since $0\not\in \supp \chi$. 
As before we complete the proof by undoing the commutators with $g(s)b^{-1}$.  

We now prove (M3).  The first condition of (M3) follows from Prop. \ref{3.19} (2). To prove the second it suffices to prove that 
\beq\label{e3.91}
[\langle b\rangle, \langle s\rangle^{\delta}]\in B(\cH), \ 0\leq \delta\leq 1,
\eeq
which has been shown in Lemma \ref{segolo}.
Finally (M4) is true since $k$ and $\jr$ commute, and the fact that 
$\tau(b)= \{m_{\infty}\}$ follows from Prop.\ \ref{3.19} (4). \qed

\appendix
\section{}\label{secapp.2}\init
\subsection{Diamagnetic and Hardy inequalities}\label{secapp.2.1}

 We start by recalling some well-known facts related to the diamagnetic inequality. We are working on the scattering manifolds introduced in 
 Subsect.\ \ref{Scattman} and set
\[
p=-\sum_{j,k=1}^d g^{-1/4}\partial_jg^{1/2}g^{jk}\partial_kg^{-1/4}, \ p_{A}:=-\sum_{j,k=1}^d g^{-1/4}(\partial_j-\i A_j) g^{1/2}g^{jk}(\partial_k-\i A_k)g^{-1/4}
\]
where $ A(s,\omega)$ satisfies (\ref{e3.0}). We use the notations of 
Subsect.\ \ref{Scattman}.
\begin{lemma}\label{diam}
Let $ V\in C^{0}(\rr^{d}, \rr)$ be a bounded potential. Then:
\[
 p+V\geq 0 \Rightarrow 	p_A+ V\geq 0,
\]
\end{lemma}
\proof 
Let us first recall the diamagnetic inequality~:
\begin{equation}
\label{diamag}
|\e^{- t (p_A+V)}u|\leq \e^{- t (p+V)}|u|, \ u\in \cH, \ t\geq 0.
\end{equation}
This inequality is well known on $\rr^d$ and also holds on scattering manifolds. Indeed it is equivalent to a certain estimate on the quadratic forms associated to the operators, which clearly also holds on scattering manifolds, see \cite{Si79} for details. 
Now recall that
\beq\label{ea.0}
a^{-\alpha}= C_{\alpha}\int_{0}^{+\infty}t^{\alpha-1}\e^{- t a}d t, \ a>0, \alpha>0,
\eeq
where $C_{\alpha}$ is a positive constant. Using (\ref{ea.0}) for $ \alpha=1$ we obtain 
\[
(u|p_A+V+ \delta)^{-1}u)\leq (|u| |(p+V+ \delta)^{-1}|u|)\leq \delta^{-1} \|u\|^{2},
\]
which by Kato-Heinz inequality implies that  $p_A+V+ \delta\geq \delta$, which proves  the lemma. \qed

\medskip

We now prove some estimates related to Hardy's inequality on the scattering manifold $(\cM, g)$  considered in Subsect.\ \ref{Scattman}.

Let $g^{0}$ be a conic metric as in Def.\ \ref{def-de-conic}, restricted to $\cM_{\infty}= ]1, +\infty[\times \cN$. The corresponding Laplace-Beltrami operator is
\[
-\Delta_{g^{0}}= - s^{(1-d)}\p_{s}s^{d-1}\p_{s}- \Delta_{h}, 
\]
which is self-adjoint on $L^{2}(\cM_{\infty}, s^{d-1}|h|^{\12}dsd\omega)$. 
The usual proof on $\rr^{d}$, which relies on the identity $s^{-2}= -\12s\p_{s}(s^{-2})$ for $s= |x|$, yields for  $u\in \coinf(\cM_{\infty})$:
\begin{equation}
\label{hardyto}
(\frac{d-2}{2})^{2}\int_{\cM_{\infty}}s^{-2}|u|^{2}s^{d-1} |h|^{\12}dsd\omega\leq -\int_{\cM_{\infty}}\overline{u}\Delta_{g^{0}}u s^{d-1} |h|^{\12}dsd\omega,
\end{equation}
where $-\Delta_{h}$ is the Laplace-Beltrami operator on $(\cN, h)$. Using the unitary map: 
\[
T: L^{2}(\cM, |g|^{\12}dsd\omega)\ni u\mapsto |g|^{1/4}u\in L^{2}(\cM, dsd\omega),
\]
this immediately implies that (\cite[Prop. 3.4]{VaWu10}) :
\beq\label{hardyti}
s^{-2}\leq C p, \ C>0, \hbox{ on }\cH= L^{2}(\cM, dsd\omega).
\eeq
\begin{proposition}\label{app.1c}
 Assume (\ref{e3.0}) and $d\geq 3$. Then:
 \ben
 \item $p_A\geq  C \langle s\rangle^{-2}$,
 \item if in addition $v_{l}(x)$ satisfies (\ref{e3.2}) then 
 \[
(1+\alpha)^{-1}p_A\geq F_{\infty}^{2}(R^{-1}|s|)v_{l}^{2}(s),\hbox{ for some }0<\alpha<1, \ R\gg 1,
\]
and  $F_{\infty}$ as in (\ref{9.4}).
 \een
\end{proposition}
\proof  Statement (1) follows from (\ref{hardyti}) and Lemma \ref{diam}.
Let us now prove (2).  Since $g$ is a long-range perturbation of $g^{0}$, we deduce from (\ref{hardyto}) that
\[
\one_{|s|\geq R}\leq (\frac{2}{d-2})^{2}(1+ O(R^{-\mu})) p,
\]
hence by Lemma \ref{diam}:
\[
\one_{\{s\geq R\}}(s)\langle s\rangle^{-2}\leq (\frac{2}{d-2})^{2}(1+ O(R^{-\mu}))p_A,
\]
which implies (2), using the estimate (\ref{e3.2}) on $v_{l}(s,\omega)$. \qed

\subsection{Proof of Prop.\ \ref{9.0}}\label{app1.1}

Let $\epsilon_{1}^{2}= \epsilon^{2}- r_{1}^{2}$. Since $\|r_{1}\epsilon^{-1}\|<1$,   we have $\epsilon_{1}^{2}\sim \epsilon^{2}$. Therefore ${\rm Ker}\,\epsilon_{1}=\{0\}$,  and $r_{2}\epsilon_{1}^{-1}\in B_{\infty}(\cH)$.
We have
\[
h= \epsilon^{2}- r^{2}= \epsilon_{1}^{2}- r_{2}^{2}.
\]
Therefore denoting $\epsilon_{1}$ again by $\epsilon$ we can assume that $r_{1}=0$ and denote $r_{2}$ by $r$, so that $r\epsilon^{-1}\in 
B_{\infty}(\cH)$, $h= \epsilon^{2}-r^{2}$.  Note that  $\sigma_{\rm ess}(h)= \sigma_{\rm ess}(\epsilon^{2})$. If $m>0$ then $\sigma_{\rm ess}(h)\subset [m_{\infty}^{2}, +\infty[$ for some $m_{\infty}>0$ hence ${\rm Tr}\one_{]-\infty, 0]}(h)<\infty$. Moreover $0\not\in \sigma(h)$ and $0\not \in \sigma(\epsilon^{2})$, thus $\sh\sim \jh\sim \epsi^{2}\sim \epsilon^{2}$.
Hence (2) in \eqref{eq:a2} also holds.

Let us now assume that $m=0$. We first prove (1) from \eqref{eq:a2}.
Noting that $r\epsi^{-1}$ is bounded, we obtain by the Birman-Schwinger principle that 
 \[
{\rm Tr}\one_{]-\infty, - \alpha]}(h)= {\rm Tr}\one_{]1, +\infty[}(K_{\alpha}),
\]
for $K_{\alpha}= r (\epsilon^{2}+ \alpha)^{-1}r\in B_{\infty}(\cH)$, $\alpha>0$.  Since $r\epsilon^{-1}\in B_{\infty}(\cH)$ we have 
 $K_{\alpha}\nearrow K_{0}= r \epsilon^{-2}r\in B_{\infty}(\cH)$, hence 
\[
{\rm Tr}\one_{]-\infty, 0[}(h)= {\rm Tr}\one_{]1, +\infty[}(K_{0})<\infty.
\]
Since $\Ker\, h=\{0\}$, this implies that ${\rm Tr}\one_{]-\infty, 0]}(h)<\infty$, which proves (1) in \eqref{eq:a2}.

We now prove (2) from \eqref{eq:a2}. 
Set $P_{\pm}:= \one_{\rr^{\pm}}(h)$. If  
$hu= (\epsilon^{2}-r^{2})u= -\lambda u$, $\lambda>0$ we have $\epsilon^{-1}u= (-\lambda)^{-1}(\one - \epsilon^{-1}r^{2}\epsilon^{-1})\epsilon u
\in \cH$.  This implies that 
\[
|u)(u|\leq C \epsilon^{2}, \ C>0.
\]
Since ${\rm Tr}P_{-}<\infty$ this implies that $P_{-}\leq C \epsilon^{2}$, for some $C>0$. Now
\[
|h|= h- 2hP_{-}\leq h + 2|\inf \sigma(h)| P_{-}\leq \epsilon^{2}+ 2C|\inf \sigma(h)|\epsilon^{2},
\]
which shows that $| h|\leq C \epsilon^{2}$ for some $C>0$. 

To prove the lower bound, we adapt some arguments in \cite{Sof}. Let \[
h_{\delta}= h -\delta r^{2}= \epsilon^{2}- (1+ \delta)r^{2}.
\]
Again by the Birman-Schwinger principle, we have 
$ {\rm Tr}\one_{]-\infty, 0[}(h_{\delta})= {\rm Tr}\one_{]-\infty, 0[}(h)$, for $\delta$ small enough.
  Therefore there exists $c_{0}, \delta_{0}>0$ such that   
 \beq\label{e1.2}
\one_{[-c_{0}, 0[}(h_{\delta})= 0,\ \forall \ 0\leq \delta\leq \delta_{0}.
\eeq
We fix cutoff functions $\chi_{\pm}$ with $\chi_{-}\in \coinf(]-\infty, -c_{0}/2[)$, $\chi_{+}\in \cinf(]-c_{0}, +\infty[)$ and $\chi^{2}_{-}(h)+ \chi^{2}_{-}
(h)=\one$.
From (\ref{e1.2}) it follows that $\chi_{+}(h_{\delta})h_{\delta}\chi_{+}(h_{\delta})\geq 0$. Therefore:
\beq\label{e1.6}
\begin{array}{rl}
P_{+}h_{\delta}P_{+}=& P_{+}\left(\chi_{-}(h_{\delta})h_{\delta}\chi_{-}(h_{\delta})+ \chi_{+}(h_{\delta})h_{\delta}\chi_{+}(h_{\delta})\right)P_{+}\\[2mm]
\geq &P_{+}(\chi_{-}(h_{\delta})h_{\delta}\chi_{-}(h_{\delta})P_{+}= P_{+}Rh_{\delta}RP_{+},
\end{array}
\eeq
for $R= \chi_{-}(h_{\delta})- \chi_{-}(h)$, using that $P_{+}\chi_{-}(h)=0$.
We claim that
\begin{equation}
\label{e1.3c}
Rh_{\delta}R\geq - C\delta^{2} \epsilon^{2}, \ C>0 \ \hbox{ uniformly for }0\leq \delta\leq \delta_{0},
\end{equation}
which follows from
\begin{equation}
\label{e1.3}
\|\epsilon^{-1}R\langle h_{\delta}\rangle^{\12}\|\leq C\delta, \hbox{ uniformly for }0\leq \delta\leq \delta_{0}.
\end{equation}
To prove (\ref{e1.3}) it suffices to check that
\beq\label{e1.3d}
\|(\epsilon+\alpha)^{-1}R\langle h_{\delta}\rangle^{\12}\|\leq C\delta, \hbox{ uniformly for }0\leq \delta\leq \delta_{0}, \alpha>0.
\eeq
We have
\beq\label{e1.3b}
\begin{array}{rl}
&(\epsilon+\alpha)^{-1}R\langle h_{\delta}\rangle^{\12}\\[2mm]
=&\delta\frac{\i}{2\pi}\int_{\cc}\frac{\p}{\p \bar z}\tilde{\chi}_{-}(z)(\epsilon+\alpha)^{-1}(z-h)^{-1}r^{2}(z-h_{\delta})^{-1}\langle h_{\delta}
\rangle^{\12} \: d z\wedge d\bar{z},
\end{array}
\eeq
where $\tilde{\chi}_{-}(z)\in \coinf(\rr)$ is an almost analytic extension of $\chi_{-}$. We write:
\[
\begin{array}{rl}
&(\epsilon+\alpha)^{-1}(z-h)^{-1}r^{2}(z-h_{\delta})^{-1}\langle h_{\delta}\rangle^{\12}\\[2mm]
=&(z-h)^{-1}(\epsilon+\alpha)^{-1}r^{2}(z-h_{\delta})^{-1}\langle h_{\delta}\rangle^{\12}\\[2mm]
+&(z-h)^{-1}[h, (\epsilon+\alpha)^{-1}](z-h)^{-1}r^{2}(z-h_{\delta})^{-1}\langle h_{\delta}\rangle^{\12}\\[2mm]
=&(z-h)^{-1}(\epsilon+\alpha)^{-1}r^{2}(z-h_{\delta})^{-1}\langle h_{\delta}\rangle^{\12}\\[2mm]
+&(z-h)^{-1}(\epsilon+\alpha)^{-1}r^{2}(z-h)^{-1}r^{2}(z-h_{\delta})^{-1}\langle h_{\delta}\rangle^{\12}\\[2mm]
-&(z-h)^{-1}r^{2}(\epsilon+\alpha)^{-1}(z-h)^{-1}r^{2}(z-h_{\delta})^{-1}\langle h_{\delta}\rangle^{\12}\\[2mm]
=:& I_{1}(z)+ I_{2}(z)- I_{3}(z).
\end{array}
\]
We write:
\beq\label{e1.4}
\begin{array}{rl}
I_{1}(z)=&(z-h)^{-1}\times (\epsilon+\alpha)^{-1}r^{2}\langle h_{\delta}\rangle^{-\12}\times (z-h_{\delta})^{-1}\langle h_{\delta}\rangle\\[2mm]
=&O(|{\rm Im}z|^{-2}), \hbox{ uniformly in }\alpha, \delta\hbox{ and }z\in \supp \tilde{\chi}_{-},
\end{array}
\eeq
using  that $r\epsilon^{-1}$ is bounded, and  $r\langle h_{\delta}\rangle^{-\12}$ is bounded uniformly in $0\leq \delta\leq \delta_{0}$.
Similarly we have:
\begin{equation}
\label{e1.5}
\begin{array}{rl}
I_{2}(z)=&(z-h)^{-1}\times(\epsilon+\alpha)^{-1}r^{2}\langle h\rangle^{-\12} \times \langle h\rangle (z-h)^{-1}\\[2mm]
&\times \langle h\rangle^{-\12}r^{2}\langle h_{\delta}\rangle^{-\12}\times (z-h_{\delta})^{-1}\langle h_{\delta}\rangle\\[2mm]
=& O(|{\rm Im}z|^{-3}), \hbox{ uniformly in }\alpha, \delta\hbox{ and }z\in \supp \tilde{\chi}_{-}.
\end{array}
\end{equation}
A similar argument shows that  $I_{3}(z)$ satisfies the same bound as $I_{2}(z)$. Therefore using (\ref{e1.3b}) we obtain (\ref{e1.3d}), hence 
(\ref{e1.3c}).

We have now 
\[
\begin{array}{rl}
P_{+}hP_{+}=& (1+\delta)^{-1}p_{+}h_{\delta}P_{+}+ \delta(1+\delta)^{-1}P_{+}\epsilon^{2}P_{+}\\[2mm]
\geq &(\delta- C \delta^{2})(1+ \delta)^{-1}P_{+}\epsilon^{2}P_{+},
\end{array}
\]
by (\ref{e1.6}) and (\ref{e1.3c}). Choosing $\delta$ small enough we obtain that
\begin{equation}
\label{e1.7}
P_{+}\epsilon^{2}P_{+}\leq CP_{+}hP_{+},\ C>0.
\end{equation}
On the other hand since ${\rm Ran}P_{-}$ is finite dimensional  and included in $\Dom\, \epsilon$, we have $P_{-}\epsilon^{2}P_{-}\leq C P_{-}^{2}$ for some
$C>0$. Using that 
\[
P_{+}\epsilon^{2}P_{-}+ P_{-}\epsilon^{2}P_{+}\leq P_{+}\epsilon^{2}P_{+}+ P_{-}\epsilon^{2}P_{-}
\]
we finally obtain
\[
\begin{array}{rl}
\epsilon^{2}=& (P_{+}+ P_{-})\epsilon^{2}(P_{+}+ P_{-})\\[2mm]
\leq & 2P_{+}\epsilon^{2}P_{+}+ 2 P_{-}\epsilon^{2}P_{-}\\[2mm]
\leq &2CP_{+}hP_{+}+2C P_{-}^{2}\leq C'|h|,
\end{array}
\]
where in the last inequality we used (\ref{e1.2}) for $\delta=0$. This proves
(2) in \eqref{eq:a2}. 
\qed

\subsection{Proof of Lemma \ref{segol}}\label{app1.10}

 Let $\tilde{f}_{-\delta}$ be an almost analytic extension of the function $ \langle\cdot \rangle^{-\delta}$ satisfying:  \[
\begin{array}{rl}
&\supp \tilde{f}_{-\delta}\subset \{z\in \cc  : |{\rm Im}z|\leq c \langle {\rm Re}z\rangle\},\\[2mm]
&|\frac{\p \tilde{f}_{-\delta}}{\p\overline{z}}(z)|\leq C_{N}\langle z\rangle^{-\delta- 1-N}|{\rm Im}z|^{N}, \ N\in \nn,
\end{array}
\]
see \cite[Appendix C.2]{DG}. We have 
\begin{align}
& [ \langle a_{\chi}\rangle^{-\delta}, \langle b\rangle -b]\langle a_{\chi}\rangle^{\delta} \label{sego} = \\[2mm]
& \frac{\i}{2\pi}\int_{\cc}\frac{\p \tilde{f}_{-\delta}}{\p\overline{z}}(z)
(z -a_{\chi})^{-1}[a_{\chi}, \langle b \rangle -b](z- a_{\chi})^{-1} \langle a_{\chi}\rangle^{\delta}
d z \wedge d \overline{z}. \nonumber
\end{align}
Since $0\not \in \supp \chi$,   there exist $g\in \coinf(\rr)$ such that 
\[
[a_{\chi}, \langle b \rangle -b]= \chi(b^{2})[a, g(b^{2})]\chi(b^{2})\in B(\cH),
\]
by (M1). Now we use  the bound $\|(z-a_{\chi})^{-1}\langle a_{\chi}\rangle^{\delta}\|\in O( \langle z\rangle^{\delta}|{\rm Im}z|^{-1})$ and the estimates satisfied by $\tilde{f}_{-\delta}$ to obtain that the integral  in (\ref{sego}) is norm convergent. This completes the proof of the lemma. \qed
\subsection{Proof of Lemma \ref{segolo}}\label{app1.11}
Note that $b^{2}$ is of the form (\ref{e3.8}). We claim  first  that
\begin{equation}
\label{e3.900}
[b^{2}, \langle s\rangle^{\delta}]\langle b\rangle^{-1}, \ \langle b\rangle^{-1}[b^{2}, [b^{2}, \langle s\rangle^{\delta}]]\langle b\rangle^{-1}\in B(\cH), \ 0\leq \delta\leq 1.
\end{equation}
In fact this follows by an easy computation using (\ref{e3.9})  and the fact that
\[
\p_{s}\langle b\rangle^{-1}, \ c_{-\12}(s, \omega) \p_{\omega}\langle b\rangle^{-1}\in B(\cH),
\]
see \cite[Lemma 4.3.1]{Ha1}.

We write now
\[
\langle b\rangle= (b^{2}+1)\langle b\rangle^{-1}, \ [\langle b\rangle, \langle s\rangle^{-\delta}]= [b^{2}, \langle s\rangle^{-\delta}]\langle b\rangle^{-1}+ (b^{2}+1)[\langle b\rangle^{-1}, \langle s\rangle^{-\delta}].
\]
The first term is bounded by (\ref{e3.900}). To estimate the second term, we introduce the function $f_{-\12}(\lambda)= (\lambda+1)^{-\12}$  and write with $\tilde{f}_{-\12}$ as in (\ref{e3.9b}):
\[
\begin{array}{rl}
&(b^{2}+1)[\langle b\rangle^{-1}, \langle s\rangle^{-\delta}]\\[2mm]
=&\frac{\i}{2\pi}\int_{\cc}\frac{\p \tilde{f}_{-\12}}{\p\overline{z}}(z)(b^{2}+1)(z-b^{2})^{-1}[b^{2}, \langle s\rangle^{\delta}](z-b^{2})^{-1}d z \wedge d \overline{z}\\[2mm]
=&(b^{2}+1)f_{-\12}'(b^{2})[b^{2}, \langle s\rangle^{\delta}]\\[2mm]
+&\frac{\i}{2\pi}\int_{\cc}\frac{\p \tilde{f}_{-\12}}{\p\overline{z}}(z)(b^{2}+1)(z-b^{2})^{-2}[b^{2},[b^{2}, \langle s\rangle^{\delta}](z-b^{2})^{-1}d z \wedge d \overline{z}.
\end{array}
\] 
The first term is again bounded using (\ref{e3.900}) and the fact that $f_{-\12}'(\lambda)\in O(\langle \lambda\rangle^{-3/2})$. The integral in the second term is norm convergent, using (\ref{e3.900}), the estimates on $\tilde{f}_{-\12}$ and the bounds $\langle b\rangle^{\alpha}(z-b^{2})^{-1}\in O(\langle z\rangle^{\alpha/2}|{\rm Im}z|^{-1})$ for $0\leq \alpha\leq 2$. This completes the proof of the fact that $[\langle b\rangle, \langle s\rangle^{\delta}]$ is bounded. \qed

\section{}\label{secapp.3}\label{proof-iaca}
\init

In this appendix we prove Thm.\ \ref{iaca}.
We first recall some standard results.

Let $X$ be a locally compact space, $C(X)$ the space of  bounded continuous functions and $B(X)$ the space of bounded Borel functions.  We recall that a sequence $(\varphi_{n})_{n\in \nn}$ in $B(X)$ is {\em b-convergent} to 
$\varphi$, written as ${\rm b-}\lim_{n}\varphi_{n}=\varphi$ if
 $\sup_{n}\| \varphi_{n}\|_{\infty}<\infty$ and $\varphi_{n}\to \varphi$ pointwise on $X$.
 
 The {\em monotone class theorem} implies that $B(X)$ is the smallest space of functions on $X$ containing $C(X)$ and stable under bounded convergence of sequences. The {\em Riesz theorem}  says that any continuous linear form on $C(X)$ uniquely extends to a linear form on $B(X)$ continuous for the b-convergence of sequences.

Recall also that a Banach space  $\cH$  is {\em weakly sequentially complete}, if for each sequence $(u_{n})_{n\in \nn}$ in $\cH$ such that  $\lim_{n}\langle f, u_{n}\rangle$ exists for each $f\in \cH^{\t}$, there exists $u\in \cH$ with $\lim_{n}\langle f, u_{n}\rangle= \langle f, u\rangle$ for each $f\in \cH^{\t}$.  Reflexive Banach spaces, hence Hilbertizable spaces, 
are weakly sequentially complete.

This property implies that if $(T_{n})_{n\in \nn}$ is a sequence in $B(\cH)$ such that $\lim_{n} \langle f, T_{n}u\rangle$ exists for each $u\in \cH, f\in \cH^{\t}$, then there is a unique $T\in B(\cH)$ such that $\wlim _{T_{n}}=T$.

From these facts, it is straightforward  to prove the following result, see eg \cite[Cor. 9.1.10]{Wora}.
\begin{theoreme}\label{arsi}
 Let $X$ a locally compact space, $\cH$ a weakly sequentially complete Banach space. Then if $F_{0}: C(X)\to B(\cH)$ is a continuous algebra morphism, there is a unique algebra morphism $F: B(X)\to B(\cH)$ such that ${\rm b-}\lim_{n}\varphi_{n}=\varphi$ implies $\wlim F(\varphi_{n})= F(\varphi)$.
\end{theoreme}

{\it Proof of Thm. \ref{iaca}.}

\smallskip

We will deduce Thm. \ref{iaca} from Thm. \ref{arsi} for a convenient choice of  the locally compact space $X$.
We use the notations in Subsects.\ \ref{seckrein.3}, \ref{borelborel}.

Let $\chi$ be a smooth function which has a zero of order
$\alpha(\xi)$ at each $\xi\in\supp\alpha$ and has no other zeros.
This means $\chi=c_{\omega}\chi_{\omega}+o(\chi_{\omega})$ if
$\omega\in\tilde\alpha$ with $c_{\omega}$ non zero numbers and
$\chi(x)\neq0$ outside $\supp\alpha$. 

Let $2\varepsilon$  be the minimal distance between two points of $\supp\alpha$
 and let $\theta_{0}$ be a smooth function with
$\theta_{0}(x)=1$ if $|x|<\varepsilon/3$ and $\theta_{0}(x)=0$ if
$|x|>\varepsilon/2$.  Then let $\theta_{1}$ be a smooth function equal to
$1$ on a neighborhood of $\infty$ and equal to $0$ at points at
distance $<\varepsilon$ from $\supp\alpha\cap\rr$. Finally, if
$\omega=(\xi,s)\in\tilde\alpha$ and $\xi\in\rr$ then we set
$\theta_{\omega}(x)=\theta_{0}(x-\xi)$, and if $\xi=\infty$ we set
$\theta_{\omega}=\theta_{1}$. Thus the functions in the family
$\{\theta_{\omega}\}_{\omega\in\tilde\alpha}$ have disjoint supports
and each of them is equal to one on a neighborhood of a unique point
from $\supp\alpha$.

Recall that for $\varphi\in\Lambda^{\alpha}$ we have
$\delta_\omega(\varphi)=\delta_\omega(\varphi^\circ)$ if
$\omega\prec\alpha$ and so $T_\omega\varphi=T_\omega\varphi^\circ$
if $\alpha\preceq\alpha$.  We associate to such a $\varphi$ a function
$\tilde\varphi\in B(\hat\rr)$ defined by
\[
\tilde{\varphi}=
\chi^{-1}\big(\varphi^{\circ}-
{\textstyle\sum_{\omega\in\tilde\alpha}}\theta_{\omega}T_{\omega}\varphi
\big)
\]
outside $\supp \alpha$,
while  at points $\xi\in\supp\alpha$ we set
$\tilde{\varphi}(\xi)=c_{\omega}^{-1}\delta_{\omega}(\varphi)$ with
$\omega=(\xi,\alpha(\xi))$.  The definition of $\tilde\varphi$ on
the support of $\alpha$ is  such that $\tilde\varphi\in C(\hat\rr)$
if $\varphi\in C^\alpha(\hrr)\subset\Lambda^\alpha$. Observe that 
\begin{equation*}
{\textstyle\sum_{\omega\in\tilde\alpha}}\theta_{\omega}T_{\omega}\varphi
={\textstyle\sum_{\mu<\omega\in\tilde\alpha}}\theta_{\omega} 
\delta_{\mu}(\varphi)\chi_{\mu}
={\textstyle\sum_{\mu<\omega\in\alpha}}\theta_{\mu} 
\delta_{\mu}(\varphi)\chi_{\mu}
={\textstyle\sum_{\omega\prec\alpha}}\theta_{\omega} 
\delta_{\omega}(\varphi)\chi_{\omega}
\end{equation*}
because for $\mu<\omega$ we have $\theta_{\mu}=\theta_{\omega}$.
Thus we have
\begin{equation}\label{eq:obs}
\varphi^{\circ}=\chi\tilde{\varphi} +
{\textstyle\sum_{\omega\in\tilde\alpha}}\theta_{\omega}T_{\omega}\varphi
=\chi\tilde{\varphi} +
{\textstyle\sum_{\omega\prec\alpha}}\theta_{\omega}\delta_{\omega}(\varphi)\chi_{\omega}
\end{equation}
Now let us denote $\widehat\alpha=\{\omega\mid\omega\prec\alpha\}$ and
let us consider the map
\[ 
A:\Lambda^{\alpha}=L^{\alpha}(\hrr)\oplus\cc^{\tilde\alpha} \to
B(\widehat\rr)\oplus\cc^{\widehat\alpha} \quad\text{defined by }
A\varphi=\big(\tilde\varphi,(\delta_{\omega}(\varphi))_{\omega\prec\alpha}\big).
\]
Then clearly $A$ is linear continuous and injective and we have
$AC^\alpha(\hrr)\subset C(\widehat\rr)\oplus \cc^{\widehat\alpha}$. On the other
hand, from \eqref{eq:obs} it follows that $A$ is bijective with
continuous inverse given by
\[
A^{-1} \big(\psi,(a_{\omega})_{\omega\prec\alpha}\big)
=\big(\chi\psi +
{\textstyle\sum_{\omega\prec\alpha}} \theta_{\omega}a_{\omega}\chi_{\omega},
(c_\omega\psi(\omega))_{\omega\in\tilde\alpha}\big)
\]
where we used the notation $\psi(\omega)=\psi(\xi)$ for
$\omega=(\xi,\alpha(\xi))\in\tilde\alpha$. It is also easy to check
that $A^{-1}$ sends $C(\widehat\rr)\oplus\cc^{\widehat\alpha}$ into
$C^\alpha(\hrr)$, hence $A:C^\alpha(\hrr)\to C(\widehat\rr)\oplus\cc^{\widehat\alpha}$
is an isomorphism. 

Summarizing  we have:
\[
\begin{array}{ll}
i)&A:\ C^{\alpha}(\hrr)\sim C(\hrr)\oplus \cc^{\widehat\alpha},\\[2mm]
ii)&A:\ \Lambda^{\alpha}\sim  B(\hrr)\oplus \cc^{\widehat\alpha}.
\end{array}
\]
Let $\hat\rr\sqcup\widehat\alpha$ be the topological disjoint union of
$\hat\rr$ with the discrete space $\widehat\alpha$. We have obvious
identifications
$C(\hat\rr)\oplus\cc^{\widehat\alpha}\sim C(\hat\rr\sqcup\widehat\alpha)$ and
$B(\hat\rr)\oplus\cc^{\widehat\alpha}\sim B(\hat\rr\sqcup\widehat\alpha)$, which in
particular induce the natural notion of b-convergence for sequences
on the space $B(\hat\rr)\oplus\cc^{\widehat\alpha}$. Then it is clear
that $A$ and $A^{-1}$ are continuous for the b-convergence.  
It suffices now to apply Thm. \ref{arsi} to $X= \hrr\cup \widehat{\alpha}$, using also Thm.\ \ref{th:main}. 
\qed

\end{document}